\documentclass[a4paper,12pt]{article}
\newcommand{\newsection}[1]{
\vspace{5mm}
\pagebreak[3]
\refstepcounter{section}
\setcounter{subsection}{0}
\begin{flushleft}
{\large\bf \thesection. #1}
\end{flushleft}
\nopagebreak
\medskip
\nopagebreak}

\newcommand{\newsubsection}[1]{
\vspace{5mm}
\pagebreak[3]
\addtocounter{subsection}{1}
\addcontentsline{toc}{subsection}{\protect
\numberline{\arabic{section}.\arabic{subsection}}{#1}}
\noindent{\em 
\thesubsection. #1}
\nopagebreak
\vspace{2mm}
\nopagebreak}
\newcommand{\ud}{\mathrm{d}}

\makeatletter

\def\clap#1{\hbox to 0pt{\hss #1\hss}}%

\def\ligne#1{%
  \hbox to \hsize{%
    \vbox{\centering #1}}}%

\def\haut#1#2#3{%
  \hbox to \hsize{%
    \rlap{\vtop{\raggedright #1}}%
    \hss
    \clap{\vtop{\centering #2}}%
    \hss
    \llap{\vtop{\raggedleft #3}}}}%

\def\maketitle{%
  \thispagestyle{empty}\vbox to \vsize{%
    \@preprint
    \vfill
    \ligne{\Large \@title}
    \vspace{5mm}
    \ligne{\@author}
    \vspace{1mm}\ligne{\texttt{<\@email>}}
    \vspace{5mm}
    \ligne{\@blurb}
    \vspace{1cm}
    \@abstract
    \vfill
    }%
  }

\def\date#1{\def\@date{#1}}
\def\author#1{\def\@author{#1}}
\def\title#1{\def\@title{#1}}
\def\location#1{\def\@location{#1}}
\def\blurb#1{\def\@blurb{#1}}
\def\email#1{\def\@email{#1}}
\def\abstract#1{\def\@abstract{#1}}
\def\preprint#1{\def\@preprint{#1}}

\date{\today}
\author{}
\title{}
\location{Durham}
\blurb{}
\email{no email address}
\makeatother
  \title{Particle Motion in the Rotating Black Ring Metric}
  \author{James Hoskisson}
  \email{James.Hoskisson@durham.ac.uk}
  \date{1st May 2007}
  \preprint{\begin{flushright}
  DCPT-07/17
  \end{flushright}}
  \location{Durham}
  \blurb{
    Department of Mathematical Sciences \\
    University of Durham \\
    Science Laboratories \\
    South Road \\
    Durham \\
    DH1 3LE
    }%
  \abstract{
  \begin{center}
	\bf{Abstract}
	\end{center}
	\vspace{3mm}
In this paper, the equations of motion for geodesics in the neutral rotating Black Ring 	metric are derived and the separability of these equations is considered. The bulk of the paper is concerned with sets of solutions where the geodesic equations can be examined analytically - specifically geodesics confined to the axis of rotation, geodesics restricted to the equatorial plane, and geodesics that circle through the centre of the ring. The geodesics on the rotational axis behave like a particle in a potential well, while the geodesics confined to the equatorial plane mimic those of the Schwarzschild metric. It is shown that it is impossible to have circular orbits that pass through the ring, but some numerical results are presented which suggest that it is possible to have bound orbits that circle through the ring.}

\usepackage{graphicx}	
\usepackage[font={footnotesize,sf},labelfont={bf,rm}]{caption}

\begin{document}

\maketitle

\newsection{Introduction}

In recent years the study of black holes in higher dimensions has gained much greater prominence due primarily to the growing interest in string theory and the associated models where more than three spatial dimensions are considered \cite{cite:Stringref1,cite:Stringref2}. The extra degrees of freedom, due to these additional dimensions, have facilitated a much broader range of possibilities for gravitational phenomena. This is exemplified by the discovery of a black hole with a ring shaped horizon that can only exist when there are more than four dimensions \cite{cite:EmparanReall2002}. This Black Ring solution was also the first solution to Einstein's equations that demonstrated black hole non-uniqueness i.e. a black hole that isn't described solely in terms of its mass and angular momentum.

It has also been suggested that gravitational systems in higher dimensions may be of use in exploring four dimensional phenomena. There are many ideas of current research interest, such as brane-worlds \cite{cite:braneref} and the holographic principle \cite{cite:hologp}, that rely on interpreting the effects of higher dimensional objects to explain apparently unrelated effects in the observable four dimensional world. Much of the most interesting gravitational phenomena occur within the vicinity of black holes, so obtaining a thorough understanding of black hole spacetimes in extended dimensions seems to be an essential step in developing these theories. The study of the geodesics of these black hole metrics gives an important insight into how matter and radiation will behave and is thus useful in developing higher dimensional theories.

The general solution for a spherical stationary axisymmetric black hole of arbitrary dimension was derived by Myers and Perry in \cite{cite:MyersPerry} and it was shown in \cite{cite:FrolovStojkovic} and \cite{cite:VasudevanStevensPage} that the geodesic equations of motion are separable. The separability of the equations of motion allows the radial and angular motion to be considered separately, giving $n$ decoupled equations in terms of the test particle's mass, total energy, angular momenta in the various planes of motion, and some separation constants. Given that the equations can be completely decoupled, it is possible to construct an effective potential in the radial direction and thus determine the allowed regions of motion for different values of the energy and angular momenta independent of the angular variables.

In order to express the equations of motion in a separable form it is necessary to use Boyer-Lindquist coordinates. These coordinates are similar to spherical polar coordinates but are ellipsoidal, rather than spherical, and thus the relationship to Cartesian coordinates is more complicated. The Kerr metric expressed in these coordinates is given by
\begin{equation}
ds^2=-\ud t^2 + \Sigma\left(\frac{\ud r^2}{\Delta}+\ud \theta^2\right)+(r^2+a^2)\sin^2{\theta}\,\ud \phi^2 +\frac{2Mr}{\Sigma}\left(a\sin^2\theta\,\ud\phi - \ud t\right)^2
\end{equation}
where $\Sigma=r^2+a^2\cos^2\theta$ and $\Delta=r^2-2Mr+a^2$. The transformations between these coordinates and Cartesian coordinates are given by: \cite{cite:BoyerLindquist}
\begin{eqnarray}
 t &= &t' \nonumber \\
 x &= &\sqrt{r^2+a^2}\sin{\theta}\cos\left[\phi - \tan^{-1}{\left(\frac{a}{r}\right)}\right] \nonumber \\
 y &= &\sqrt{r^2+a^2}\sin{\theta}\sin\left[\phi - \tan^{-1}{\left(\frac{a}{r}\right)}\right] \nonumber \\
 z &= &r\cos{\theta}
\label{BLtrans}
\end{eqnarray}
where ($t',r,\theta,\phi$) are the Boyer-Lindquist coordinates and $a$ is the angular momentum per unit mass of the Kerr black hole.

\begin{figure}[htbp]
\center
\includegraphics[viewport=50 100 400 540,width=6cm,angle=270,clip]{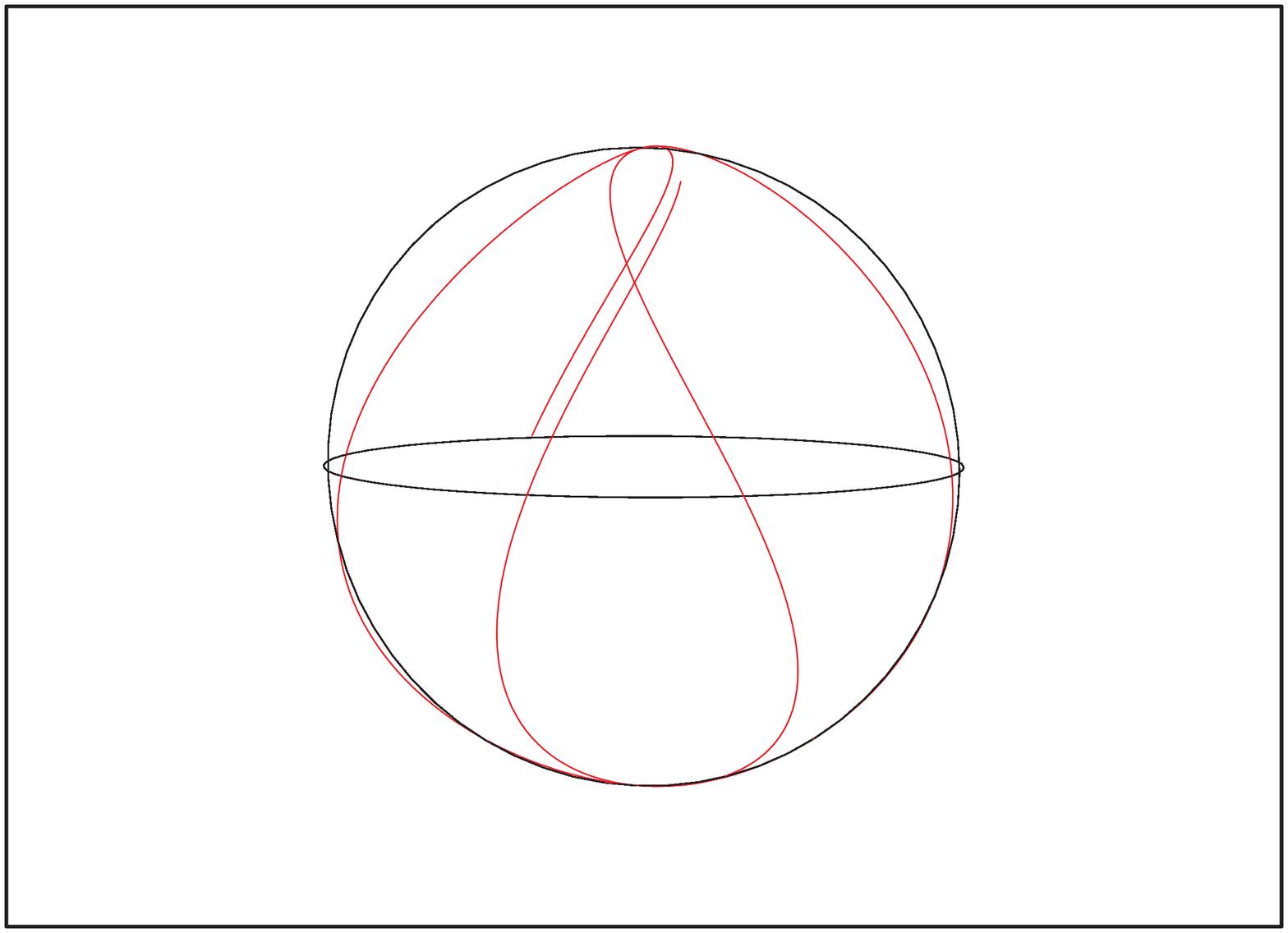}
\includegraphics[viewport=50 100 400 540,width=6cm,angle=270,clip]{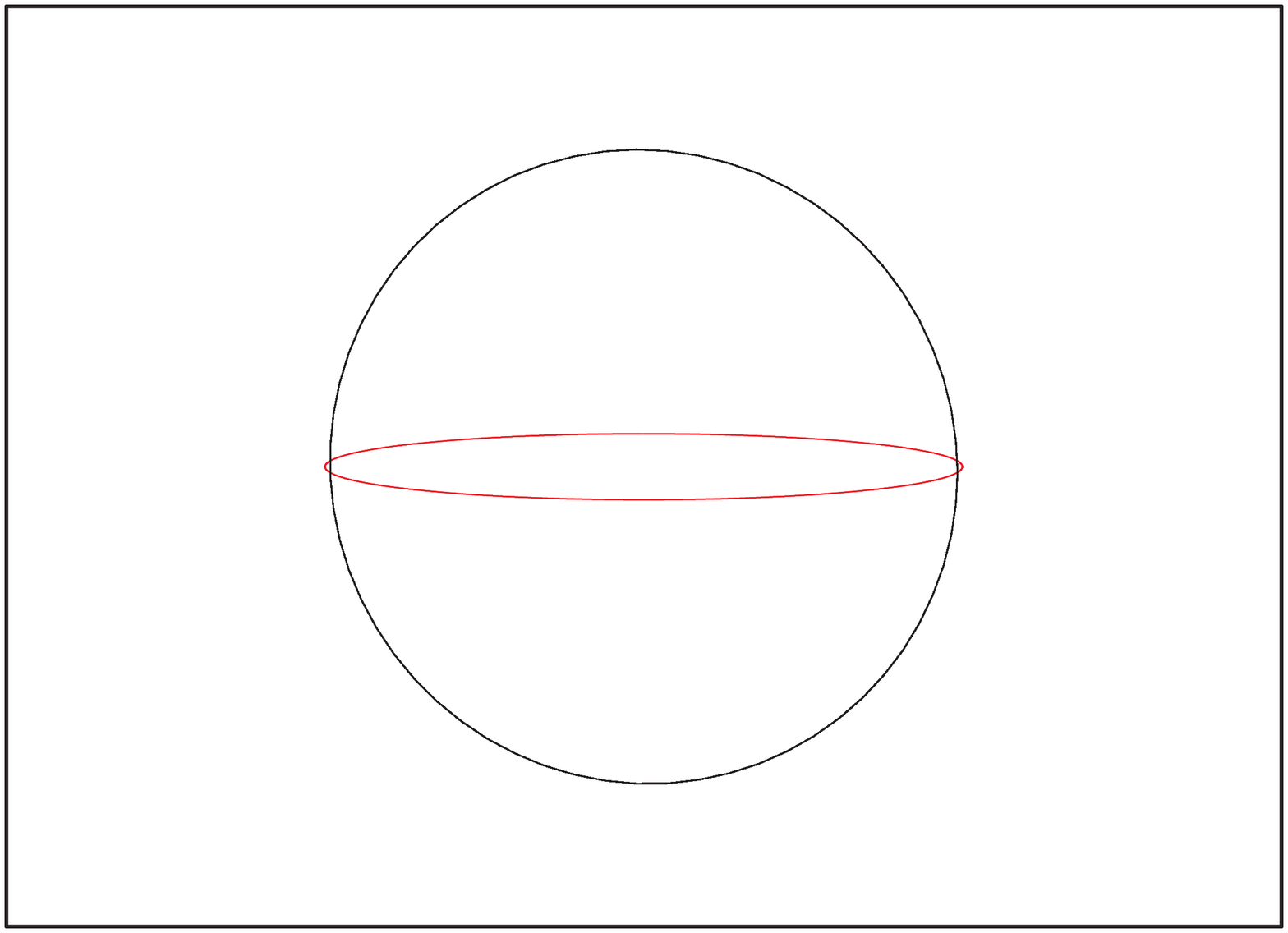}
\caption{The left-hand plot gives an example of the orbit traced out in three dimensions by a photon moving with constant radial coordinate. The right-hand plot shows the orbit when the angular momenta are chosen so that it is confined to a plane.}
\label{fig:photonorbit}
\end{figure}

In this metric it is possible to have timelike and null geodesics that orbit at constant radial distance with the angular coordinates varying. This gives spherical orbits with the radius determined by the associated conserved quantities i.e. the energy, angular momenta, particle mass, and separation constants. An example of this sort of orbit is shown in the left hand plot of figure \ref{fig:photonorbit}. In \cite{cite:Teo}, Teo takes advantage of the separability of the equations of motion to calculate some examples of spherical photon motion for the 4D Kerr black hole. The dynamics of the higher dimensional Kerr metrics are qualitatively similar, so Teo's work on spherical photon orbits could, in principle, be extended to higher dimensional black holes, although one would expect the extra degrees of freedom, afforded by the extra dimensions, to give more complicated classes of motion.

It is also possible to choose the angular momenta so that the geodesic is confined to a single plane of motion at constant radius. This forces the geodesics to orbit within the plane, as in the second plot of figure \ref{fig:photonorbit}. This plot has the separation constant chosen in such a way that any geodesic, with $\theta_0=0$, will orbit in the plane i.e. only $\phi$ will vary with time.

\begin{figure}[htbp]
\center
\includegraphics[viewport=20 20 460 640,width=6cm,angle=270,clip]{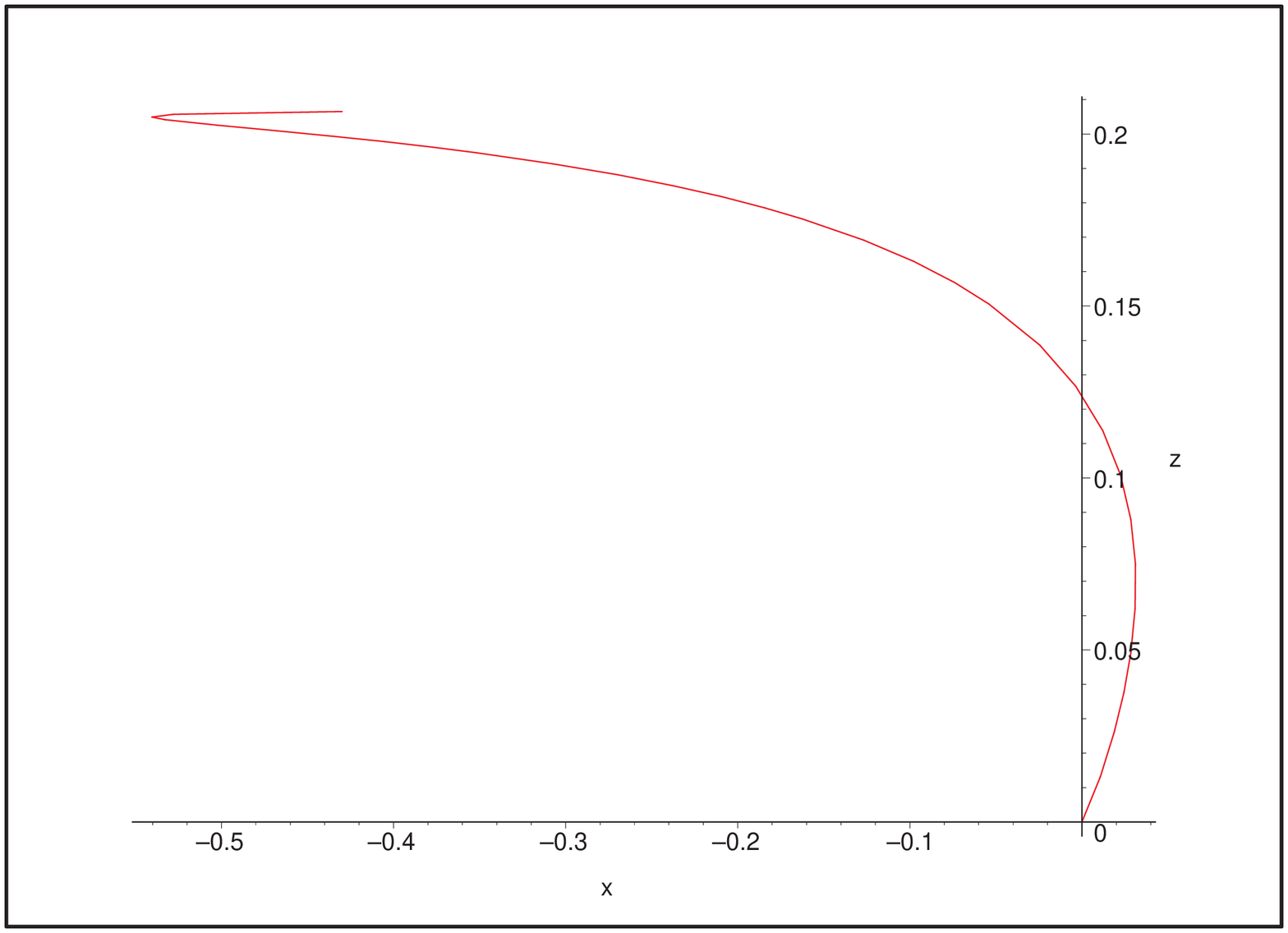}
\includegraphics[viewport=20 20 460 640,width=6cm,angle=270,clip]{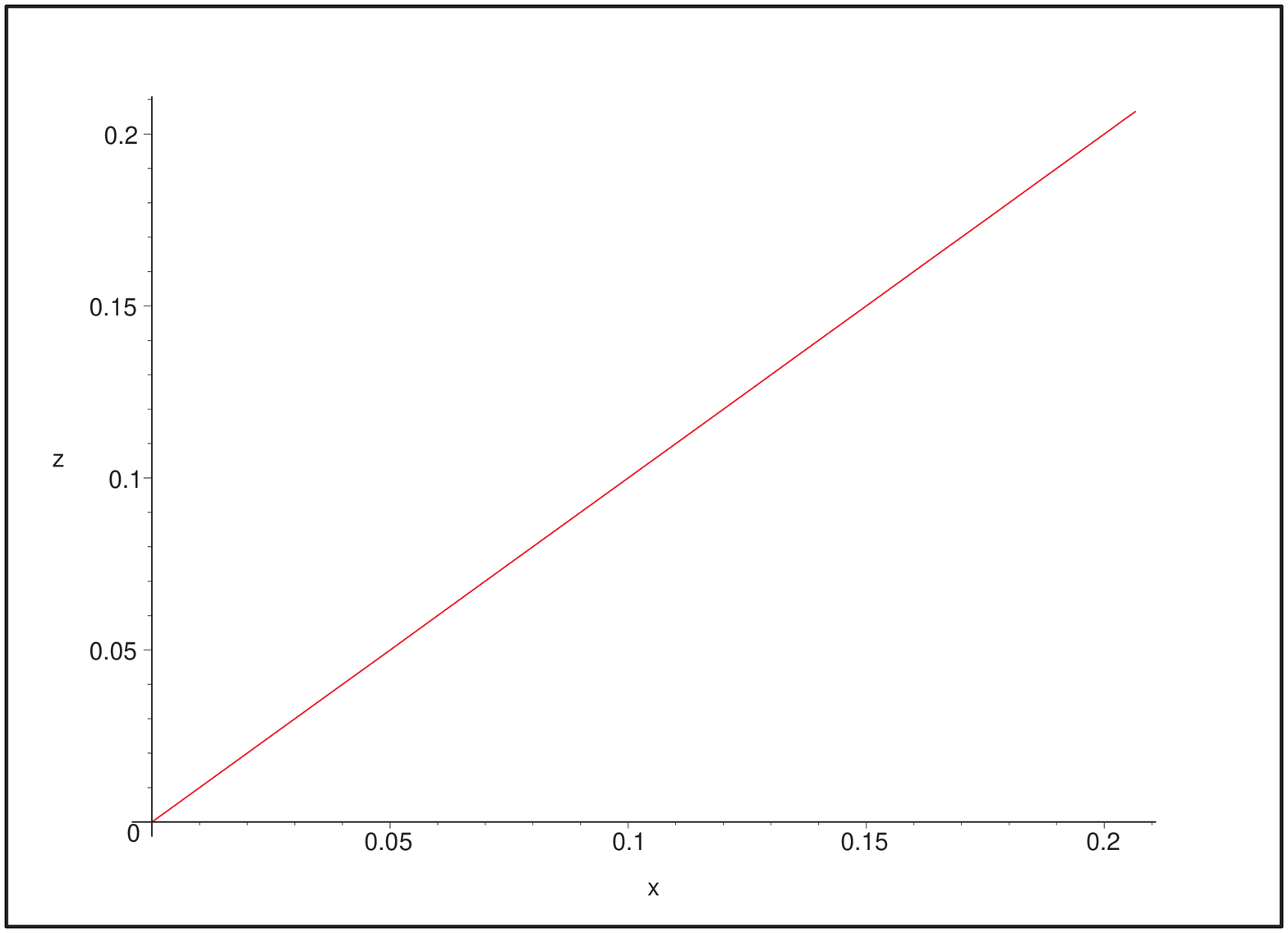}
\caption{An example of particle motion along a ``pseudo-radial'' geodesic in the Kerr metric, with constant $\theta=\frac{\pi}{4}$. The left hand graph plots the projection of a ``pseudo-radial'' geodesic on to a constant $\phi$ cross section in Cartesian coordinates. The right hand graph plots the same geodesic when the $\phi$ variation is suppressed, thus demonstrating that the geodesic is radial.}
\label{fig:radmotion}
\end{figure}

The geodesics of the Kerr metric also demonstrate ``pseudo-radial'' motion, as shown in figure \ref{fig:radmotion}. This is where the test particle moves along a line of constant $\theta$ but $r$ and $\phi$ vary. In some spacetimes it is possible to choose the angular momentum so that $\phi$ is also constant for all values of $r$, but the Lense-Thirring effect precludes this in the Kerr spacetime, hence the term ``pseudo-radial'' to describe these geodesics. The motion of the test particle in the left hand plot of figure \ref{fig:radmotion} appears to describe a curve but this is because the plot is a projection of the geodesic on to a constant $\phi$ cross-section. If the coordinate system is made to rotate with the test particle, so that $\phi$ is effectively constant, then it becomes obvious that the test particle's motion is radial. This is shown in the right hand plot of figure \ref{fig:radmotion}. The angles $\theta_0$ for which radial motion can occur are given by the equation
\begin{equation}
\Theta(\theta_0)=\left.\frac{d\Theta}{d\theta}\right|_{\theta_0} = 0
\end{equation}
where $\Theta(\theta)=0$ is the equation of motion describing the variation of $\theta$ with time. If the energy $E>m^2$, where $m$ is the test particle mass, the geodesics are unbounded and can reach infinity, whilst the geodesics with $E<m^2$ are bounded from above and cannot escape to infinity.

Given all that is known about the geodesics of the Kerr black hole it seems logical to go on and investigate the corresponding situation for Black Rings to see whether any of the properties of the Kerr geodesics are shared by those of the Black Ring. The physical properties of the neutral rotating Black Ring solution have been extensively studied in papers such as \cite{cite:ElvangEmparan2003} and \cite{cite:EmparanReall2002} but very little is known about the geodesics associated with this metric. There has been some rudimentary work in some papers such as \cite{cite:NozawaMaeda}, \cite{cite:SenguptaSahay} and \cite{cite:ElvangEmparanVirmani}, but the geodesic calculations are very much secondary to the other facets under consideration. The calculations in these papers were very restricted in their application, so one would expect more general classes of geodesics to be more complicated.

The Black Ring has horizon topology $S^2\times S^1$, as opposed to the $S^3$ topology of the 5D Kerr horizon, giving greater scope for interesting classes of geodesics. This paper will explore these different situations as well as seeing whether any of the special classes of Kerr geodesics described above can be reproduced in the Black Ring metric. The toroidal nature of the Black Ring means that there won't be any spherical orbits, but the analogous situation, where the geodesic remains at a constant distance from the event horizon, will be examined.

This paper is divided into 6 main sections. The Black Ring metric and some of its properties are investigated in section \ref{ch:Metcoords}. It also describes the toroidal coordinate system for the metric and plots the contour lines for these coordinates. Section \ref{ch:geoequs} derives the geodesic equations for the Black Ring metric and presents the conserved quantities associated with the symmetries of the metric. The remaining sections \ref{ch:axisgeos} through \ref{ch:radgeos} investigate some specific classes of geodesics where the equations of motion become separable.

Section \ref{ch:axisgeos} investigates the case where the geodesics are confined to the rotational axis of the ring. In this case the geodesics are restricted to have angular momentum in only one plane and their motion can broadly be divided into three separate classes. In the first case, when the angular momentum is zero, the geodesics pass through the origin and, in the timelike case, oscillate back and forth through the ring. If the angular momentum is small but non-zero, then both the null and timelike geodesics can be made to oscillate back and forth through the ring. In the final case, where the angular momentum is large, both the timelike and null geodesics are repelled by the Black Ring and thus can only ever be made to pass through the ring once.

Section \ref{ch:plangeos} describes geodesics that orbit at a constant radius in the equatorial plane of the ring. In this case the geodesics can only orbit in the same plane as that of the ring, and maintain a constant radius from the origin. These geodesics can also only have angular momentum in one direction in order for them to remain on the plane. It is shown that there are no geodesics of this type in the interior equatorial plane of the ring, meaning that the only non-trivial geodesics are in the exterior equatorial plane. The geodesics in the exterior plane can be further sub-divided into null and timelike geodesics, with the only valid timelike geodesic being on the ergosurface. There are less constraints for the null geodesics of this type, so it is sometimes possible to have two planar orbits for the same angular momentum. In general the number of solutions for the null curves varies between zero and two, depending on the particular angular momentum of the geodesic.

The \ref{ch:ringgeos}th section looks at the case where the geodesics are restricted to move along circles of constant radius through the ring. These geodesics are constrained so that they follow contour lines that circle around the ring, much like a wire wrapped around a circular solenoid. In this case, the calculations show that there are no physically valid geodesics that execute this motion. However, numerical simulations of the geodesics suggest that it is possible to have bound orbits where the particle's radius varies as it orbits the ring.

The final part, section \ref{ch:radgeos}, calculates which classes of geodesics can perform ``pseudo-radial'' motion. In the Black Ring coordinates general ``pseudo-radial'' lines become curves because the coordinate system is symmetric about the origin, rather than the edge of the $S^1$ which defines the ring. However, the derivation given in this section shows that this particular type of radial geodesic can only exist in the equatorial planes where the geodesics are straight lines.

\newsection{The Metric and Black Ring Coordinates}
\label{ch:Metcoords}

The metric for the rotating Black Ring solution is given in terms of the toroidal coordinates $(t,x,y,\phi,\psi)$, where
\begin{equation}
-\infty\le y\le -1 \hspace{3cm} -1\le x\le 1
\label{xyrange}
\end{equation}
and the metric is given by \cite{cite:EmparanReall2006}
\begin{equation}
ds^2 = -\frac{F(y)}{F(x)}\left(\ud t-CR\frac{1+y}{F(y)}\ud\psi\right)^2 + \frac{R^2F(x)}{(x-y)^2}\left[-\frac{G(y)}{F(y)}\ud\psi^2 -\frac{\ud y^2}{G(y)} +\frac{\ud x^2}{G(x)} + \frac{G(x)}{F(x)}\ud\phi^2\right]
\label{BRmetric}
\end{equation}
where
\begin{equation}
F(\xi)=1+\lambda\xi \hspace{3cm} G(\xi)=(1-\xi^2)(1+\nu\xi)
\end{equation}
and
\begin{equation}
C=\sqrt{\lambda(\lambda-\nu)\frac{1+\lambda}{1-\lambda}}
\end{equation}
$R$ has the dimensions of length and can be interpreted as the radius of the ring. The parameters $\lambda$ and $\nu$ lie in the range
\begin{equation}
0<\nu\le\lambda<1
\end{equation}
and describe the shape and rotation velocity of the ring, as described in \cite{cite:EmparanReall2006}. The ring rotates in the $\psi$ direction and in order for the string-like tension of the ring \cite{cite:ElvangEmparanVirmani} to be balanced by the centrifugal force, $\lambda$ has to be chosen such that
\begin{equation}
\lambda_c=\frac{2\nu}{1+\nu^2}
\end{equation}
All of the examples shown in this paper will consider Black Rings that are in equilibrium, so $\lambda$ will always satisfy the above equation.

In (\ref{BRmetric}) the event horizon is given by $y=-\frac{1}{\nu}$ and there is a spacelike curvature singularity at $y=-\infty$. The Black Ring is also rotating so it has an ergosurface where $\partial_t$ becomes spacelike. The ergosurface is given by $y=-\frac{1}{\lambda}$ and, like the event horizon, has topology $S^2\times S^1$.

The coordinates used in (\ref{BRmetric}) are adapted to the shape of a ring with radius $R$, where $\psi$ parameterises the $S^1$ and $(x,y,\phi)$ parameterise the remaining space. In these toroidal coordinates flat space is given by
\begin{eqnarray}
ds^2=\frac{R^2}{(x-y)^2}\left[(y^2-1)\ud\psi^2+\frac{\ud y^2}{y^2-1}+\frac{\ud x^2}{1-x^2}+(1-x^2)\ud\phi^2\right]
\end{eqnarray}
It isn't immediately obvious that this metric has no curvature, but a change of coordinates allows the metric to be transformed to something more familiar. The above metric can be transformed to polar coordinates using
\begin{equation}
y=-\frac{R^2+r_1^2+r_2^2}{\sqrt{(r_1^2+r_2^2+R^2)^2-4R^2r_2^2}} \hspace{3cm} x=\frac{R^2-r_1^2-r_2^2}{\sqrt{(r_1^2+r_2^2+R^2)^2-4R^2r_2^2}}
\label{polartrans}
\end{equation}
This re-casts the metric in the form
\begin{equation}
ds^2=dr_1^2+r_1^2d\phi^2+dr_2^2+r_2^2d\psi^2
\end{equation}
which describes flat space using two radial coordinates and two angular coordinates. The transformations given in (\ref{polartrans}) give limits on the $(x,y)$ coordinates, as $r_1$ and $r_2$ have to remain real and non-negative. This is why $x$ and $y$ are restricted to the ranges given by (\ref{xyrange}).

To understand how $(x,y)$ span a constant $\psi$ cross-section it is useful to consider the inverse transformations of (\ref{polartrans}) given by
\begin{equation}
r_1=R\frac{\sqrt{1-x^2}}{(x-y)} \hspace{3cm} r_2=R\frac{\sqrt{y^2-1}}{(x-y)}
\label{r1r2trans}
\end{equation}
This allows the three dimensional cross-section of constant $\psi$ to be parameterised by a single radial coordinate and two angular coordinates. Combining the two transformations from (\ref{r1r2trans}) gives
\begin{equation}
r=\sqrt{r_1^2+r_2^2}=\frac{R\sqrt{y^2-x^2}}{x-y}
\label{rtrans}
\end{equation}
and
\begin{equation}
\tan{\theta}=\frac{r_2}{r_1}=\sqrt{\frac{y^2-1}{1-x^2}}
\label{thetatrans}
\end{equation}
which are combined with $\phi$ to span the 3D cross-section of constant $\psi$.

\begin{figure}[htbp]
\center
\includegraphics[viewport=20 100 430 550,width=10cm,angle=270,clip]{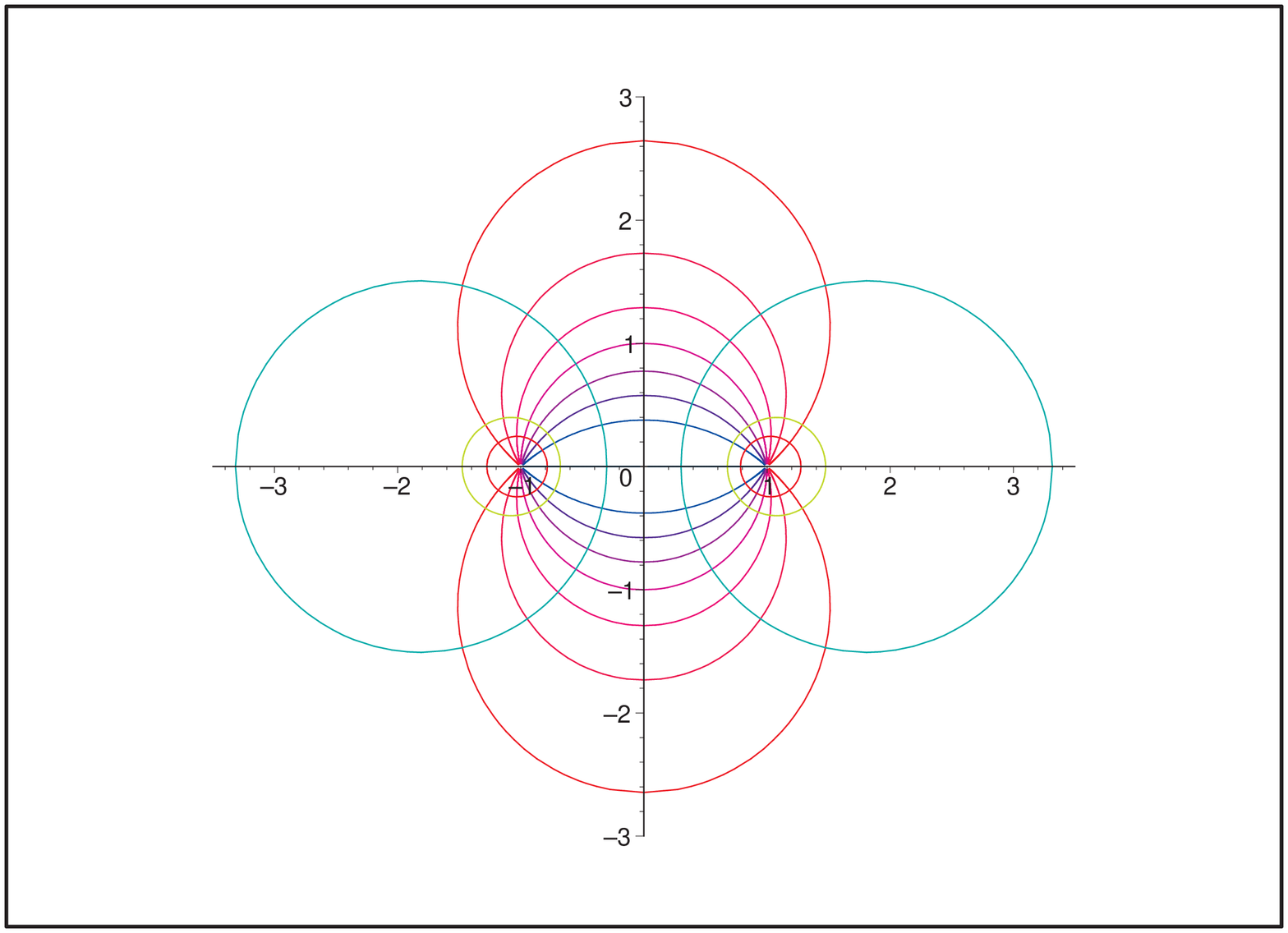}
\caption{A two dimensional cross-section of constant $\phi$ and $\psi$ (as well as the antipodal points $\phi+\pi$ and $\psi+\pi$) of the $(x,y)$ coordinates. The red-turquoise circles (centred on the vertical axis) are lines of constant $y$ and the blue-magenta circles (centred on the horizontal axis) are lines of constant $x$. The vertical axis on this plot corresponds to $y=-1$ and the horizontal axis corresponds to $x=\pm 1$, where $x=+1$ corresponds to the centre of the ring up to the inner edge, and $x=-1$ corresponds to the region from the outer edge of the ring to infinity.}
\label{fig:xyXsec}
\end{figure}

Figure \ref{fig:xyXsec} uses these transformations to show the lines of constant $y$ and $x$ when $\phi$ and $\psi$ are held constant. The $\psi$ coordinate would define the plane coming out of the page, perpendicular to the vertical axis. As can be seen from the plot, the lines of constant $y$ define circles (or spheres when the $\phi$ coordinate is included) that foliate the space, with the $x$ coordinate varying around the circle. The circles get bigger as $y$ increases towards $-1$ with $y=-1$ defining the axis of rotation of the ring, and $x=y=-1$ being equivalent to $r=\infty$ in polar coordinates. It is worth pointing out that the $(x, y)$ coordinates only cover a semi-circle in one of the quadrants, depending on the values of $\phi$ and $\psi$. Figure \ref{fig:xyXsec} plots the contour lines for all four quadrants to emphasise how well adapted the coordinates are to the shape of the Black Ring.

\newsection{Geodesic Equations and Conserved Quantities}
\label{ch:geoequs}

To obtain the geodesic equations of motion, the following lagrangian is formed
\begin{equation}
{\cal L} = \frac{1}{2}\left\{-\frac{F(y)}{F(x)}\left(\dot t-\frac{CR(1+y)}{F(y)}\dot\psi\right)^2 + \frac{R^2F(x)}{(x-y)^2}\left[-\frac{G(y)}{F(y)}\dot \psi^2-\frac{\dot y^2}{G(y)}+\frac{\dot x^2}{G(x)}+\frac{G(x)}{F(x)}\dot\phi^2\right]\right\}
\label{BRlag}
\end{equation}
The conjugate momenta can then be calculated using
\begin{equation}
p_\mu=g_{\mu\nu}\dot x^\nu
\end{equation}

It is obvious from the form of (\ref{BRmetric}) that there are three Killing vectors given by $\partial_t$, $\partial_\psi$, and $\partial_\phi$. This means that the momenta in these directions will be conserved, giving
\begin{eqnarray}
p_t &= -E = & -\frac{F(y)}{F(x)}\dot t + \frac{CR(1+y)}{F(x)}\dot\psi \label{ptequ} \label{tequ} \\
p_\phi &= \;\;\ell \;\; = & \frac{R^2G(x)}{(x-y)^2}\dot\phi \label{phiequ} \\
p_\psi &= \; \Psi \; = & \frac{CR(1+y)}{F(x)}\dot t - \frac{C^2R^2(1+y)^2}{F(x)F(y)}\dot\psi - \frac{G(y)R^2F(x)}{F(y)(x-y)^2}\dot \psi \label{ppsiequ} \label{psiequ}\\
p_x &= \;\; &\frac{R^2F(x)\dot x}{G(x)(x-y)^2} \\
p_y &= \;\; &-\frac{R^2F(x)\dot y}{G(y)(x-y)^2} \label{pyequ}
\end{eqnarray}
where $E$ is the energy, and $\ell$ and $\Psi$ are the constants associated with the angular momenta in the $\phi$ and $\psi$ directions respectively. The momenta in the $x$ and $y$ direction have also been included for future reference.

Unfortunately, these conserved quantities aren't sufficient to allow the equations of motion to be separated immediately. The easiest way to check whether the equations can be separated is to consider the Hamilton-Jacobi equation
\begin{equation}
g^{\mu\nu}p_\mu p_\nu=-m^2
\label{HamiltonJacobi}
\end{equation}
Calculating the inverse metric components from (\ref{BRmetric}) and substituting into (\ref{HamiltonJacobi}) gives
\begin{equation}
Y(y)+X(x)=\frac{F(x)}{(x-y)^2}\left(\frac{E^2F(x)}{F(y)}-m^2\right)
\label{BRHamJac}
\end{equation}
where $Y(y)$ and $X(x)$ are given by
\begin{eqnarray}
Y(y) &= &-\frac{F(y)}{G(y)}\left(\frac{\Psi}{R}-\frac{EC(1+y)}{F(y)}\right)^2 - \frac{G(y)p_y^2}{R^2} \\
X(x) &= &\frac{\ell^2 F(x)}{R^2G(x)}+\frac{G(x)p_x^2}{R^2}
\end{eqnarray}
The right hand side of equation (\ref{BRHamJac}) indicates that the Hamilton-Jacobi equation can't be separated for arbitrary values of the constants of motion but if $E=0$ and $m=0$ then the function on the right hand side will go to zero, allowing the equation to be separated out into terms involving only $x$ and $y$.

These particular values of $E$ and $m$ correspond to null geodesics that don't move relative to the space. In the case of the black ring, this means that the geodesics will rotate with the ring, but won't move in the $x$, $y$, or $\phi$ directions. Null geodesics with $E=0$ aren't physically realisable, but this does provide a way to check whether the numerical solutions to the equations of motion are consistent.

Applying the variational principle to (\ref{BRlag}) gives three equations of motion for the Killing directions as per equations (\ref{tequ})-(\ref{psiequ}). The remaining two equations of motion are calculated by varying with respect to $x$ and $y$ respectively
\begin{eqnarray}
H(x,y)-J(x) &= & \frac{\ell^2(x-y)^2G'(x)}{2R^2G(x)^2} \label{xELeqn} \\
H(y,x)-J(y) &= & \frac{(x-y)^2\left[ECR(1+y)+\Psi F(y)\right]^2}{RF(x)^2G(y)F(y)}\left[\frac{G'(y)F(y)}{2RG(y)} - \frac{EC(1-\lambda)}{ECR(1+y)+\Psi F(y)} \right] \label{yELeqn}
\end{eqnarray}
where
\begin{eqnarray}
H(\zeta,\eta) & = & \frac{R^2F(\zeta)}{(\zeta-\eta)^2} \left[\frac{\ddot \zeta}{G(\zeta)} - \frac{G'(\zeta)\dot \zeta^2}{2G(\zeta)^2} - \frac{[F(\zeta)+F(y)]\dot\zeta^2}{2F(\zeta)G(\zeta)(\zeta-\eta)}\right. \nonumber \\
&& \left. \hspace{2cm} + \; \frac{[F(x)+F(\zeta)]\dot \eta\dot \zeta}{G(\zeta)F(x)(\zeta-\eta)} - \frac{[F(\zeta)+F(y)]\dot\eta^2}{2F(\zeta)G(\eta)(\zeta-\eta)}\right] \\
J(\zeta) &= & \frac{E^2\lambda}{2F(y)} + \frac{x-y}{R^2F(x)}\left[\frac{\left[ECR(1+y)+\Psi F(y)\right]^2[F(y)+F(\zeta)-\lambda(x-\zeta)]}{2F(x)F(y)G(y)} - \frac{\ell^2F(\zeta)}{G(x)}\right]
\end{eqnarray}
These two equations have been expressed in terms of the conserved quantities by substituting for $\dot \phi$, $\dot \psi$, and $\dot t$ from equations (\ref{tequ})-(\ref{psiequ}). In certain circumstances it is also useful to use the first integral of motion, which is given by
\begin{equation}
\frac{R^2F(x)}{(x-y)^2}\left(\frac{\dot x^2}{G(x)}-\frac{\dot y^2}{G(y)}\right)+\frac{\ell^2(x-y)^2}{R^2G(x)}-\frac{E^2F(x)}{F(y)}-\frac{(x-y)^2\left[RE(1+y)C+\Psi F(y)\right]^2}{F(x)F(y)R^2G(y)}=\epsilon
\label{firstint}
\end{equation}
where $\epsilon$ determines the nature of the geodesics as
\[ \epsilon = \left\{
										\begin{array}{rl}
													-1 & \mbox{timelike} \\
													0  & \mbox{null} \\
													+1 & \mbox{spacelike}
										\end{array}
							\right.
\]

As previously mentioned, equation (\ref{BRHamJac}) isn't separable in the general case, so the following four sections consider some special cases where either $x$ or $y$ remain constant throughout the geodesic's motion. These specific cases give the limiting behaviour of geodesics in different parts of the space with varying angular momenta and energy. They can then be used to give a good idea of how the geodesics behave when their initial conditions are similar to any of the cases examined in sections \ref{ch:axisgeos}-\ref{ch:radgeos}. As the initial conditions are varied away from these limiting cases, the behaviour of the geodesics gradually breaks down until the motion is completely dissimilar.

\newsection{Geodesics Along the Rotational Axis of the Ring}
\label{ch:axisgeos}

The geodesic equations, as they are presented in equations (\ref{tequ}-\ref{yELeqn}), are too complicated to analyse straight away. In order to reduce the complexity of the problem, it is necessary to look for certain values of the initial conditions and conserved quantities that simplify the equations. The most obvious way to do this is to look for initial values of $y$ that solve $G(y_0)=0$. The reasoning behind this is easiest to see by multiplying (\ref{firstint}) by $G(y)$ and then choosing $y=y_0$ to be a root of $G(y)$, so that $G(y)\rightarrow 0$. The remaining terms are then given by
\begin{equation}
\frac{R^2F(x)}{(x-y_0)^2}\dot y^2+\frac{(x-y_0)^2\left[RE(1+y_0)C+\Psi F(y_0)\right]^2}{F(x)F(y_0)R^2}=0
\label{yonaxis}
\end{equation}
From this equation it is obvious that the velocity in the $y$ direction, given by $\dot y$, will be zero if the second term is zero. $G(y)$ is already a fully factored cubic function with three real roots, so the solutions of $G(y)=0$ are given by: $y_0=\pm1$,$-\frac{1}{\nu}$. The $y$ coordinate is necessarily constrained such that $-\infty\le y\le -1$, which reduces the possible values for $y_0$ to $-1$ or $-\frac{1}{\nu}$. Fortunately, $y_0=-1$ will cause the second term in (\ref{yonaxis}) to go to zero, provided $\Psi=0$\footnote{It also appears that $\Psi$ and $E$ can be chosen to effect the same outcome for $y=-\frac{1}{\nu}$ but, as will be seen later, $\Psi$ has to be set to zero if $G(y)=0$.}. The line $y=-1$ represents the axis of rotation of the ring, so it is not surprising that the equations of motion become considerably simpler along this line.

Ensuring that the initial value of $\dot y$ is zero will make sure that the geodesic doesn't move away from $y=-1$ immediately but, for the geodesic to remain on the line $y=-1$, $\ddot y$ also has to be zero for all subsequent times. To check that $\ddot y=0$, multiply (\ref{yELeqn}) by $G(y)$ and substitute $\dot y=0$. This gives
\begin{eqnarray} 
&& \frac{F(y)G'(y)(x-y)^2\left[RE(1+y)C+\Psi F(y)\right]^2}{2R^2F(x)G(y)} - \frac{EC(x-y)^2(1-\lambda)[RE(1+y)C+\Psi F(y)]}{RF(x)} \nonumber \\
&& -\frac{R^2F(y)^2F(x)\ddot y}{(x-y)^2} + \frac{(x-y)\left[RE(1+y)C+\Psi F(y)\right]^2[2F(y)-\lambda(x-y)]}{2R^2F(x)}=0
\end{eqnarray}
At first glance it looks like substituting $y=-1$ and $\Psi=0$ will ensure that $\ddot y=0$ but the $G(y)$ factor in the denominator of the first term causes problems because it doesn't cancel with all the terms in the numerator. More specifically, there will be a term of the form\footnote{The following analysis is slightly cavalier. A more detailed analysis of the singularities caused when $y=-1$ is given in Appendix A}
\begin{equation}
\frac{\left[(1+y)+\Psi F(y)\right]^2}{G(y)}
\end{equation}
which will be indeterminate when $y=-1$, due to $G(y)\rightarrow 0$. To take the limit as $y\rightarrow -1$, it is necessary to express $G(y)$ explicitly as $G(y)=(1-y)(1+y)(1+\nu y)$. Expanding the numerator as well gives
\begin{equation}
\frac{(1+y)^2}{(1-y)(1+y)(1+\nu y)} + \frac{2(1+y)\Psi F(y)}{(1-y)(1+y)(1+\nu y)} + \frac{\Psi^2 F(y)^2}{(1-y)(1+y)(1+\nu y)}
\label{yaxissing}
\end{equation}
It is now obvious that the first two terms will go to zero in the limit as $y\rightarrow -1$ but the only way to ensure that the third term doesn't blow up is to define $\Psi=0$. Since, $y=-1$ corresponds to the axis of rotation of the ring, one would expect that $\Psi$ would have to be zero because the angular momentum of the particle is zero when it is on the axis. This can be quickly verified by substituting $y=-1$ into (\ref{ppsiequ}).

Equation (\ref{yaxissing}) also explains why $y=-\frac{1}{\nu}$ can't be used as an initial condition because, from (\ref{yonaxis}), $\dot y$ can't be made to go to zero while $\Psi=0$, unless $y=-1$. Furthermore, the line $y=-\frac{1}{\nu}$ corresponds to the event horizon, so this possibility can be excluded on physical grounds.

Having ensured that $\dot y=0$ for all motion along the axis of rotation, it is now much simpler to calculate how the geodesic varies in the $x$ direction. The $x$ evolution is calculated by substituting $\dot y=0$ and $y=-1$ into (\ref{xELeqn}) and then integrating it numerically. Before doing that, it is helpful to calculate an effective potential for the motion along the axis to get some idea of the allowed motion.

To calculate the effective potential substitute $\dot y=0$ and $y=-1$ into (\ref{firstint}). Rearranging and expressing in terms of $p_x$ now gives
\begin{equation}
p_x^2+\frac{\ell^2F(x)}{G(x)^2}-\frac{\epsilon R^2F(x)}{(x+1)^2G(x)} -\frac{E^2R^2F(x)^2}{(x+1)^2G(x)(1-\lambda)}=0
\label{axispot}
\end{equation}
This form of the equation can be compared with the equation of motion for a classical particle, with unit mass, in a one dimensional potential i.e.
\begin{equation}
\frac{1}{2}p_x^2+V(x)-{\cal E}=0
\end{equation}
where ${\cal E}$ is the total energy of the particle. In this case the effective potential $V(x)$ can be found by solving for $\cal E$ when $p_x=0$ i.e. when the total energy is the same as the effective potential. Equation (\ref{axispot}) is not quite of this form, since the equation is quadratic in $E$, but it is possible to construct an effective potential in an analogous way by setting $p_x=0$ and then solving for $E$ to find two solutions $V_\pm(x)$. The motion of a particle in this potential is now possible only when $E\ge V_+(x)$ or $E\le V_-(x)$.

The effective potential for (\ref{axispot}) can now be calculated, giving
\begin{equation}
V_\pm(x)=\pm\sqrt{\frac{\ell^2(x+1)^2(1-\lambda)}{G(x)R^2F(x)}-\frac{\epsilon (1-\lambda)}{F(x)}}	
\label{yeffpot}
\end{equation}
In the following it is assumed that $E\ge 0$ so, in this case, the only relevant potential is $V(x)\equiv V_+(x)$.

The position of the turning points in the potential is given by the solution to
\begin{equation}
\frac{dV(x)}{dx}=0
\label{ytrnpt}
\end{equation}
The general form of this equation when $R=1$ and $\lambda=\lambda_c$ is
\begin{equation}
Sx^4 + Tx^3 + Ux^2 + Vx + W = 0
\label{barrierequ}
\end{equation}
where
\begin{eqnarray}
S & =& 2\epsilon\nu^3 \\
T & =& (4\ell^2\nu^2+4\epsilon\nu^2-4\epsilon\nu^3) \\
U & =& (4\ell^2\nu^2+3\ell^2\nu+\ell^2\nu^3+2\epsilon\nu^3+2\epsilon\nu-8\epsilon\nu^2) \\
V & =& (4\epsilon\nu^2+6\ell^2\nu+2\ell^2\nu^3-4\epsilon\nu-4\ell^2\nu^2) \\
W & =& (2\ell^2+2\ell^2\nu^2-\ell^2\nu^3-3\ell^2\nu+2\epsilon\nu)
\end{eqnarray}
In general this is a quartic equation so it is best to solve it for specific values of $\nu$ and $\ell$. It reduces to a cubic for $\epsilon=0$ but the general solution is still too cumbersome to manipulate algebraically. Solving (\ref{barrierequ}) gives the value of $x$ for which a test particle will remain stationary. To calculate the minimum energy a particle can have, the solution to equation (\ref{barrierequ}) has to be substituted back into (\ref{yeffpot}).

\newsubsection{Timelike Geodesics on the Rotational Axis}

\begin{figure}[htbp]
\begin{center}
\includegraphics[viewport=0 5 450 550,width=7cm,angle=270,clip]{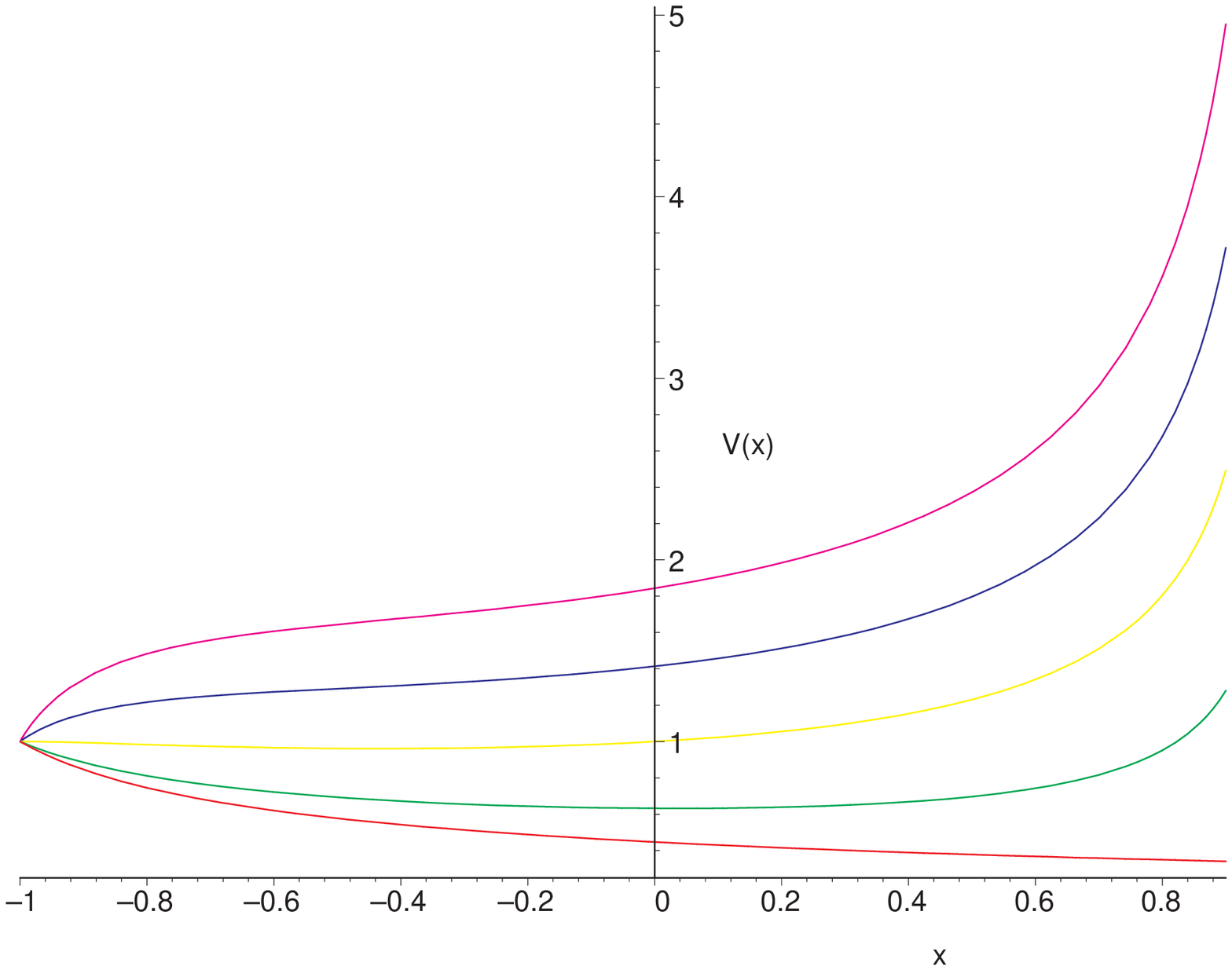}
\includegraphics[viewport=0 5 450 550,width=7cm,angle=270,clip]{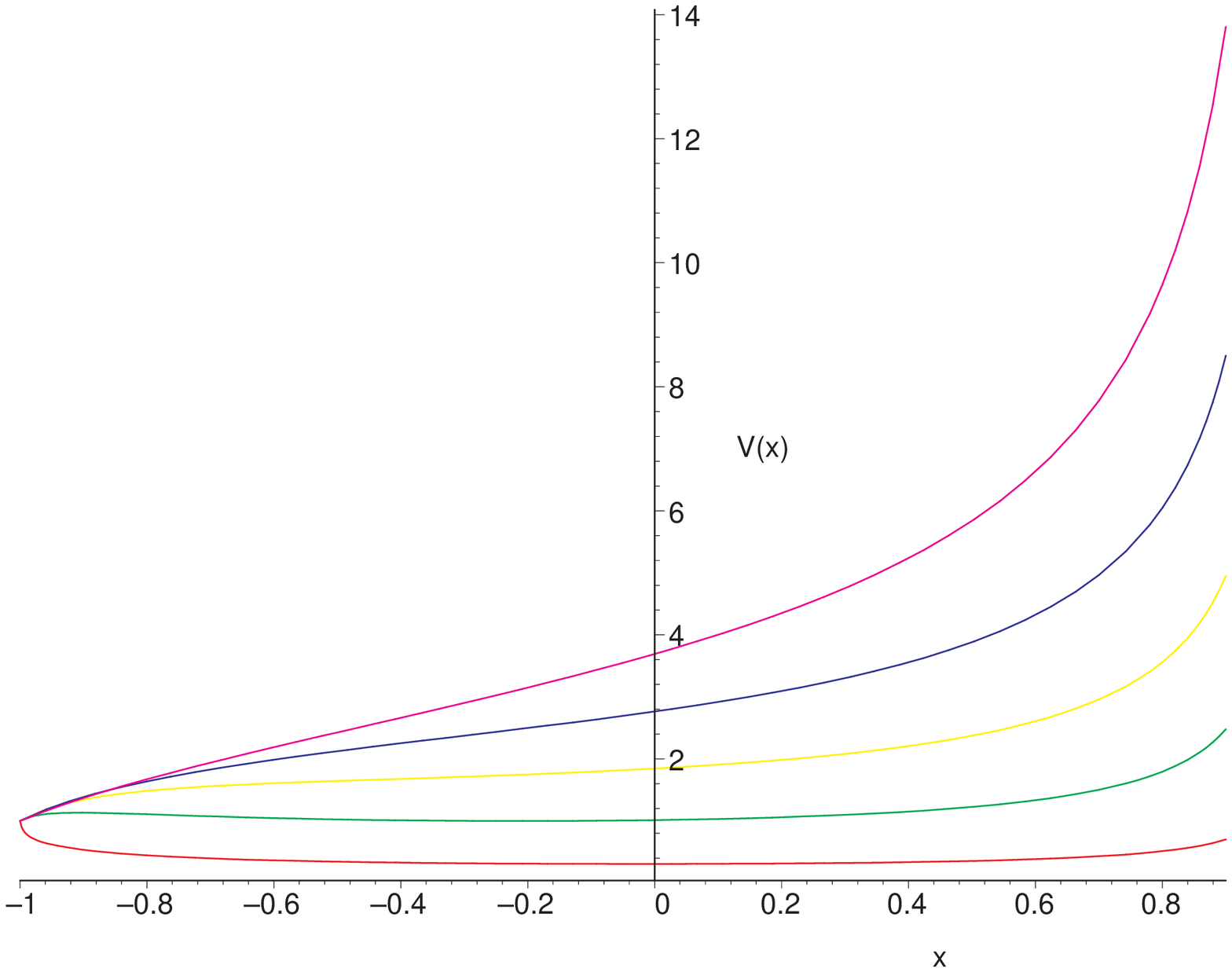}
\caption{The left hand plot shows the effective potential for timelike geodesics with $\ell=0,1,2,3,4$, when $\nu=0.5$. The lowest curve corresponds to $\ell=0$ and $\ell$ increases with each consecutive curve. The right hand plot show how the same potential varies for $\nu=0.1,0.3,0.5,0.7,0.9$, when $\ell=4$. The lowest curve corresponds to $\nu=0.9$ and $\nu$ decreases with each consecutive curve. $R=1$ in both of these plots.}
\label{fig:AMypot}
\end{center}
\end{figure}

Having calculated the effective potential and its turning points, it is now simple to deduce the shape of the geodesics on the $y=-1$ axis. Substituting values for $\ell$ and $\nu$ into (\ref{yeffpot}) gives the effective potentials shown in figure \ref{fig:AMypot}. These plots show that, in general
\begin{itemize}
\item the geodesics with low angular momentum penetrate further toward the origin at $x=1$.
\item the potential goes from being attractive at $\ell=0$ to wholly repulsive for large values of $\ell$.
\item the potential has a local minimum when $\nu$ is large.
\item there is a local maximum for certain values of $\nu$.
\item when $\ell\ne 0$, the centrifugal barrier is infinite at $x=1$.
\end{itemize}

\begin{figure}[htbp]
\begin{center}
\includegraphics[viewport=0 0 450 550,width=7cm,angle=270,clip]{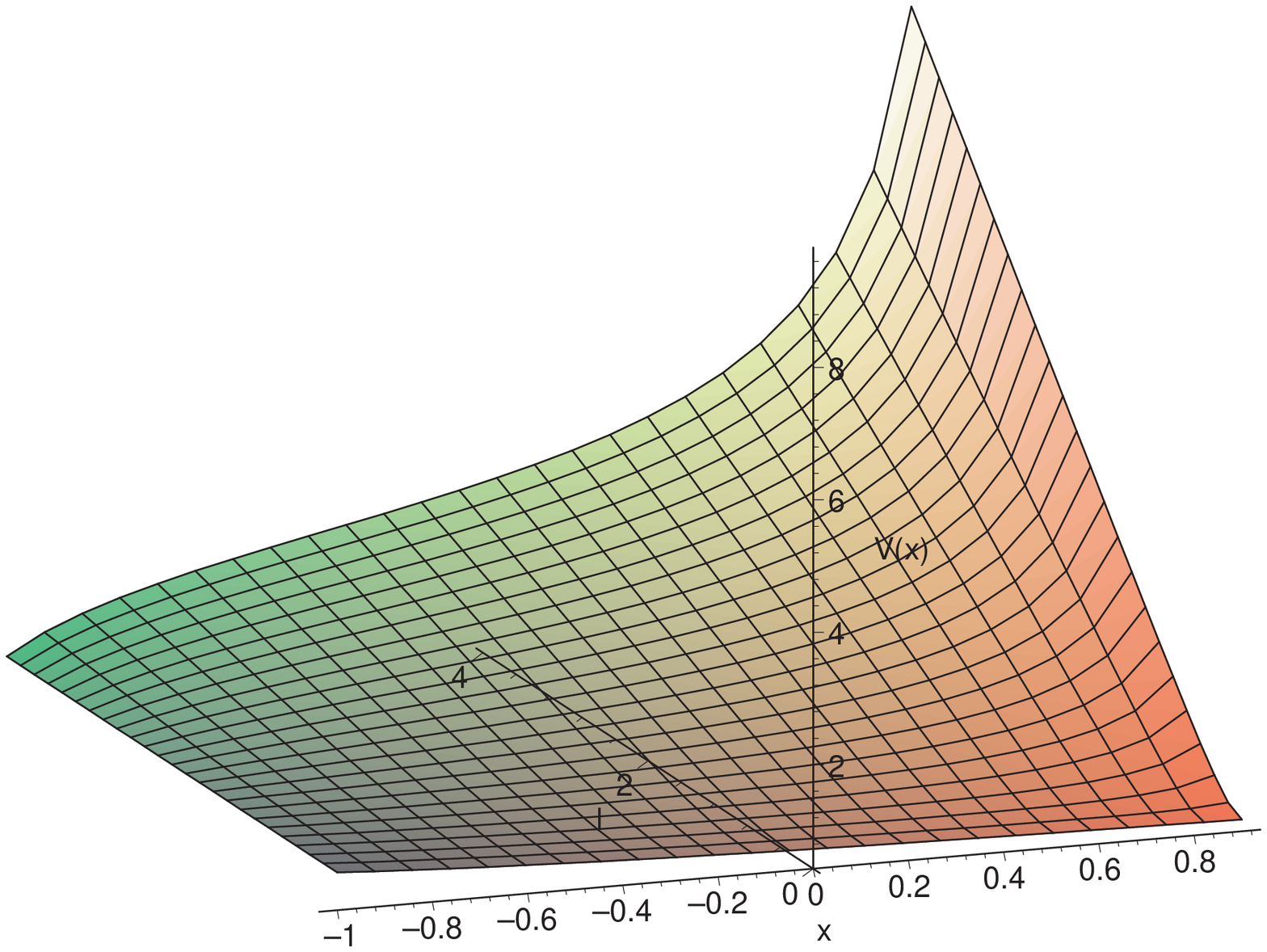}
\includegraphics[viewport=0 0 440 550,width=7cm,angle=270,clip]{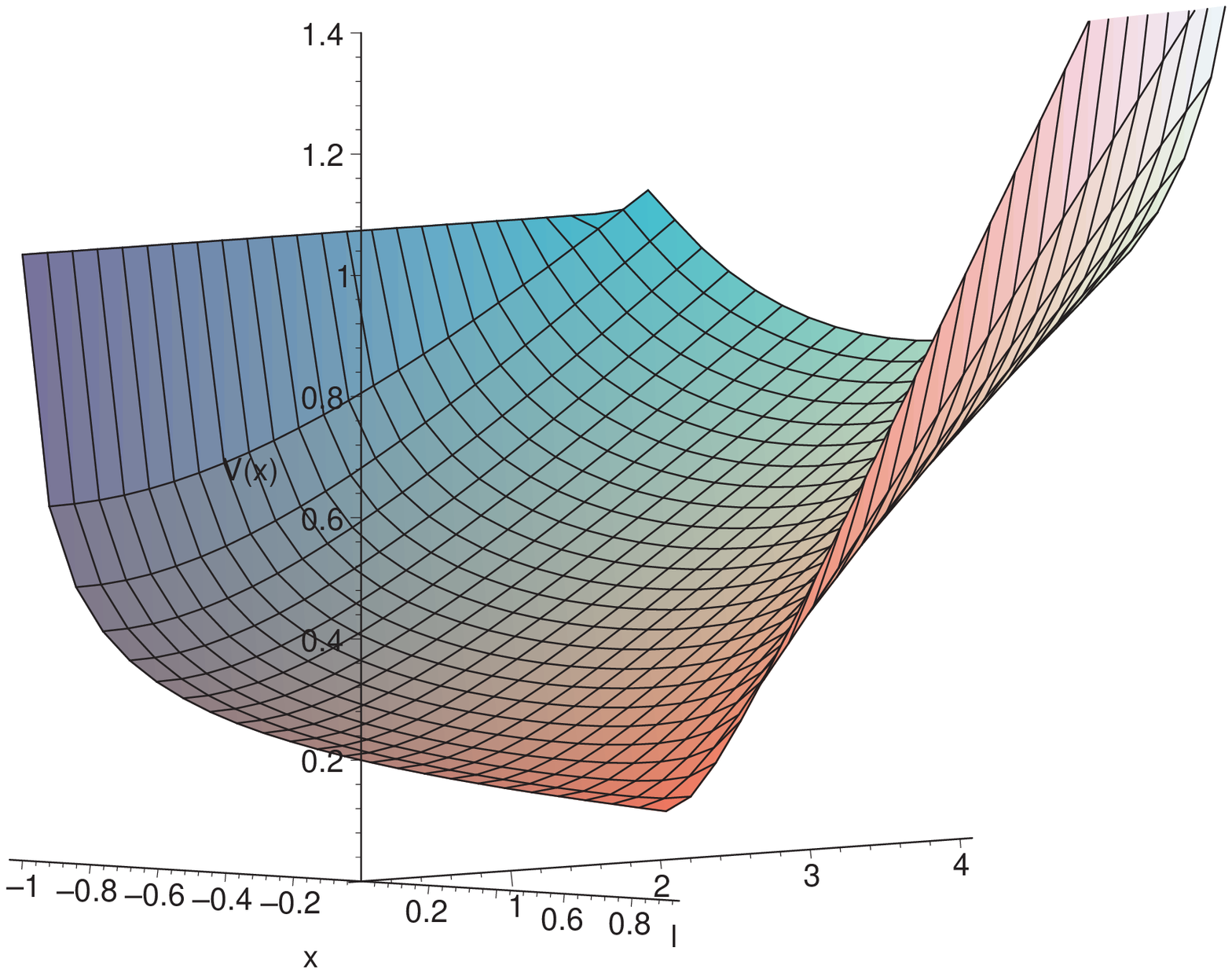}
\caption{3D plots showing the variation of the timelike effective potential with $\ell$ and $x$ for $\nu=\frac{1}{4}$ and $\nu=\frac{3}{4}$ respectively. In both of these plots $R=1$.}
\label{fig:AMypot3d}
\end{center}
\end{figure}

Figure \ref{fig:AMypot3d} shows how the effective potential varies for timelike geodesics when the angular momentum is varied between 0 and 4 for two different values of $\nu$. In the case of $\nu=\frac{1}{4}$ and $R=1$, the potential is initially attractive for $\ell=0$ but as $\ell$ is increased the centrifugal barrier at $x=1$ becomes infinite and then widens toward $x=-1$. This also shifts the minimum toward $x=-1$ and makes the well shallower. This process continues until the centrifugal barrier cancels out the potential well completely and the potential becomes repulsive for all values of $x$. Although the centrifugal barrier, on the right of the plots in figure \ref{fig:AMypot}, widens with $\ell$, $V(-1)=1$ for all values of $\ell$ and $\nu$. This means that the potential well can usually only trap particles with $E<1$ but, as can be seen from the blue line in the right hand plot of figure \ref{fig:AMypot}, for some values of $\nu$ there is a local maximum near $x=-1$. This maximum is more apparent in the right hand plot of figure \ref{fig:AMypot3d} where a ridge appears at $l\approx 3.5$ and $x\approx -0.9$. This maximum only exists for certain values of $\nu$ and $l$, which can be determined for balanced rings by analysing (\ref{yeffpot}) when $\epsilon=-1$, and $\lambda=\lambda_c$.

The limits on $\ell$ are in general dependent upon $\nu$, so the upper limit of $\ell$ is given by
\begin{equation}
\ell_+=\frac{R(\Upsilon^2+\sqrt{2}\Upsilon+2)\left[1-\nu(\Upsilon^2+\sqrt{2}\Upsilon+1)\right]}{\sqrt{\frac{1}{2}(\nu^2-32\nu+3)\Upsilon^4-2\nu\Upsilon^6-6\sqrt{2}\nu\Upsilon^5+\sqrt{2}(\nu^2-12\nu+3)\Upsilon^3+(\nu^2-8\nu+3)\Upsilon^2-\frac{1}{\nu}(\nu-1)^3}}
\label{maxell}
\end{equation}
where $\Upsilon^3=\sqrt{2}(\nu-1)\nu^{-1}$. The lower limit on $\ell$ is given by
\begin{equation}
\ell_-=\frac{2R\sqrt{\nu}}{\sqrt{1-\nu}}
\label{minell}
\end{equation}
Both of these equations assume that $\ell$ is positive but it is always possible for $\ell$ to be negative, in which case the lower and upper limits swap over and both acquire an overall minus sign.

Having found the limits on $\ell$ it is necessary to consider the possible values of $\nu$ that produce a local maximum in the potential. Considering the roots of (\ref{barrierequ}) shows that the minimum value that $\nu$ can take, for which there can be a local maximum in the potential, is given by $\nu=\frac{1}{3}$. Below this value it is impossible to have a local maximum near $x=-1$ and the potential will increase continuously with increasing $x$. Plugging $\nu=\frac{1}{3}$ into (\ref{maxell}) and (\ref{minell}) shows that there is only one possible value of $\ell$ at this point given by $\ell_+=\ell_-=\sqrt{2}$.

One other interesting feature of equation (\ref{maxell}) is that the denominator becomes imaginary for $\nu>0.654$. This indicates that the potential will always have a local maximum for all values of $\ell$, so all rings with $\nu>0.654$ will be able to capture particles along the axis of rotation with $V_{max}>E>1$, no matter how large their angular momentum. For rings with $\nu<0.654$, increasing $\ell$ will eventually smooth out the potential well and cause the potential to be continuously increasing with increasing $x$.

\begin{figure}[htbp]
\begin{center}
\includegraphics[width=6.2cm,angle=270]{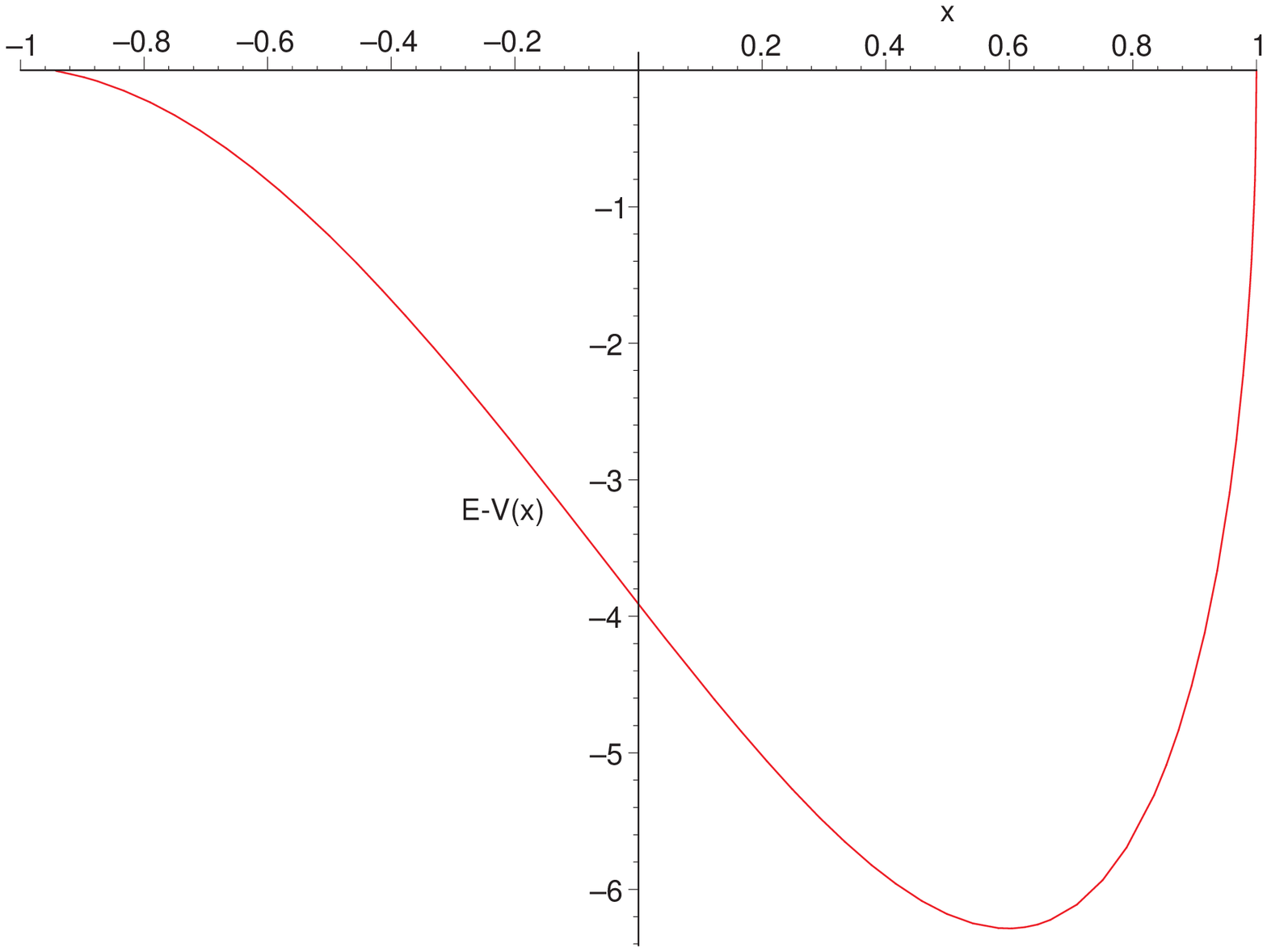}
\includegraphics[width=6.2cm,angle=270]{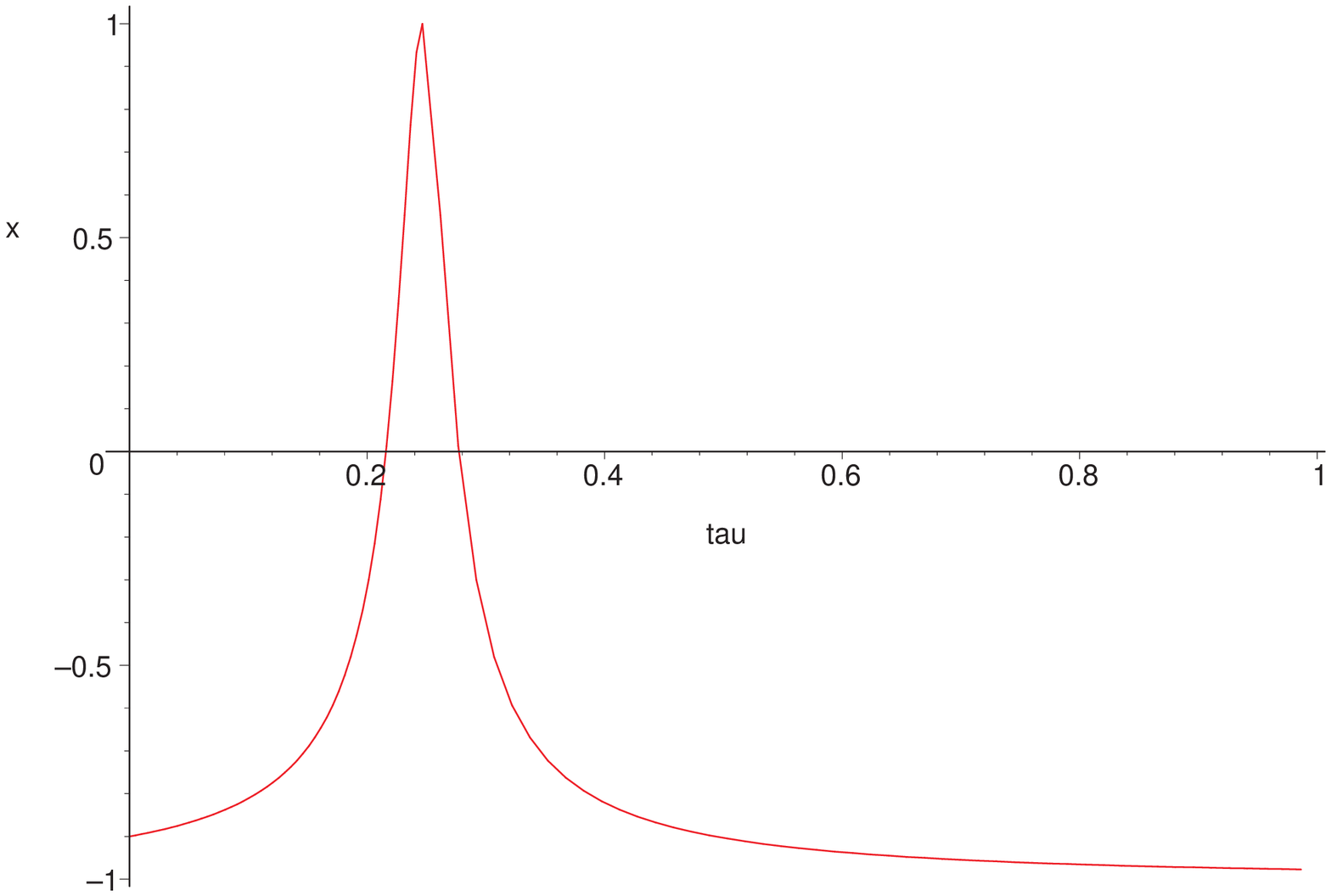}
\caption{The plot on the left shows $E-V(x)$ for a timelike geodesic with $\ell=0$, $\nu=0.9$ and $R=1$. The right hand plot shows the evolution of $x$ with $\tau$ when the particles are started at $x_0=-0.900$ with $E=2$. The initial velocity is chosen so that $\dot x=0.508$.}
\label{fig:0AMygeo}
\end{center}
\end{figure}

Figure \ref{fig:0AMygeo} shows plots of $E-V(x)$ vs $x$ and also $x$ vs $\tau$, where $\tau$ is the affine parameter, for a timelike geodesic with $\ell=0$. The left hand plot is now dependent on the energy and gives a better indication of how the velocity of the test particle changes as it moves along the path, particularly at $x=+1$. The allowed region of motion is now given by the area under the $x$ axis, so the plot indicates that the test particle will approach $x=1$, pass through the origin, and then continue out to infinity at $x=-1$. This behaviour is confirmed by the numerical simulation shown in the right hand plot.

The numerical simulation has a discontinuity at $x=1$ i.e. at the origin. This is a consequence of the coordinate system, since there is a singularity in the equations of motion at $x=1$ and $y=-1$. Naively substituting $x=+1$ into (\ref{xELeqn}) causes some problems because the terms which have $G(x)^2$ in the denominator will blow up but, so long as $\ell=0$, it can be shown that these terms are zero when transformed into polar coordinates. The fact that $\ell$ has to be zero for the particle to go through the origin is obvious when one considers that $\ell$ measures the angular momentum around the axis perpendicular to $y=-1$, so any point on this axis (including the origin) will automatically have zero angular momentum.

Unfortunately, at the origin, the transformations to polar coordinates become undefined, meaning that the previous analysis isn't valid and a further coordinate transformation is required. To analyse the behaviour of the geodesics at the origin it is necessary to transform to Cartesian coordinates, for which the transformations are given in Appendix A. These transformations show that for Cartesian coordinates, given by $(z_0,z_1)$, ${\dot z}_0\not\rightarrow 0$ when $\dot x\rightarrow 0$ at the origin. This means that the test particle is still moving, even though $\dot x$ appears to be zero, so the particle will pass through the origin and out into the other side of the ring. On the other side of the ring, the potential is exactly the same but the particle is moving in the opposite direction in the potential, so $\dot x$ becomes negative.

Rearranging (\ref{xELeqn}) to give $\ddot x$ in terms of the other quantities and substituting $x=1$, $y=-1$, $\ell=0$, and $\dot y=0$ leaves\footnote{Although $\dot y$ is technically zero at the origin, the term in question is $\frac{\dot y^2}{G(y)}$, so slightly more care has to be taken. It is shown in Appendix A that choosing $\frac{\dot y^2}{G(y)}=0$ is equivalent to ensuring that the particle remains on the rotational axis.}
\begin{equation}
\ddot x=-\frac{(\nu+1)\dot x^2}{G(1)}+\frac{[(1+\lambda)+(1-\lambda)]\dot x^2}{4(1+\lambda)}
\label{ddotxorig}
\end{equation}
Transforming $\frac{\dot x^2}{G(x)}$ to Cartesian coordinates and taking the appropriate limit, gives
\begin{equation}
\frac{\dot x^2}{G(1)}=\lim_{z_0\rightarrow 0}{\left[\lim_{z_1\rightarrow 0}{\frac{\dot x^2}{G(x)}}\right]}=\frac{4{{\dot z}_0}^2}{R^2(1+\nu)}
\label{xdotgxcart}
\end{equation}
The line $z_1=0$ corresponds to the rotational axis $y=-1$, where $z_0$ parameterises points along this line and $z_0=z_1=0$ gives the origin. Transforming $\dot x$ into these coordinates and taking the limit as $z_1\rightarrow 0$ gives
\begin{equation}
\lim_{z_1\rightarrow 0}{\dot x}=-\frac{4R^2z_0{\dot z}_0}{(R^2+{z_0}^2)^2}
\label{xdotcart}
\end{equation}
It is obvious from this expression that $\dot x\rightarrow 0$ as $z_0\rightarrow 0$ but there is no requirement that ${\dot z}_0\rightarrow 0$. This shows that even though $\dot x\rightarrow 0$, it doesn't necessarily mean that the test particle is at rest. The fact that $\dot x=0$ when $x=1$ is purely an artefact of the Black Ring's toroidal coordinate system.

Using (\ref{xdotgxcart}) and (\ref{xdotcart}) to express (\ref{ddotxorig}) in Cartesian coordinates gives
\begin{equation}
\ddot x=-\frac{4{{\dot z}_0}^2}{R^2}
\end{equation}
where the second term in (\ref{ddotxorig}) goes to zero because $z_0=0$ at the origin. This shows explicitly that the point at the origin is just a coordinate singularity and that the test particle passes through it without anything out of the ordinary happening.

The centrifugal barrier exhibited in the potential plots of figure \ref{fig:AMypot} shows that it is possible to have low energy geodesics that oscillate back and forth along the rotational axis. When the angular momentum $\ell$ is non-zero, the geodesics never reach the origin at $x=1$, meaning that the particles will orbit in the $x$-$\phi$ plane, with the minimum and maximum distances away from the origin determined by the potential barriers on the right and left of the potential respectively. The ability of the geodesics to move in the $\phi$ direction means that the geodesics still pass through the centre of the ring but don't reach the origin. This is because the rotational axis $y=-1$ is actually a plane when the particles are allowed to move in the $\phi$ direction, thus allowing the particles to pass through the ring without going through the origin where $x=1$.

\begin{figure}[htbp]
\begin{center}
\includegraphics[viewport=5 2 450 530,width=4.7cm,angle=270,clip]{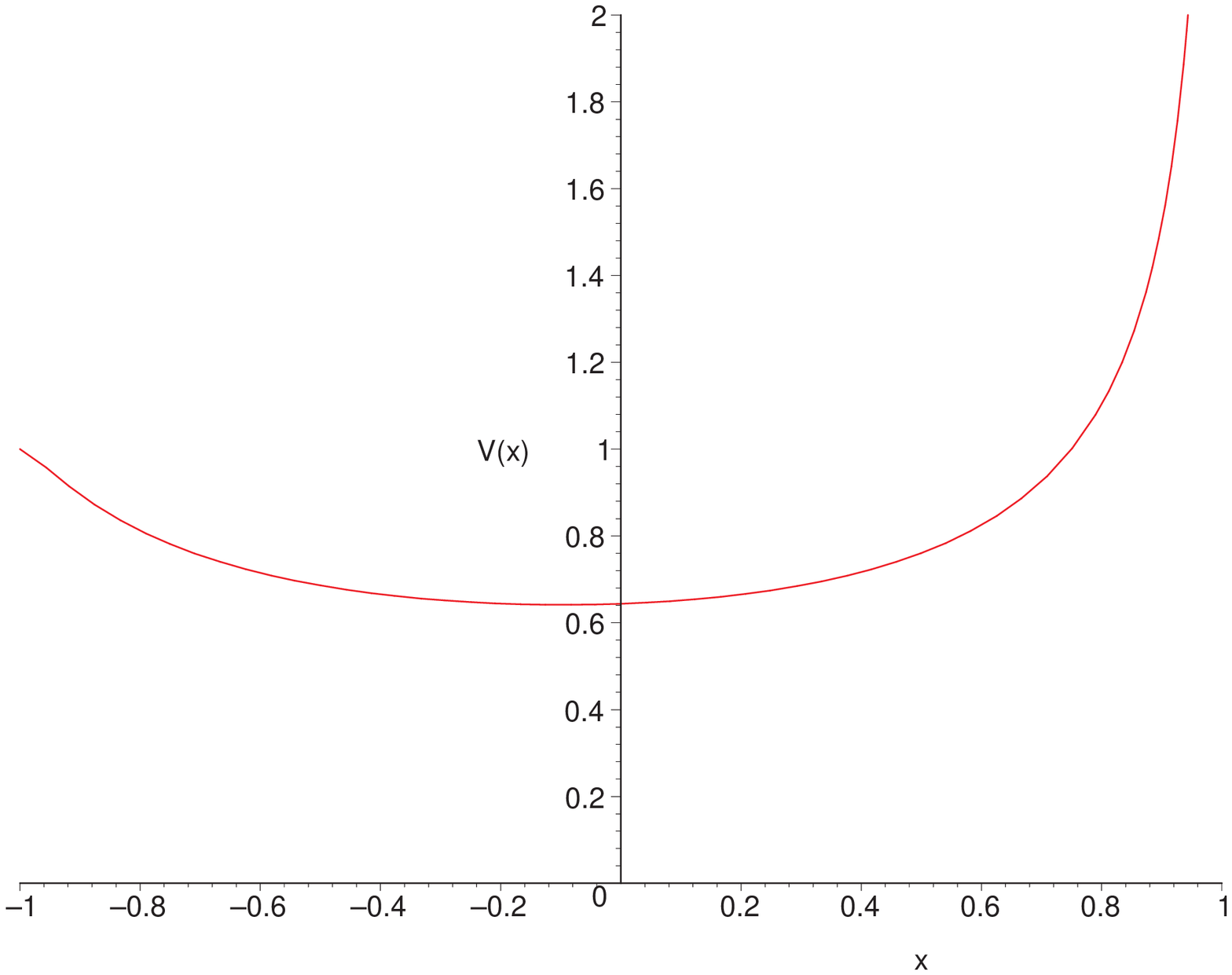}
\includegraphics[viewport=5 2 450 570,width=4.7cm,angle=270,clip]{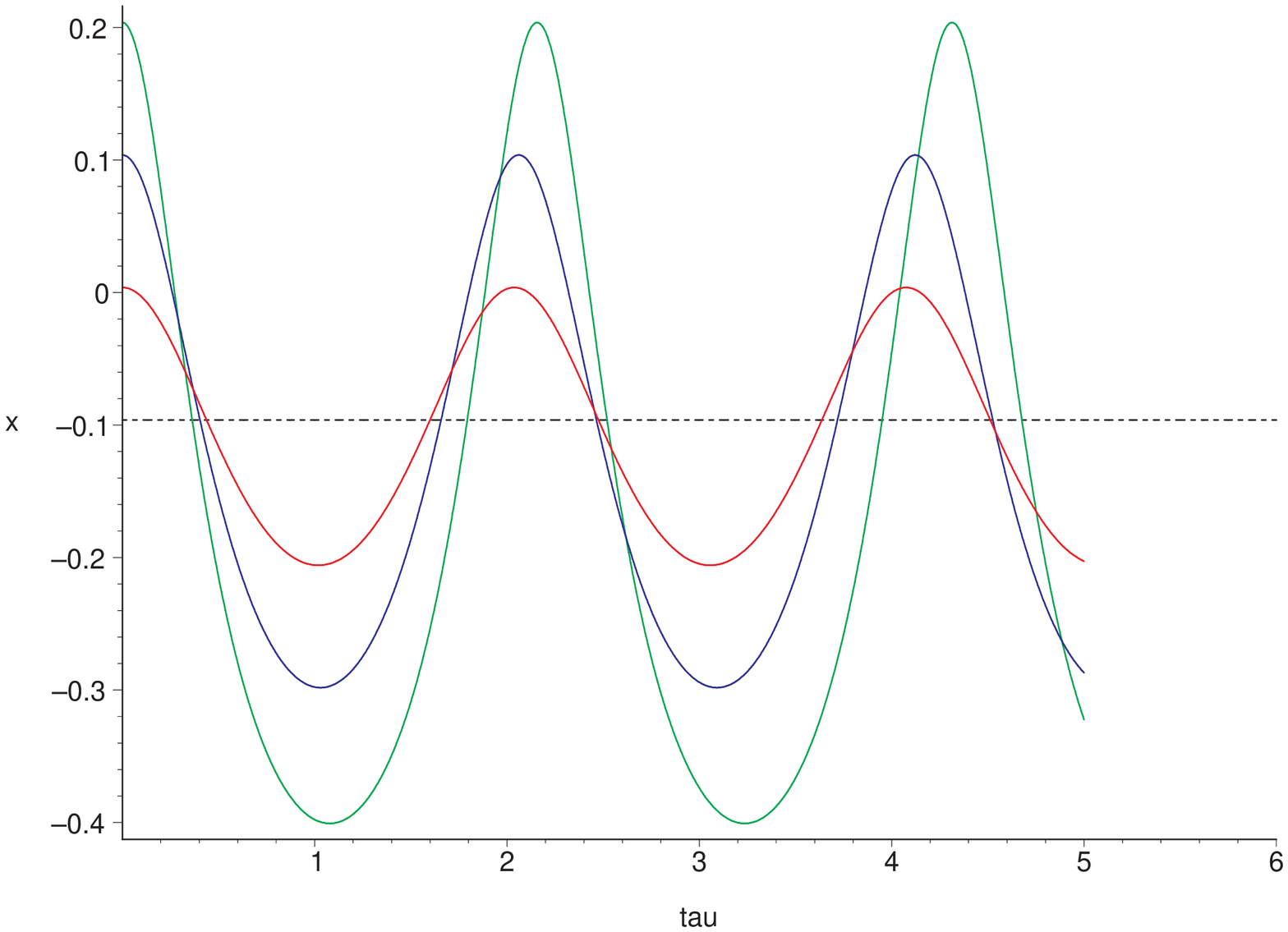}
\includegraphics[viewport=5 2 450 570,width=4.7cm,angle=270,clip]{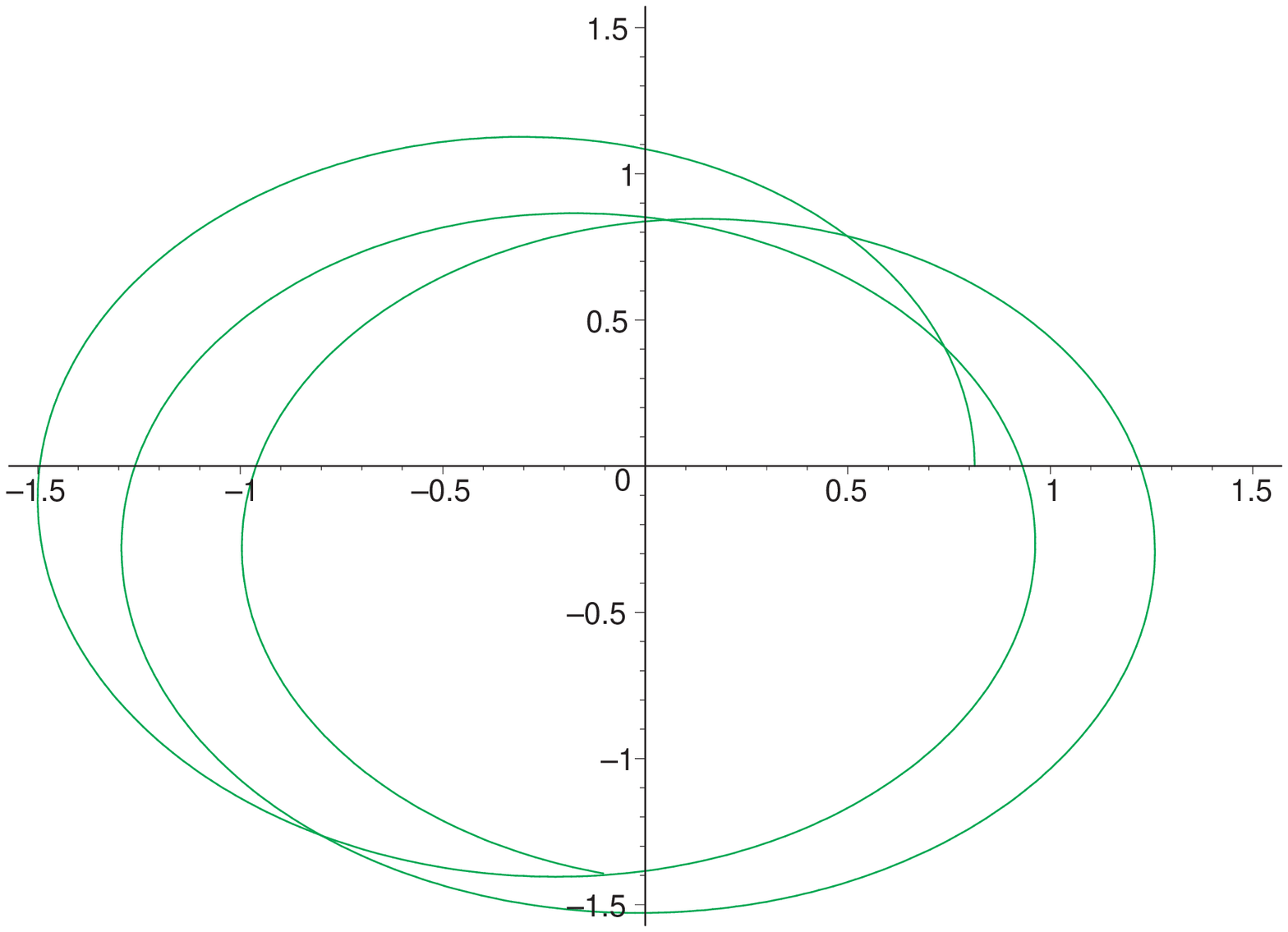}
\caption{These plots show the motion of a massive particle when it is started at different distances away from the minimum. The initial conditions were set up so that $\nu=0.8$, $\ell=4$, and $R=1$. The left hand plot shows the potential for the timelike geodesics. The middle plot shows the motion of massive particles when started at 0.1, 0.2, and 0.3 away from the minimum of the potential in red, blue, and green respectively, with the dotted black line indicating the position of the minimum. The right hand graph gives a specimen polar plot showing how the distance from the origin varies with $\phi$. The initial conditions are the same as for the green curve in the middle plot and $\tau$ ranges from 0 to 5. In all cases $\dot x=0$ and $\phi=0$ initially.}
\label{fig:oscpt}
\end{center}
\end{figure}

Figure \ref{fig:oscpt} gives an example of this motion when low energy particles are placed within the potential well. The minimum of the potential is at $x_{min}=-0.096$ and has value $E=0.642$. The middle plot clearly exhibits periodic motion, but the period is dependent upon the amplitude. As the amplitude is increased the period is also increased. This is most easily seen by comparing the period of the green curve (with the largest amplitude) with that of the red one (with the smallest amplitude.) The left hand plot indicates why this happens. The unsymmetrical shape of the potential is more marked further away from the minimum, so only the curves with the larger amplitude will show this effect.

The potential is steeper on the right hand side of the minimum potential line than it is on the left. This causes the maximum displacement to be greater to the left than it is to the right, meaning that the particle spends longer on the left hand side of the minimum potential. This gives the $x$ displacement plot a slightly ``bottom-heavy'' appearance, with the maximum displacement being greater for negative $x$. The effect is most apparent for small displacements from the minimum, with the maximum displacement becoming more equal as the initial displacement is increased.

When this is interpreted in terms of the physical motion of a test particle, it means that the particle orbits slowly when it is in the exterior of the ring and accelerates as it moves through the centre of the ring, before decelerating again on the other side. The acceleration is most marked when the geodesic passes close to the origin, so the particles with the highest energies will move very rapidly through the centre of the ring on a flat trajectory and those with lower energy will move through the ring on more of a curved orbit.

The polar plot on the right hand side of figure \ref{fig:oscpt} shows how the particle, corresponding to the green plot, moves in the $x$-$\phi$ plane, with the angle from the horizontal axis given by $\phi$ and the distance from the origin calculated using (\ref{rtrans}). This gives a clearer picture of the unsymmetric nature of the potential as each orbit is oblate with the trace precessing anti-clockwise after every revolution. If the trace is plotted over a longer time period then it does eventually return to its starting point. The polar plots corresponding to the red and blue curves (with smaller amplitudes) in the middle graph of figure \ref{fig:oscpt} show qualitatively similar behaviour, but the precession of the orbits isn't as large, due to the potential becoming more asymmetric further away from the minimum point.

\begin{figure}[htbp]
\begin{center}
\includegraphics[viewport=0 0 420 540,width=6.6cm,angle=270,clip]{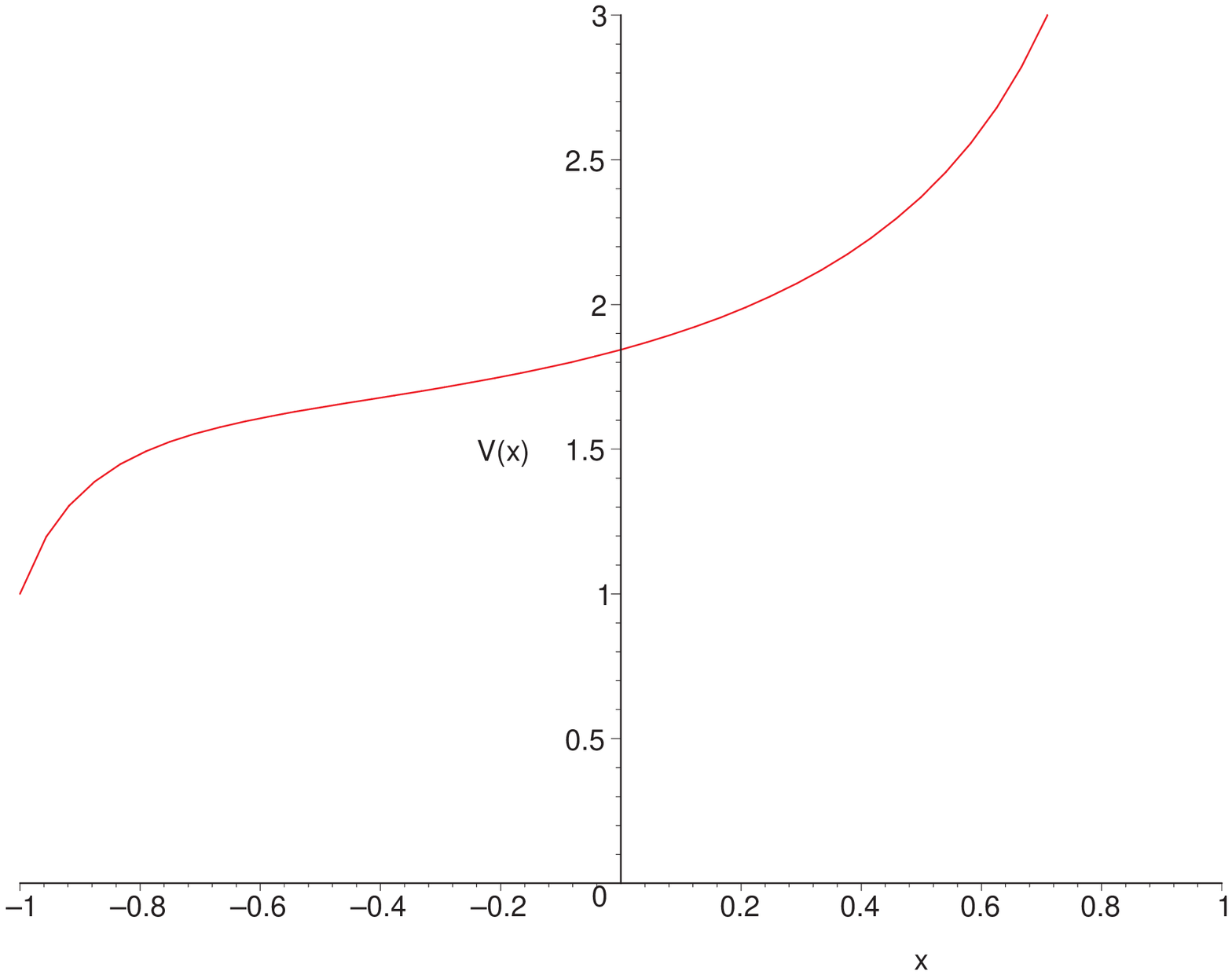}
\includegraphics[viewport=0 0 420 580,width=6.6cm,angle=270,clip]{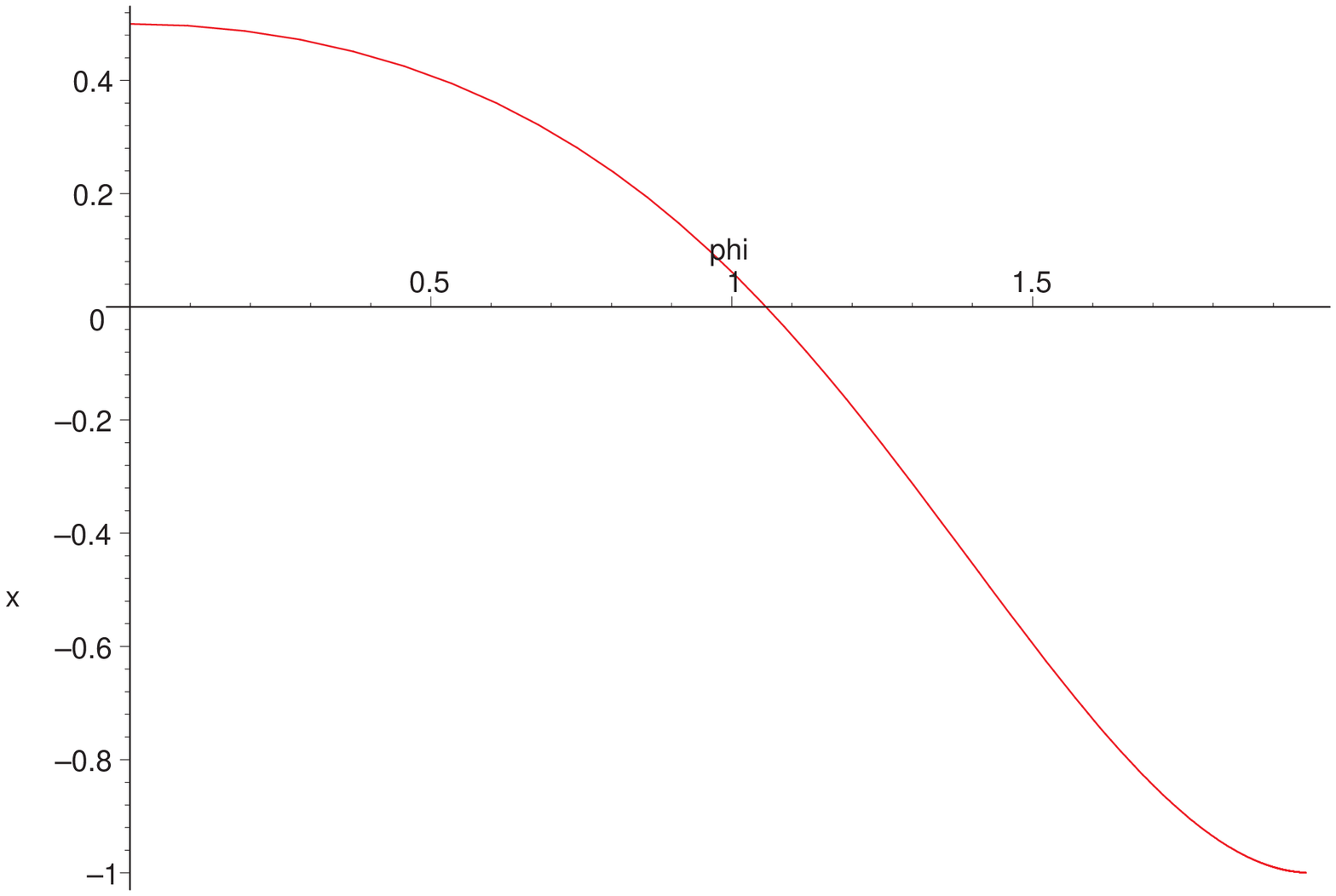}
\caption{The left hand plot shows the potential for $\ell=4$, $\nu=\frac{1}{2}$, and $R=1$. The right hand plot shows the motion of the timelike particle when it is started from rest at $x=0.5$ and $\phi=0$.}
\label{fig:repulsion}
\end{center}
\end{figure}

Figure \ref{fig:repulsion} gives a sample of the behaviour of a timelike geodesic when the angular momentum is large. In this case the potential is repulsive for all values of $x$, so a particle initially at rest will head off to infinity. If the initial velocity is increased then the particle can pass through the ring, with the minimum approach to the origin dependent on the energy of the particle. The particle will then go off to infinity on the other side of the ring.

The second plot in figure \ref{fig:repulsion} shows how $x$ varies with $\phi$ i.e. the path of the test particle in the $x$-$\phi$ plane. In this case $\frac{\ud\phi}{\ud\tau}$ is given by
\begin{equation}
\frac{\ud\phi}{\ud\tau}=\frac{\ell (1+x)}{R^2(1-x)(1+\nu x)}
\end{equation}
This shows that $\phi$ initially varies rapidly, when the particle is close to the origin, and then asymptotically approaches a constant as $x\rightarrow-1$. This is reflected in the right hand plot of figure \ref{fig:repulsion}, where the curve levels off at $\phi\approx 2$. Physically $\frac{\ud\phi}{\ud\tau}\rightarrow 0$ because the particle is approaching infinity and thus travels further and further for each interval in $\phi$.

\newsubsection{Null Geodesics on the Rotational Axis}

\begin{figure}[htbp]
\begin{center}
\includegraphics[viewport=0 5 412 550,width=7.4cm,angle=270,clip]{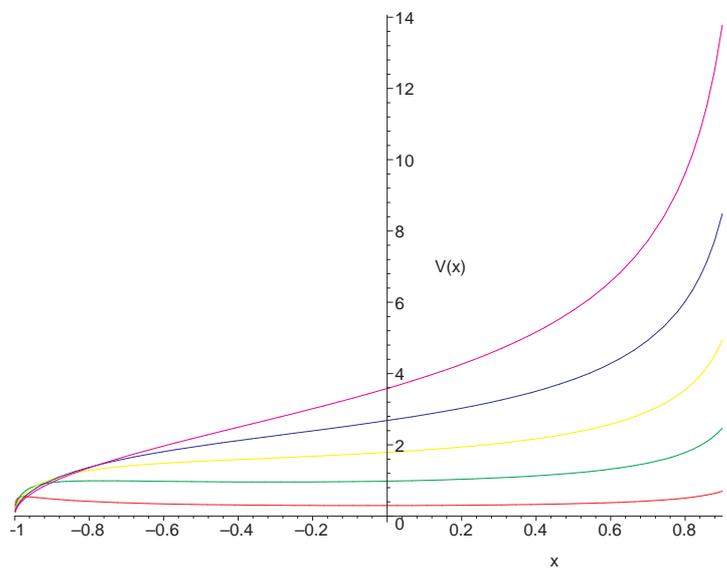}
\caption{This plot shows how the effective potential for null geodesics varies for $\nu=0.1,0.3,0.5,0.7,0.9$, when $\ell=4$. The lowest curve corresponds to $\nu=0.9$ and $\nu$ decreases with each consecutive curve. $R=1$ in both plots.}
\label{fig:nuypot}
\end{center}
\end{figure}

The potential for null geodesics is shown in figure \ref{fig:nuypot}, with the variation of the potential for different values of $\nu$ plotted for permissible values of $x$. These graphs show most of the properties of the null geodesics, principally:
\begin{itemize}
\item geodesics of low angular momentum have a closer minimum approach to the origin.
\item there is an infinite centrifugal barrier at $x=1$.
\item only geodesics with $\ell=0$ are able to pass through the origin.
\item the potential for geodesics with large $\ell$ is repulsive for all $x$.
\item the potential can have a local maximum near $x=-1$ for large values of $\nu$.
\end{itemize}

\begin{figure}[htbp]
\begin{center}
\includegraphics[viewport=0 0 330 550,width=9cm,angle=270,clip]{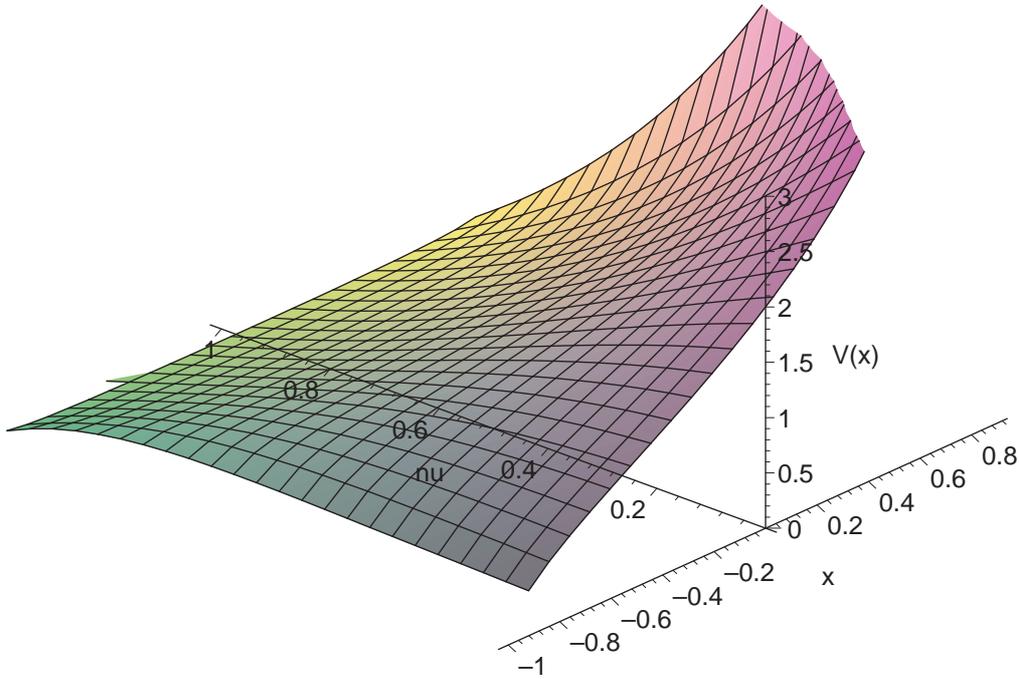}
\caption{3D plot showing how the effective potential for a null geodesic varies with $\nu$ and $x$. In this plot $\ell=4$ and $R=1$.}
\label{fig:nullAMypot3d}
\end{center}
\end{figure}

The variation of the potential with $\nu$ and $x$ is shown in figure \ref{fig:nullAMypot3d} for null geodesics. In this case the angular momentum (and indeed $R$) is purely a scale factor, as can be seen by substituting $\epsilon=0$ into (\ref{yeffpot}), so the plot shown in figure \ref{fig:nullAMypot3d} gives the variation of the potential with $\nu$ rather than $\ell$. This potential shares many of the same traits as that of the timelike geodesics, even exhibiting a local maximum near $x=-1$, as can be seen in figure \ref{fig:nullAMypot3d} when $\nu$ is large. The surface near $x=-1$ and $\nu\approx 1$ is where the difference between the local maximum and minimum is most pronounced, but the width of the peak is the smallest, so it isn't very visible in figure \ref{fig:nullAMypot3d}. The maximum height of the peak is when $\nu=0.653$, which explains why it appears more marked at this point on the 3D plot.

For the null geodesics, $\ell$ and $E$ have no independent meaning, since the test particles on null geodesics are massless, so it is only the ratio $\frac{\ell}{E}$ that is important. This is the reason why only one potential plot is given in figure \ref{fig:nuypot}. When $\ell=0$ the potential for the null geodesics is identically zero, so the null particles move as if in flat space when they go directly through the centre of the ring.

For the null geodesics, $\ell$ has no bearing on whether the maximum exists (unless of course $\ell=0$), so the lower limit on $\nu$ for a potential barrier to exist, is given by the solution to
\begin{equation}
{\nu_-}^4-26{\nu_-}^3+36{\nu_-}^2-54\nu_-+27=0
\end{equation}
This has four solutions but $0\le\nu\le 1$ for the equilibrium ring, so the only pertinent solution is given by $\nu_-= 0.653$. Values of $\nu$ less than $\nu_-$ will give potentials with no potential barrier, like the upper curves in the right hand plot of figure \ref{fig:nuypot}. For values of $\nu$ greater than $\nu_-$ there will always be a local maximum, with the position of the peak given by the solution to (\ref{ytrnpt}).

\begin{figure}[htbp]
\begin{center}
\includegraphics[viewport=5 2 450 530,width=4.7cm,angle=270,clip]{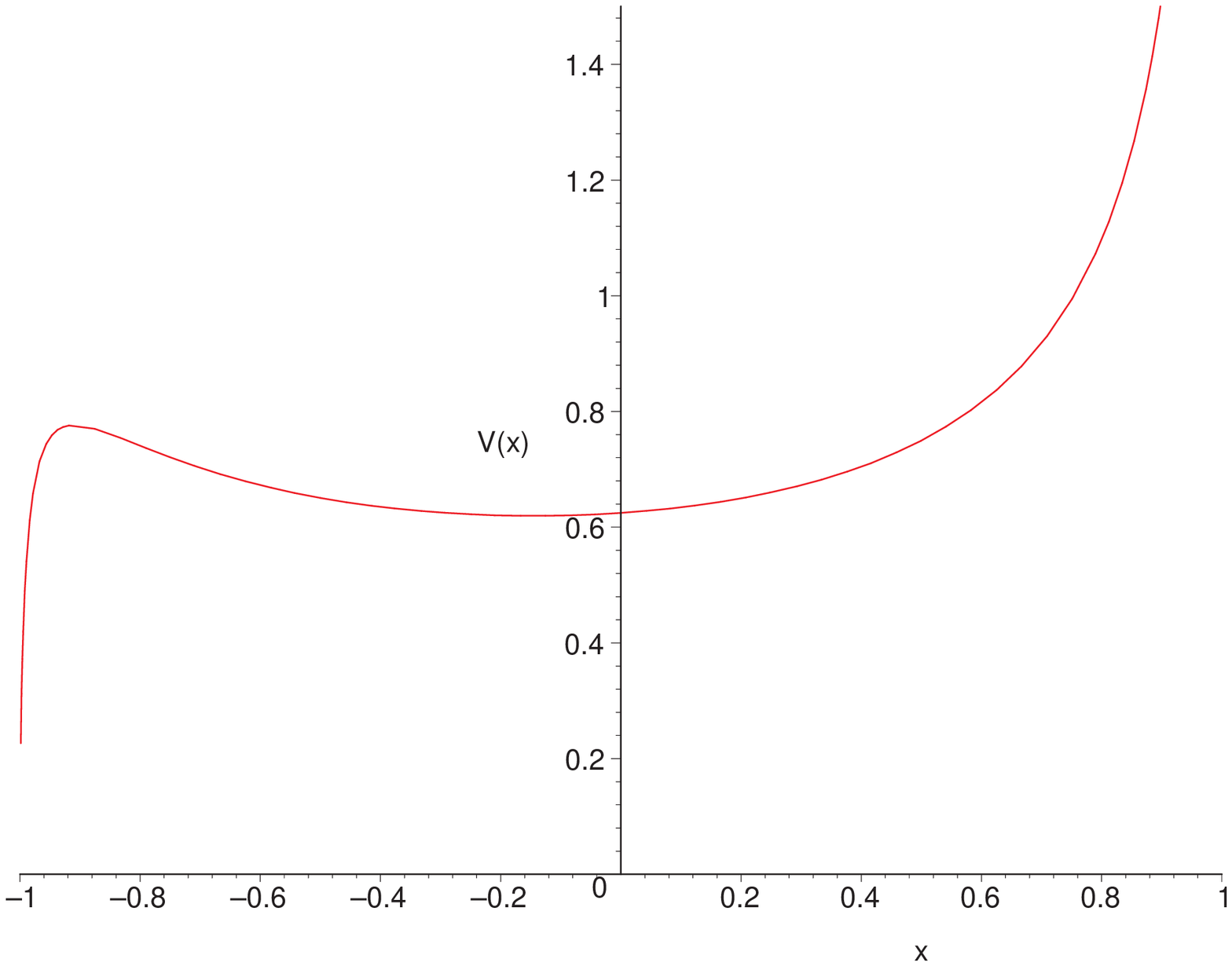}
\includegraphics[viewport=5 2 450 570,width=4.7cm,angle=270,clip]{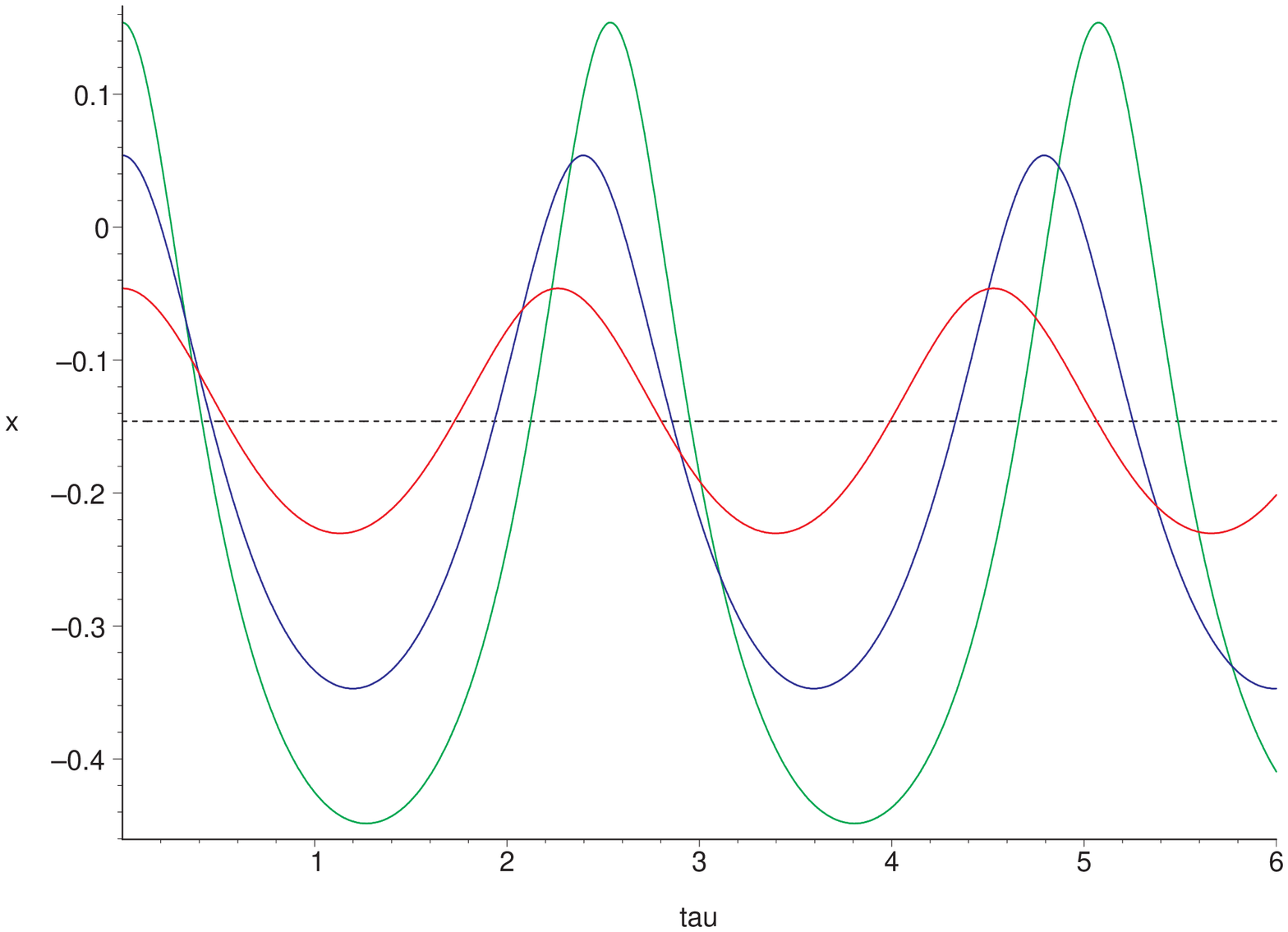}
\includegraphics[viewport=5 2 450 570,width=4.7cm,angle=270,clip]{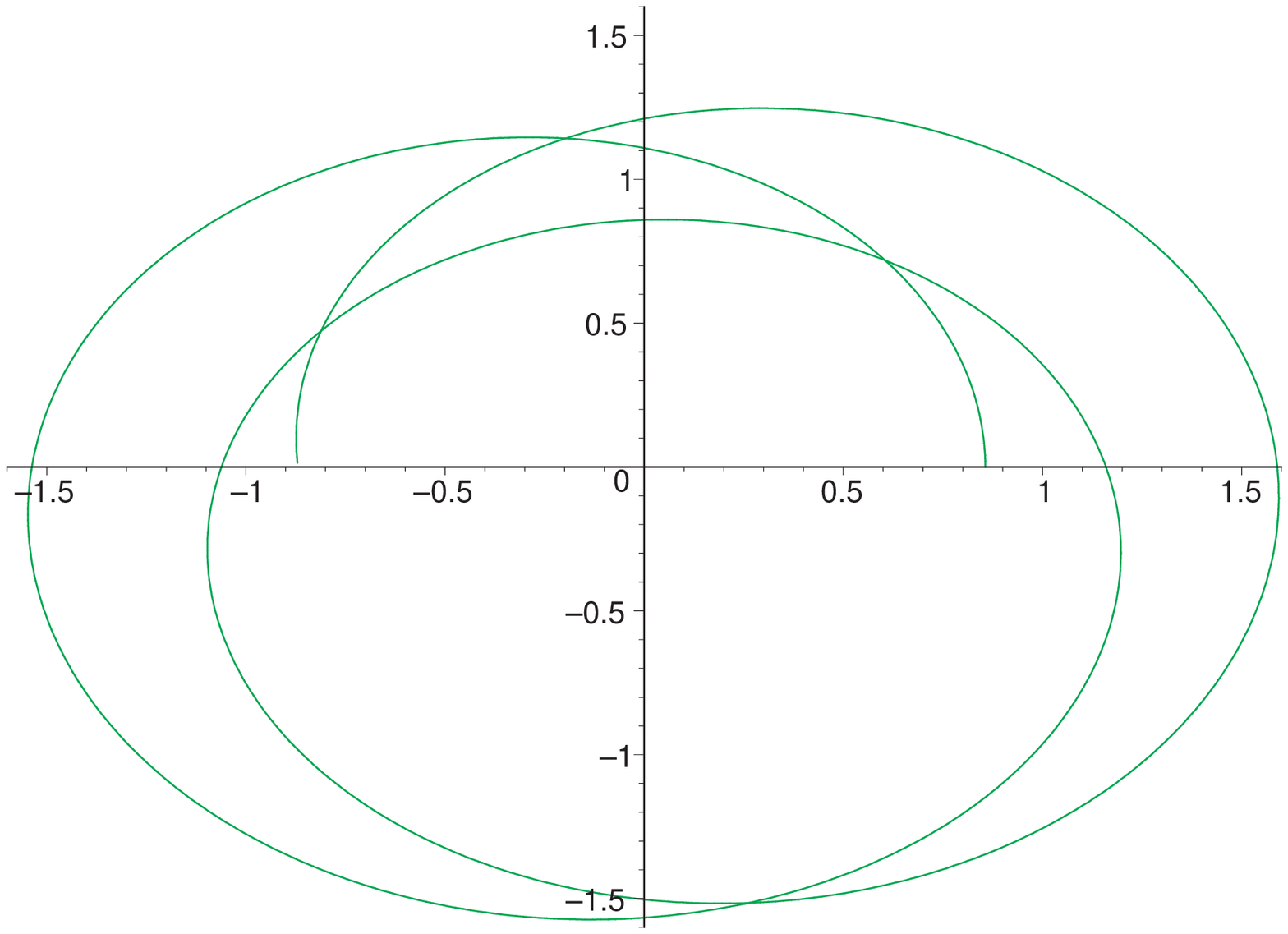}
\caption{These plots show the motion of a null particle when it is started at different distances away from the minimum. The initial conditions were set up so that $\nu=0.8$, $\ell=4$, and $R=1$. The left hand plot shows the potential for the null geodesics. The middle plot shows the motion of null particles when started at 0.1, 0.2, and 0.3 away from the minimum of the potential in red, blue, and green respectively, with the dotted black line indicating the position of the minimum. The right hand graph gives a specimen polar plot showing how the distance from the origin varies with $\phi$. The initial conditions are the same as for the green curve in the middle plot and $\tau$ ranges from 0 to 6. In all cases $\dot x=0$ and $\phi=0$ initially.}
\label{fig:nulloscpt}
\end{center}
\end{figure}

If $\ell=0$ then the null geodesics pass through the origin of the ring with the motion being almost identical to that of the timelike geodesic shown in figure \ref{fig:0AMygeo}. The major difference between the null and timelike geodesics, in this case, is that the null geodesics can't oscillate back and forth through the ring as the low energy timelike geodesics do. These null geodesics, in the toroidal coordinates, have the same problem at the origin as the timelike geodesics but the coordinate singularity is resolved in a similar manner.

Physically, null geodesics with large $E$ relative to $\ell$ behave in a similar manner to the timelike geodesics with large $E$. When the timelike geodesics have large energy the difference between the null and timelike potentials is largely irrelevant, so high energy massive particles and the corresponding massless particles aren't affected by the black ring at large distances. As these particles approach the centre of the ring at $x=1$, the centrifugal barrier comes into play, meaning that the distance of closest approach increases with $\ell$, as per spherical black holes.

If the null geodesics have small $E$ relative to $\ell$ and $\nu>0.653$, then the curvature of the space allows for the null geodesics to be captured. In this case there are two orbits at a constant distance from the ring. These are found by solving (\ref{ytrnpt}), the largest solution giving the position of the stable orbit and the next largest giving the position of the unstable orbit i.e. at the local maximum of the potential.

The local minimum in the potential for $\nu>0.653$ means that it is also possible to have null geodesics that oscillate through the ring. The plots of the potential and some examples of test particle motion are given in figure \ref{fig:nulloscpt}. The motion in this case is similar to the timelike case but the potential is less symmetrical, so the differences in the period of the motion are more pronounced. The period of the oscillations is also longer for the null geodesics than for their timelike counterparts.

\begin{figure}[htbp]
\begin{center}
\includegraphics[viewport=0 0 420 540,width=6.6cm,angle=270,clip]{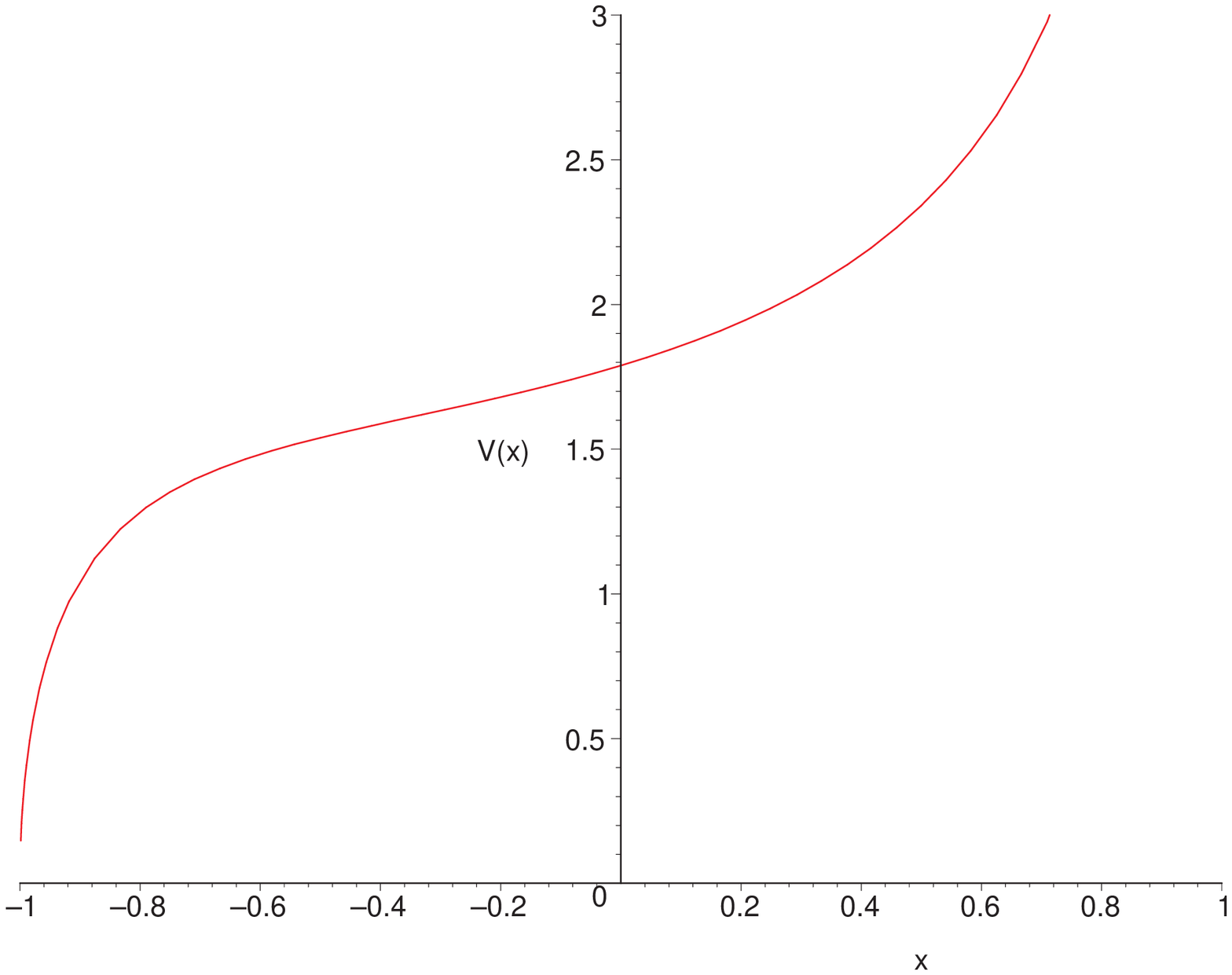}
\includegraphics[viewport=0 0 420 580,width=6.6cm,angle=270,clip]{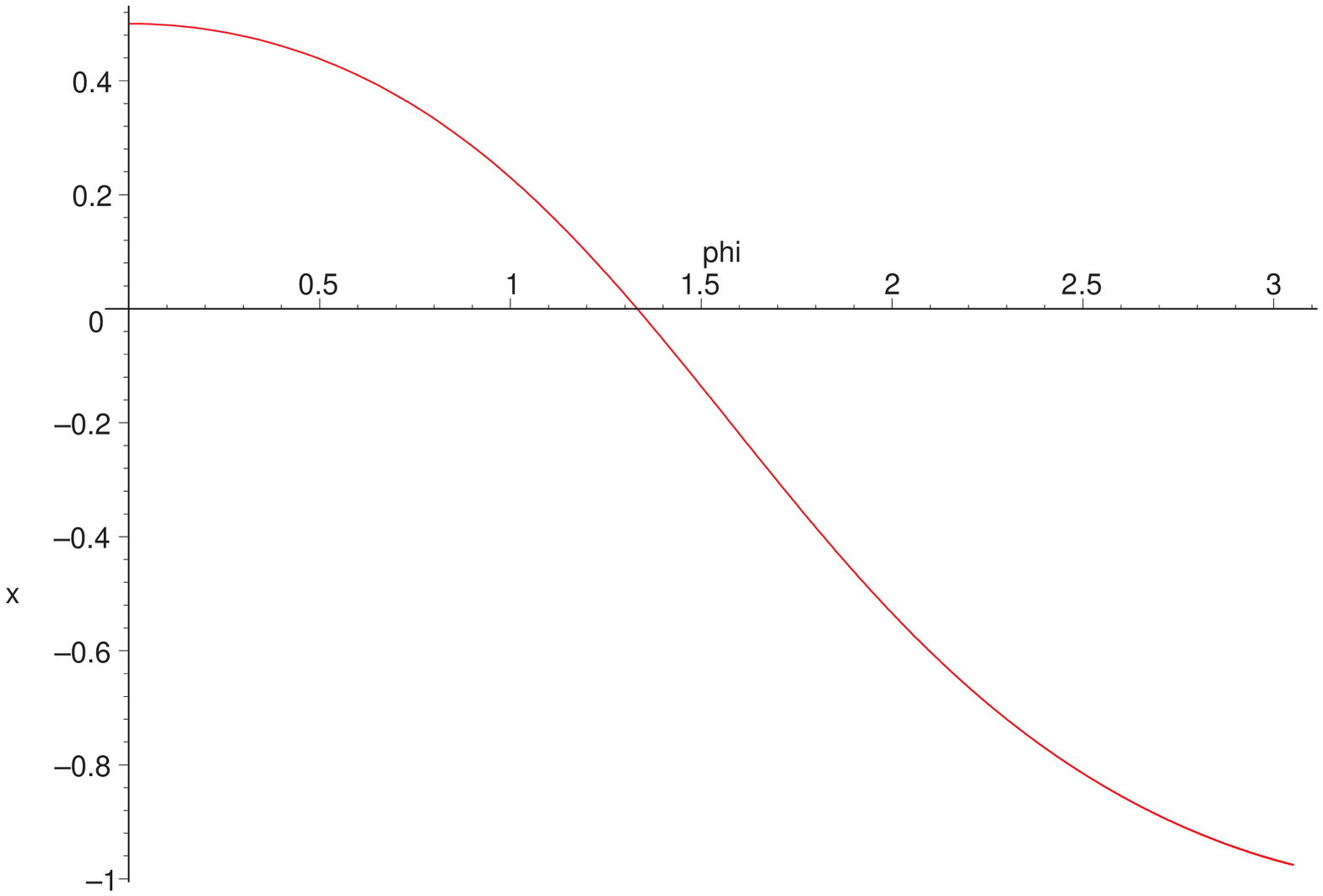}
\caption{The left hand plot shows the potential for $\ell=4$, $\nu=\frac{1}{2}$, and $R=1$. The right hand plot shows the motion of the null particle when it is started at $x=0.5$ and $\phi=0$, with $\dot x=0$ initially.}
\label{fig:nullrepulsion}
\end{center}
\end{figure}

As $\ell$ is increased the potential well is gradually smoothed out until the potential is always repulsive and takes the form given in figure \ref{fig:nullrepulsion}. The plot shown in this figure shows the path of the null geodesic when it is moving away from the centre of the ring. The null particles in this potential behave in a similar way to the massive particles shown in figure \ref{fig:repulsion} but the null particles approach $x=-1$ in a much shorter time. The null geodesics can pass through the centre of the ring but, as in the $\ell=0$ case they will then continue off to infinity.

\newsection{Planar Circular Geodesics}
\label{ch:plangeos}

To find circular orbits in the Black Ring metric it is necessary to solve the equations of motion so that $x$ and $y$ are constant for all of the motion and thus form an orbit by rotating in the $\psi$ and $\phi$ directions. In practice, this means solving (\ref{xELeqn}) and (\ref{yELeqn}) so that $\ddot x=\ddot y=0$. In general (\ref{xELeqn}) and (\ref{yELeqn}) are dependent on $\dot x$, $\dot y$, $x$, and $y$, so finding initial values for these variables that solve $\ddot y=0$ from (\ref{yELeqn}) won't guarantee that $\ddot x=0$ when they are substituted in (\ref{xELeqn}). If $\ddot x\ne 0$ then this will cause $\dot x$ to vary, which will in turn cause $\ddot y$ to vary. This means that both equations have to be solved simultaneously to find values of $x$ and $y$ that will give $\ddot y=\ddot x=0$ when $\dot x=\dot y=0$ but attempting to do this in general leads to intractable expressions that are of very high order.

The simplest way to avoid this problem is to look for values of $x$ which give $\ddot x=0$ when $\dot y=\dot x=0$ for all values of $y$, thus negating the need to consider both of the geodesic equations simultaneously, and allowing the circular orbits to be found by solving (\ref{yELeqn}) when $\ddot y=\dot y=0$. It turns out that the only way to achieve this is to set $x=\pm 1$, therefore ensuring that $\ddot x=\dot x=0$, no matter what happens to $y$. This choice of $x=\pm 1$ confines the geodesics to the plane perpendicular to the axis of rotation and also forces $\ell$ to be zero, since the geodesics can't simultaneously remain on this plane and have angular momentum with respect to it. This equatorial plane is split into two sections with $x=-1$ being the region ``outside'' of the ring and $x=+1$ being the region ``inside'' the ring.

\begin{figure}[htbp]
\begin{center}
\includegraphics[viewport=5 0 450 560,width=7cm,angle=270,clip]{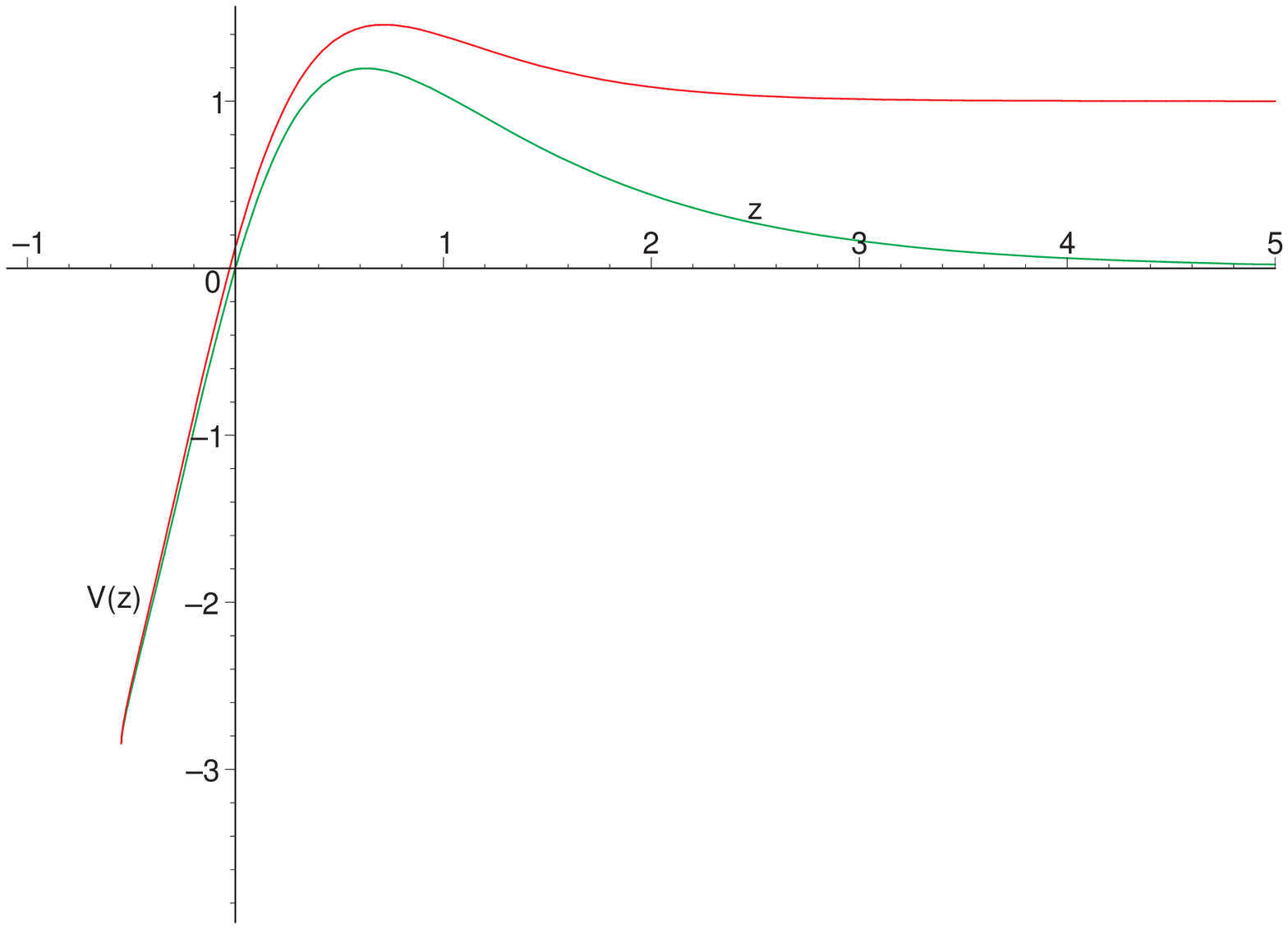}
\includegraphics[viewport=5 0 450 560,width=7cm,angle=270,clip]{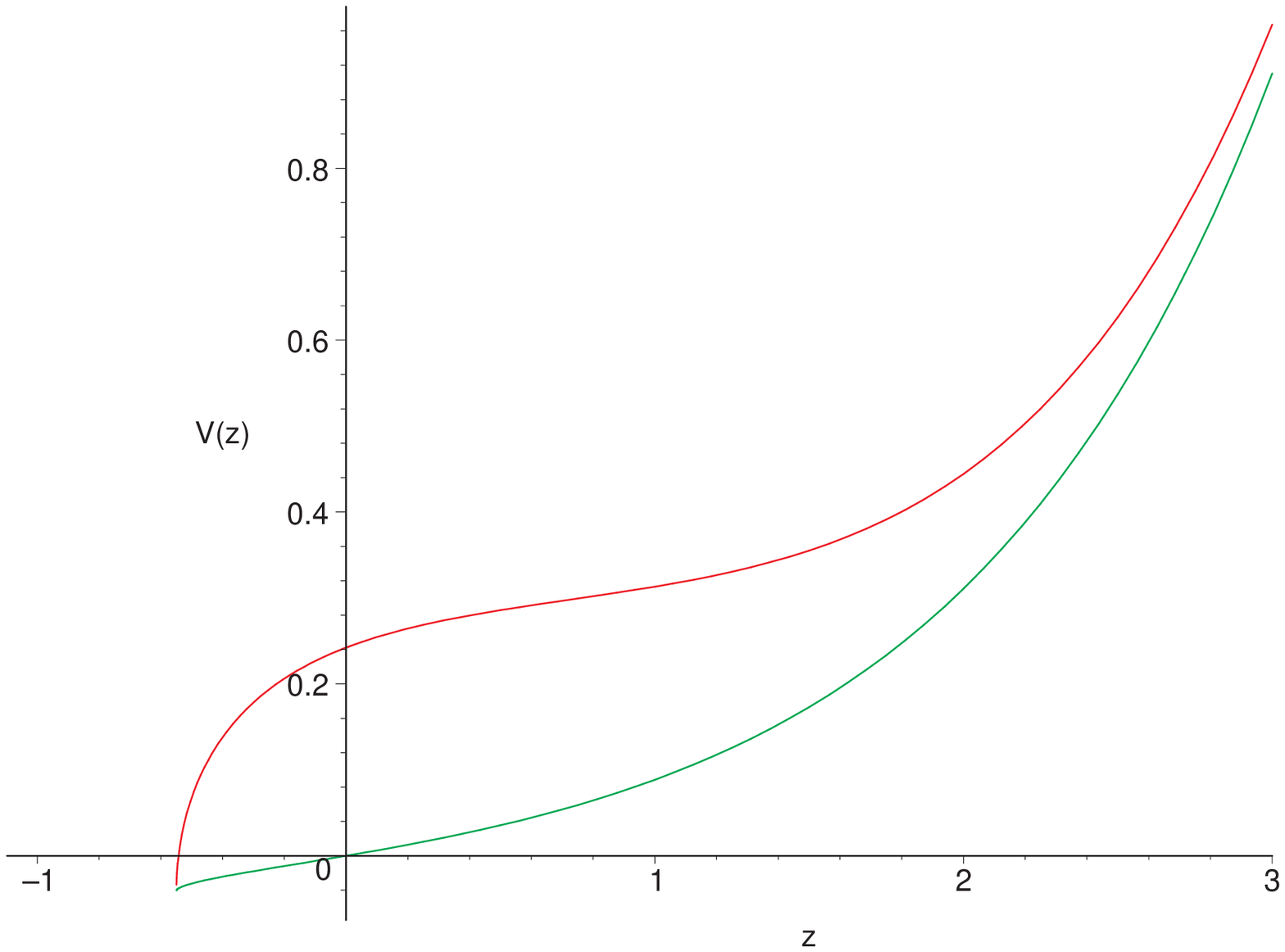}
\caption{These plots show some sample effective potentials $V_+(z)$ for red timelike and green null geodesics that are constrained to the equatorial plane. The left hand plot is the potential for the outer equatorial plane given by $x=-1$, when $\Psi=7$. The right hand plot is for the inner plane given by $x=+1$, when $\Psi=0.1$. Both plots are for $\nu=\frac{1}{2}$, and $R=1$.}
\label{fig:xeqmplane}
\end{center}
\end{figure}

The constraints discussed above are equivalent to demanding that the test particle be at a stationary point on the effective potential, with $\ddot x=0$. The effective potentials for particles on the inner and outer equatorial planes are given in figure \ref{fig:xeqmplane}, with the derivation of the effective potentials given in Appendix B. The potentials plotted in figure \ref{fig:xeqmplane} are in terms of a transformed coordinate $z$, which can be expressed in terms of $y$ using
\begin{equation}
z=-\tanh^{-1}\left(\frac{1+\lambda y}{y+\lambda}\right)
\end{equation}

This coordinate transformation is used to avoid the singularity caused when $y\rightarrow-\frac{1}{\lambda}$ i.e. the ergosurface, due to the $F(y)$ terms in the denominator of (\ref{firstint}). In terms of $z$, the ergosurface is at $z=0$, with $z=\infty$ corresponding to asymptotic infinity when $x=-1$ or the rotational axis when $x=1$. The event horizon is at $z=\tanh^{-1}{\frac{\lambda-\nu}{\lambda\nu-1}}$ and the curvature singularity is then reached at $z=-\tanh^{-1}\lambda$. Thus the range of $z$ is given by
\begin{equation}
-\tanh^{-1}\lambda\leq z \leq \infty
\end{equation}

The example effective potentials shown in figure \ref{fig:xeqmplane} plot $V_+(z)$ for the inner and outer equatorial planes, and are roughly indicative of all of the effective potentials. Varying the angular momentum for the $x=-1$ potential increases or decreases the height of the centrifugal barrier, as one would expect. In this case, the only circular orbits will be at the peak of the potential, and thus will be unstable. If the angular momentum, given by $\Psi$, is decreased to zero then the centrifugal barrier disappears and the timelike potential is strictly decreasing as $z\rightarrow-\tanh^{-1}\lambda$, whilst the null potential is identically zero for all values of $z$. This indicates that circular orbits can only exist when $\Psi\ne 0$.

The sample plot for the $x=+1$ potential allows one to immediately deduce that the black ring won't support circular orbits in the inner equatorial plane. When $\Psi\ne 0$ both the null and timelike potentials increase with increasing $z$ and have a centrifugal barrier at $z=\infty$ i.e. as they approach the axis of rotation $y=-1$. When $\Psi=0$ the centrifugal barrier disappears, allowing both null and timelike geodesics to reach $z=\infty$ and thus go through the origin. In this case the null potential is everywhere zero and the timelike potential levels off at $V_+(z)=\frac{1}{3}$. As the angular momentum is increased the centrifugal barrier dominates both the timelike and null potentials, with the two potential curves converging rapidly as $z$ increases. This has the effect of making the null and timelike potentials look identical for large $\Psi$.

When $\Psi$ is negative, both the potentials for the inner and outer equatorial planes are qualitatively similar to those given in figure \ref{fig:xeqmplane}. The major difference is for the potential of the outer equatorial 	plane, where the centrifugal barrier moves from being outside the ergosurface to being between the ergosurface and the event horizon. This is due to the frame dragging effect, whereby an incoming particle can make a closer approach to the event horizon if it is moving in the opposite direction to the rotation of the black hole.

Having chosen the values of $x$ and $\dot x$ such that the geodesics are confined to the plane, it is necessary to choose values of $y$ that ensure that $\dot y=0$ and $\ddot y=0$ i.e. the particle has to be on the peak of the centrifugal barrier. This ensures that the orbits will close up after each rotation. To find the values of $y$ that solve $\dot y=0$ it is sufficient to consider the first integral equation given by (\ref{firstint}). Substituting $\ell=0$, $x=\pm 1$ and $\dot x=0$ into this equation simplifies it substantially. After rearranging to isolate $\dot y$ and solving for $\dot y=0$, the equation becomes
\begin{equation}
\frac{E^2(1 \pm \lambda)G(y_0)}{F(y_0)}+\frac{(\pm 1-y_0)^2\left[RE(1+y_0)C+\Psi F(y_0)\right]^2}{(1 \pm \lambda)F(y_0)R^2}+\epsilon G(y_0)=0
\label{circcond}
\end{equation}
Solving this equation for $y_0$ gives values of $y_0$ that are on the effective potential line when $x=\pm 1$. Equation (\ref{circcond}) is a cubic in $y_0$ due to the $G(y_0)$ coefficients, but for $x=\pm 1$ a factor of $(y\mp 1)$ can be removed. This reduces the equation to a quadratic making it much easier to analyse. Unfortunately, the coefficients of $y$ are very complicated, so it is easiest to choose particular values for the conserved quantities and then solve the equation.

To find the values of $y$ for which the particle is at a turning point in the potential, it is necessary to simultaneously solve (\ref{yELeqn}) for $\ddot y=\dot y=\dot x=\ell=0$. Substituting in $x=\pm 1$ leaves
\begin{eqnarray}
&& \frac{E(\pm 1-y_0)^2C(1-\lambda)\left[RE(1+y_0)C+\Psi F(y_0)\right]}{R(1 \pm \lambda)G(y_0)}  -\frac{E^2\lambda(1 \pm \lambda)}{2} \nonumber \\
&& - \frac{(\pm 1-y_0)^2\left[RE(1+y_0)C+\Psi F(y_0)\right]^2}{2R^2(1 \pm \lambda)G(y_0)}\left[ \frac{F(y_0)G'(y_0)}{G(y_0)} + \frac{2F(y_0)-\lambda(\pm 1 - y_0)}{(\pm 1-y_0)}\right]=0
\label{circcond2}
\end{eqnarray}
This is a quartic equation in $y$ but, once again, it is possible to factor out $F(y_0)^2$, leaving a quadratic in $y$. Expanding the functions in (\ref{circcond2}) and re-writing in fully factored form gives
\begin{equation}
\frac{(1+\lambda y_0)^2(1-\nu)(\alpha_\pm y_0^2 +\beta_\pm y_0 + \gamma_\pm)}{2R^2(1+\nu^2)^2(\lambda-1)(1+\nu y_0)^2(1\pm y_0)^2(\pm 1+\lambda)}=0
\label{faccirccond2}
\end{equation}
where
\begin{eqnarray}
\alpha_+ & =& 2\nu(1+\nu)^3R^2E^2-2\nu(1+\nu)^2\sqrt{2(1-\nu^2)}R\Psi E+\nu\Psi^2(\nu-1)(\nu^2-4\nu-1) \\
\alpha_- & =& 2\nu(7\nu^3+\nu^2+\nu-1)R^2E^2-2\nu(1+\nu)(3\nu-1)\sqrt{2(1-\nu^2)}R\Psi E- \nonumber \\
&& \nu\Psi^2(\nu-1)(\nu^2+4\nu-1) \\
\beta_+ & =& 4\nu(1+\nu)^3R^2E^2 - 4\nu(1+\nu)^2\sqrt{2(1-\nu^2)}R\Psi E - 2\nu\Psi^2(\nu-1)(\nu^2+3) \\
\beta_- & =& 4\nu(\nu^3+7\nu^2-\nu+1)R^2E^2 - 4\nu(1+\nu)^2\sqrt{2(1-\nu^2)}R\Psi E - 2\nu\Psi^2(\nu-1)(\nu^2+3) \\ 
\gamma_+ & =& 2\nu(1+\nu)^3R^2E^2 - 2\nu(1+\nu)^2\sqrt{2(1-\nu^2)}R\Psi E - \Psi^2(\nu-1)(\nu^3+2\nu^2-\nu+2) \\
\gamma_- & =& 2\nu(-\nu^3+\nu^2+9\nu-1)R^2E^2 - 2\nu(1+\nu)(3-\nu)\sqrt{2(1-\nu^2)}R\Psi E + \nonumber \\
&& \Psi^2(\nu-1)(\nu^3-2\nu^2-\nu-2)
\end{eqnarray}
It is obvious from (\ref{faccirccond2}) that $y_0=-\frac{1}{\lambda}$ is always going to be a solution to this equation, so the roots of the quadratic part will give the non-trivial solutions to (\ref{circcond2}). The roots are given by
\begin{equation}
y_0=\frac{-\beta_\pm\pm\sqrt{{\beta_\pm}^2-4\alpha_\pm\gamma_\pm}}{2\alpha_\pm}
\end{equation}
At this point it is worth checking that these roots can become real for $x=1$ and $x=-1$ because the imaginary results are unphysical. The condition for the roots to be real is given by
\begin{equation}
{\beta_\pm}^2-4\alpha_\pm\gamma_\pm\ge 0
\label{realcond}
\end{equation}
For $R=1$ and $x=+1$ this becomes
\begin{equation}
-8\Psi^2\nu(1+\nu)^2(\nu-1)^4\left[2(1+\nu)E^2-2\Psi\sqrt{2(1-\nu^2)}E+\Psi^2(1-\nu)\right]\ge 0
\end{equation}
The factor in front of the square brackets is always negative so, apart from the trivial solutions at $\nu=\pm 1$ and $\Psi=0$, there can only be real solutions when the term in the square brackets is negative. The coefficient of the $E^2$ term is always positive, so the term will become negative for values of $E$ between the two roots. Solving for $E$ gives a repeated root at
\begin{equation}
E_\pm=\frac{\Psi\sqrt{2(1-\nu^2)}}{2(1+\nu)}
\end{equation}
This means that the term in the square brackets will never become negative and thus there will only ever be a single real solution to (\ref{faccirccond2}) for $y$. Substituting $E_\pm$ into the quadratic part of (\ref{faccirccond2}) gives
\begin{equation}
2(\nu-1)^2(1+\nu y)^2\Psi^2=0
\end{equation}
which has only one solution at $y=-\frac{1}{\nu}$ i.e. on the event horizon. Equation (\ref{circcond}) can only be satisfied for null geodesics at this point, so it is a trivial solution. The solution where $\Psi=0$ is equally trivial because (\ref{faccirccond2}) then reduces to
\begin{equation}
2E^2\nu(1+\nu)^3(1+y)^3=0
\end{equation}
This only has a solution at $y=-1$, which corresponds to asymptotic infinity. Therefore there are no non-trivial solutions for geodesics inside the ring at $x=+1$. This is in agreement with the conclusions drawn from the potential plot given in figure \ref{fig:xeqmplane}.

When $x=-1$, (\ref{realcond}) is quartic in $E$ and thus has real solutions for various values of $E$, $\Psi$, and $\nu$. These solutions are explored for timelike and null geodesics in the following two sub-sections.

\newsubsection{Circular Orbits for Timelike Geodesics}

For timelike geodesics, one of the non-trivial solutions to (\ref{circcond2}) has to be discounted because it is always less than $y=-\frac{1}{\nu}$, meaning that this solution is always inside the event horizon for all values of $\nu$, $E$, and $\Psi$. Geodesics starting at this point can't form circular orbits because they are compelled to move toward the curvature singularity by virtue of the $\partial_t$ Killing vector being spacelike. This leaves two possible solutions to (\ref{circcond2}) and (\ref{circcond}) respectively.

\begin{figure}[htbp]
\begin{center}
\includegraphics[viewport=0 0 410 560,width=6.2cm,angle=270,clip]{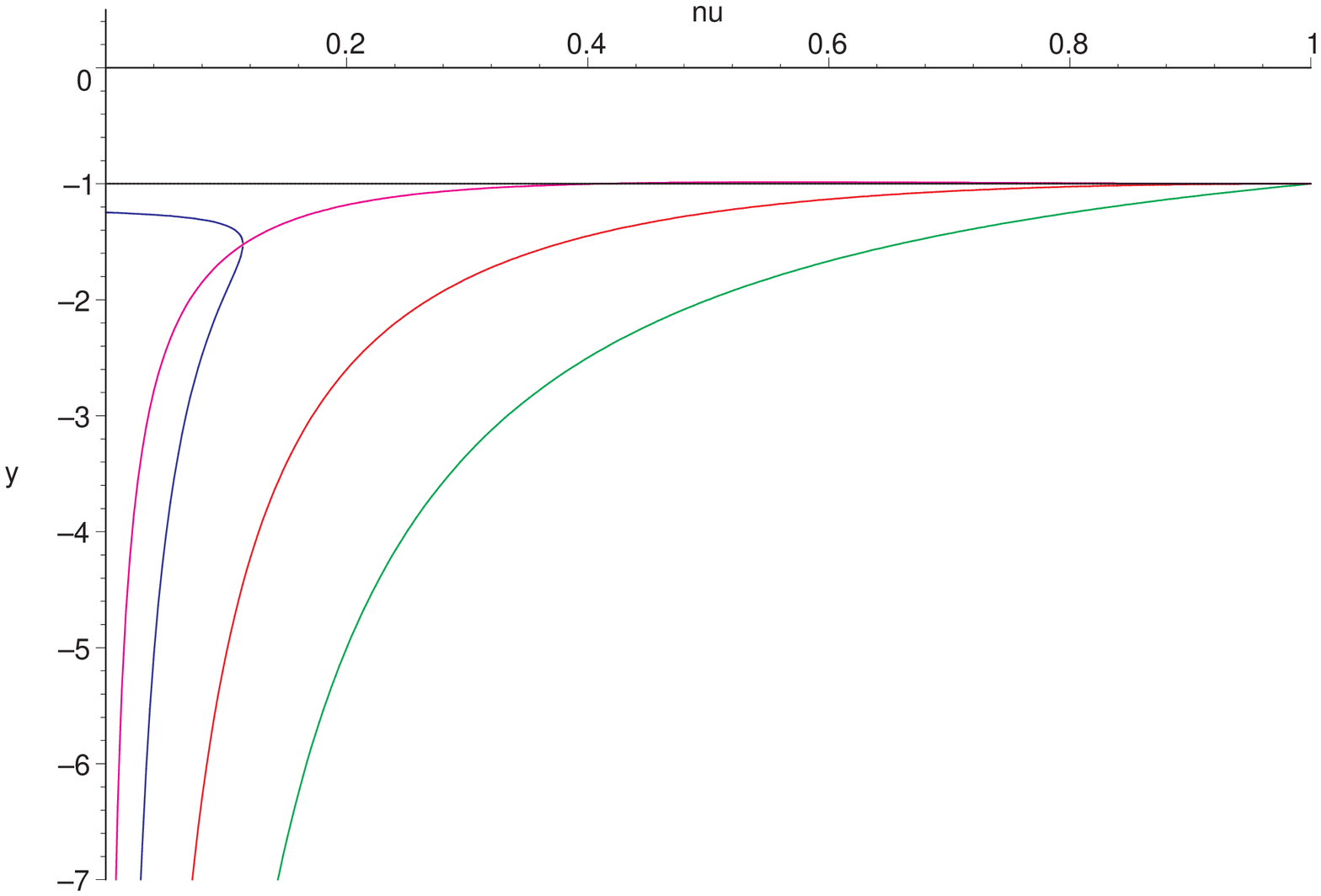}
\includegraphics[viewport=0 0 410 560,width=6.2cm,angle=270,clip]{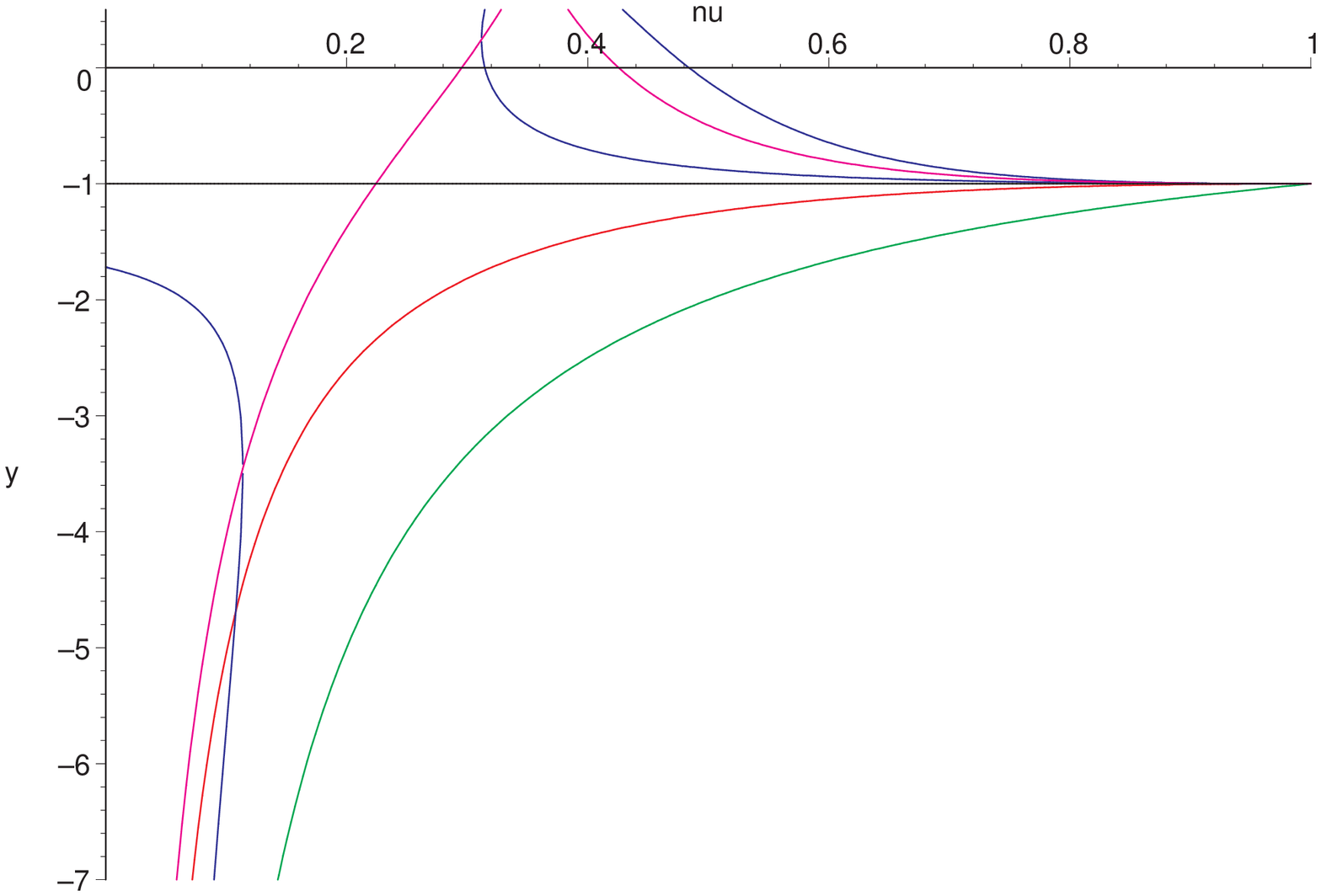}
\caption{These plots show the solution curves for the outer equatorial plane when $\ddot y=0$ in purple and red and those when $\dot y=0$ in blue. In this instance the red line also indicates the ergosurface, since this is also a solution to (\ref{circcond2}). The green line gives the position of the event horizon at $-\frac{1}{\nu}$ and the black line shows asymptotic infinity at $y=-1$. The left hand plot is for $\Psi=2.00$ and the right hand plot is for $\Psi=-1.29$. Both these plots use $E=1.20$.}
\label{fig:timeynu}
\end{center}
\end{figure}

After discounting the invalid solution curves, the circular orbits are given by the points where the solution curve for (\ref{circcond}) intersects with that of (\ref{circcond2}). This is equivalent to finding a point where both $\dot y$ and $\ddot y$ are zero. Figure \ref{fig:timeynu} shows how these solution curves vary with $\nu$ for some specific values of $E$ and $\Psi$.

The plots shown in figure \ref{fig:timeynu} show that the blue curve, where $\dot y=0$, intersects the purple curve, where $\ddot y=0$, at $\nu=0.11385$. The values of $\Psi$ were chosen specifically so that the value of $\nu$ where the curves intersect is the same for both of the plots. The purple curve corresponds to the non-trivial solution to (\ref{circcond2}), so the particular set of values used for this plot will form a circular orbit at $y=-1.52508$. The left hand plot indicates that there will only be one circular orbit when $\Psi>0$ but when $\Psi$ is negative, i.e. when the particle is moving in the opposite direction to the rotation of the ring, there can be two solutions, as indicated in the right hand plot. These solution are represented by the points where the blue curve intersects the purple and red curves. The blue and red curves can only intersect when the particle is rotating counter to the ring because the red curve represents the ergosurface, and at this point a circular orbit can only be formed if the particle is moving against the rotation of the ring, otherwise the frame dragging effect causes the particle to rotate too quickly.

\begin{figure}[htbp]
\begin{center}
\includegraphics[viewport=0 0 410 580,width=8cm,angle=270,clip]{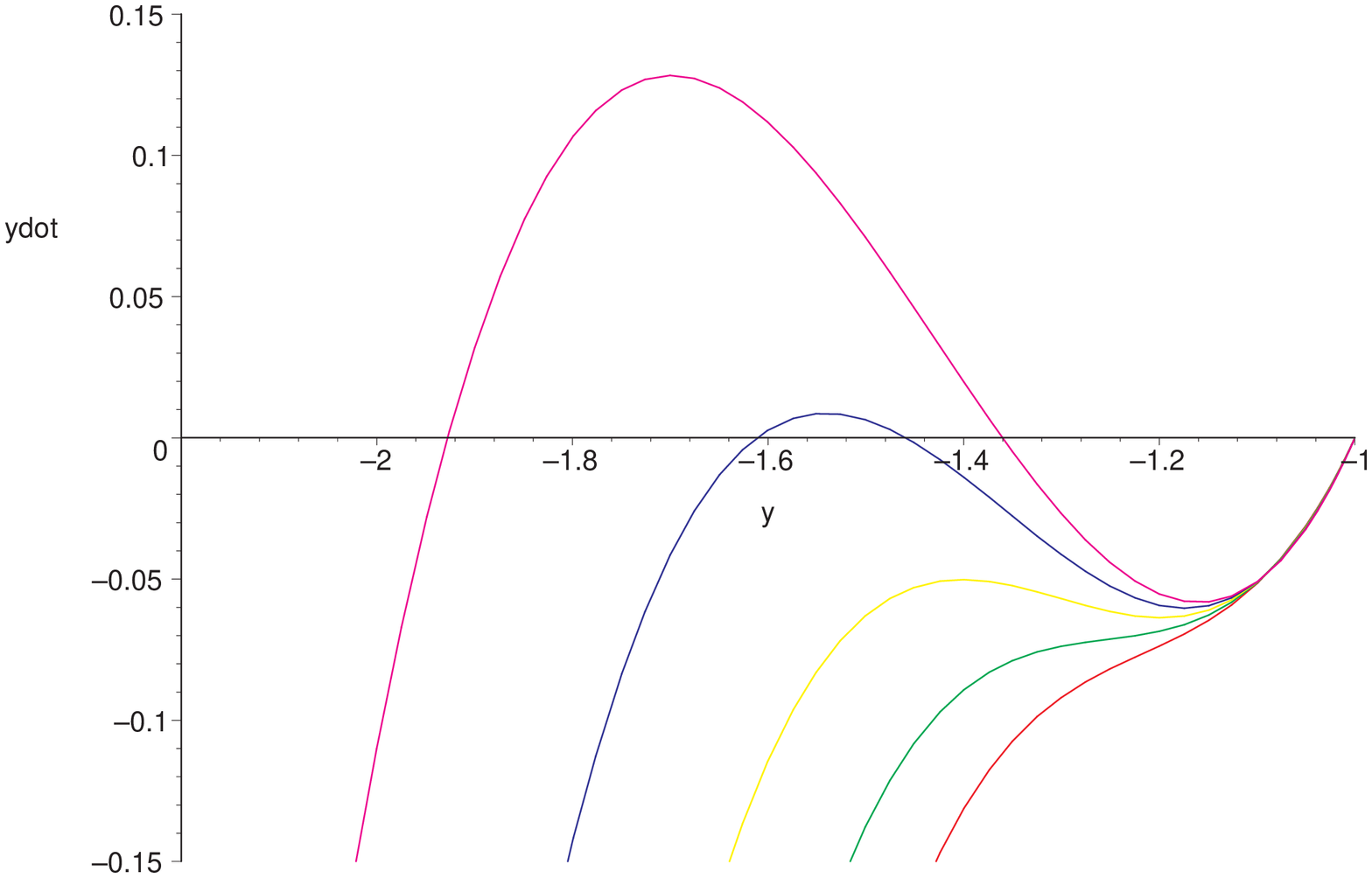}
\caption{This graph plots $\dot y$ against $y$ for values of $\nu$ from 0.1, for the uppermost curve, to 0.15 for the lowest curve, in 0.0125 increments. All the curves are plotted for $E=1.2$ and $\Psi=2$, so that they correspond to the left hand plot of figure \ref{fig:timeynu}. It indicates how the point where the curves cross in figure \ref{fig:timeynu} is where (\ref{circcond}) has a double root. A similar plot is formed if the corresponding negative value of $\Psi$ is used.}
\label{fig:timeEynu}
\end{center}
\end{figure}

In general, the point where the blue curve intersects the purple curve will always be at the turning point in the blue line. This point is where both of the roots of (\ref{circcond}) converge. This can be understood by considering the plot shown in figure \ref{fig:timeEynu}. In this plot the points where the curves cross the horizontal axis is where $\dot y=0$ and the turning points of the curves are where $\ddot y=0$. In order for $\ddot y=\dot y=0$ for the same value of $y$, the turning point has to be where the curve intersects the horizontal axis i.e. the curve must have a repeated root. The plot in figure \ref{fig:timeEynu} gives some example curves for different values of $\nu$ either side of the critical value. It verifies that there is a curve with a double root between 0.1125 and 0.1250, since the blue curve has it's turning point just above the axis and the yellow curve doesn't quite reach the axis. This agrees with the value of $\nu$ indicated by the intersection of the two curves shown in figure \ref{fig:timeynu}.

Armed with this knowledge, the point where the curves intersect can be calculated by solely considering equation (\ref{circcond}). Once the trivial solution at $y=-1$ is factored out the remaining equation is a quadratic, so the double root will be where \\
\begin{math}
(\nu-1)^2\Psi^4+2R^2(\nu-1)[(E^2+\epsilon)(3\nu^2+4\nu+1)+4\nu E^2]\Psi^2
+16R^3E\nu\sqrt{2(1-\nu^2)}(E^2+\epsilon)(1+\nu)\Psi
\end{math}
\vspace{-0.3cm}
\begin{equation} +R^4\left[E^4(\nu^4-24\nu^3-18\nu^2-8\nu+1)+2\epsilon(1+\nu)(\nu^3-13\nu^2-5\nu+1)E^2+\epsilon^2(1-\nu^2)^2\right]=0
\label{circgeocond}
\end{equation}
This is a quartic in $\Psi$ so will technically have four solutions, but only the largest and smallest roots are pertinent since the intermediate solutions are always for positive values of $y$. Solving this for $\Psi$ in terms of $E$ and $\nu$, when $\epsilon=-1$ gives
\begin{equation}
\Psi=\pm\frac{R}{\sqrt{1-\nu}}\left(2E\sqrt{\nu}+\sqrt{\left(E^2-1\right)\left(3\nu+1\pm 2\sqrt{2}\sqrt{\nu(1+\nu)}\right)\left(1+\nu\right)}\right)
\label{Psisol}
\end{equation}
This equation shows how the angular momentum has to be varied for different ring geometries (different values of $\nu$) and different particle energies (given by $E$). The two solutions for $\Psi$ represent the circular orbits when the particle is rotating with and against the motion of the ring respectively. The positive $\Psi$ solution will always be outside the ergosurface but for larger $\nu$ the negative $\Psi$ solution can give a circular orbit within the ergosurface.
\begin{figure}[htbp]
\begin{center}
\includegraphics[width=8cm,angle=270]{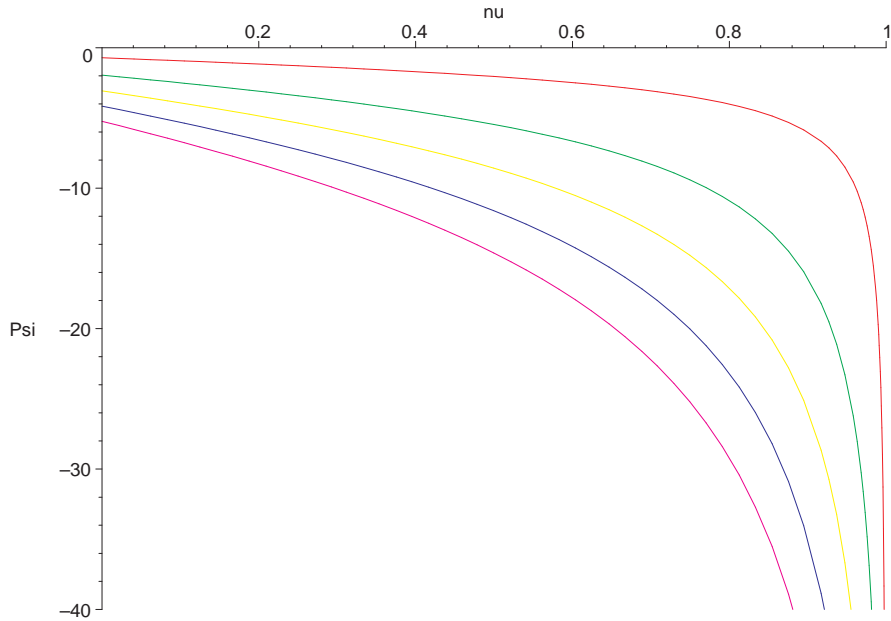}
\caption{This graph shows how $\Psi$ varies with $\nu$ to produce circular orbits on the ergosurface. The value of $\Psi$ has been plotted for 5 different energies ranging from $E=1$ to $E=5$. The graph shows that the circular orbits of fatter rings, given by larger values of $\nu$, have to have higher angular momentum in order to produce a circular orbit.}
\label{fig:timenuPsi}
\end{center}
\end{figure}

In figure \ref{fig:timeynu}, the red curve represents the ergosurface, given by $y_0=-\frac{1}{\lambda}=-\frac{(1+\nu^2)}{2\nu}$. When the blue line intersects this line the circular orbit is on the ergosurface. This circular orbit will always exist for all values of $E$ and $\nu$, unlike the solution given by (\ref{Psisol}), which is complex for $E<1$.

To find the value of $\Psi$ for which the circular orbit exists, substitute $y=-\frac{1}{\lambda}$ into (\ref{circcond}) for $\epsilon=-1$ and $x=-1$. This allows $\Psi$ to be expressed in terms of $E$ and $\nu$ as
\begin{equation}
\Psi=\frac{R\left[\nu^2(1-E^2)+2\nu(1-4E^2)+1-3E^2\right]}{2E\sqrt{2(1-\nu^2)}}
\end{equation}
Figure \ref{fig:timenuPsi} gives some examples of the permissible values of $\Psi$, the angular momentum about the rotational axis of the ring, as $\nu$ varies for a range of different energies. In general $|\Psi|$ increases with the energy and also with $\nu$. The sign of $\Psi$ remains constant as $\nu$ is varied, so the trace shown in figure \ref{fig:timenuPsi} will never cross the axis. This confirms that the particle's angular momentum has to always be in the opposite direction to the rotation of the ring for circular orbits on the ergosurface.

\newsubsection{Circular Orbits for Null Geodesics}

\begin{figure}[htbp]
\begin{center}
\includegraphics[width=6.3cm,angle=270]{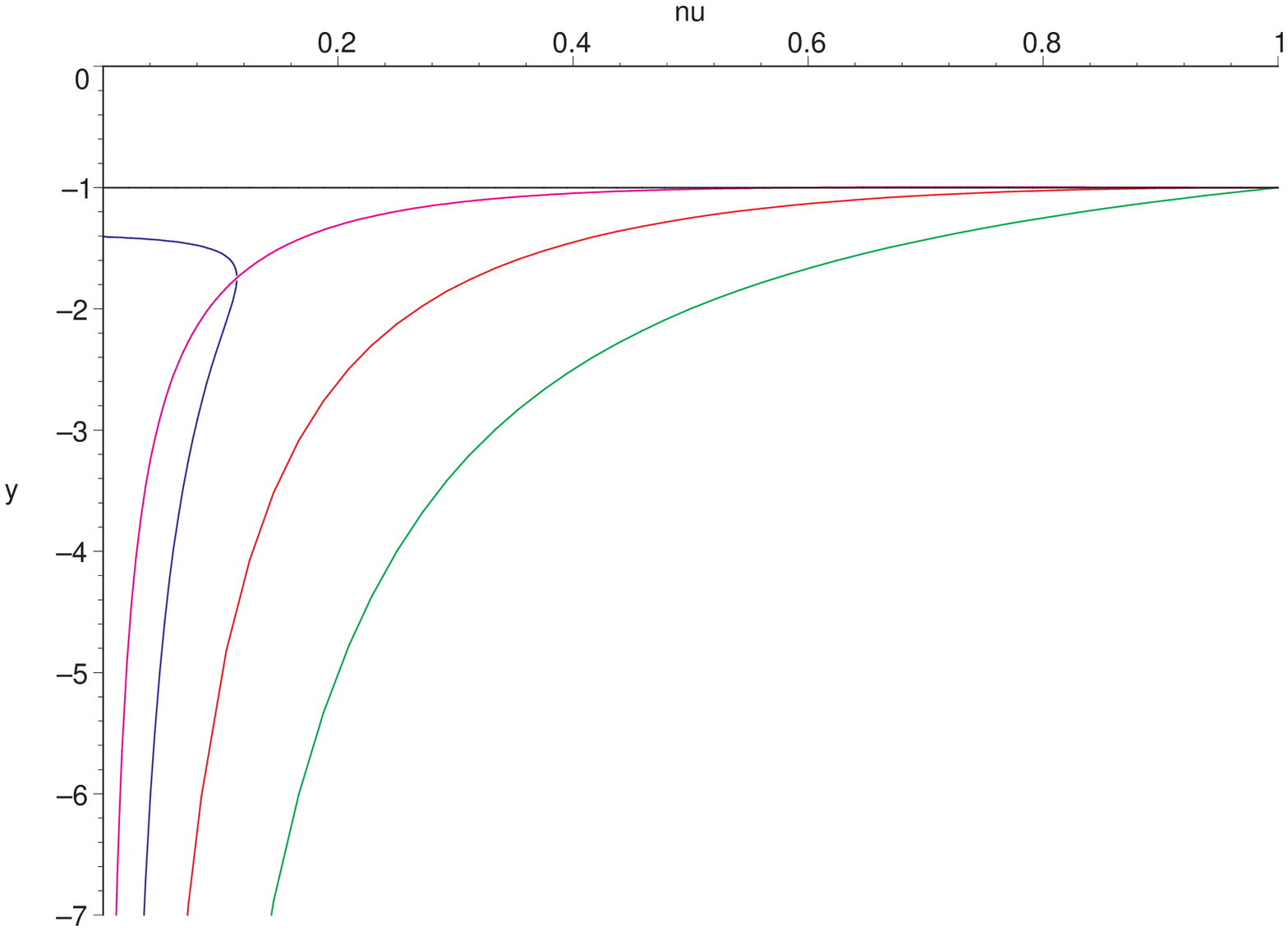}
\includegraphics[width=6.3cm,angle=270]{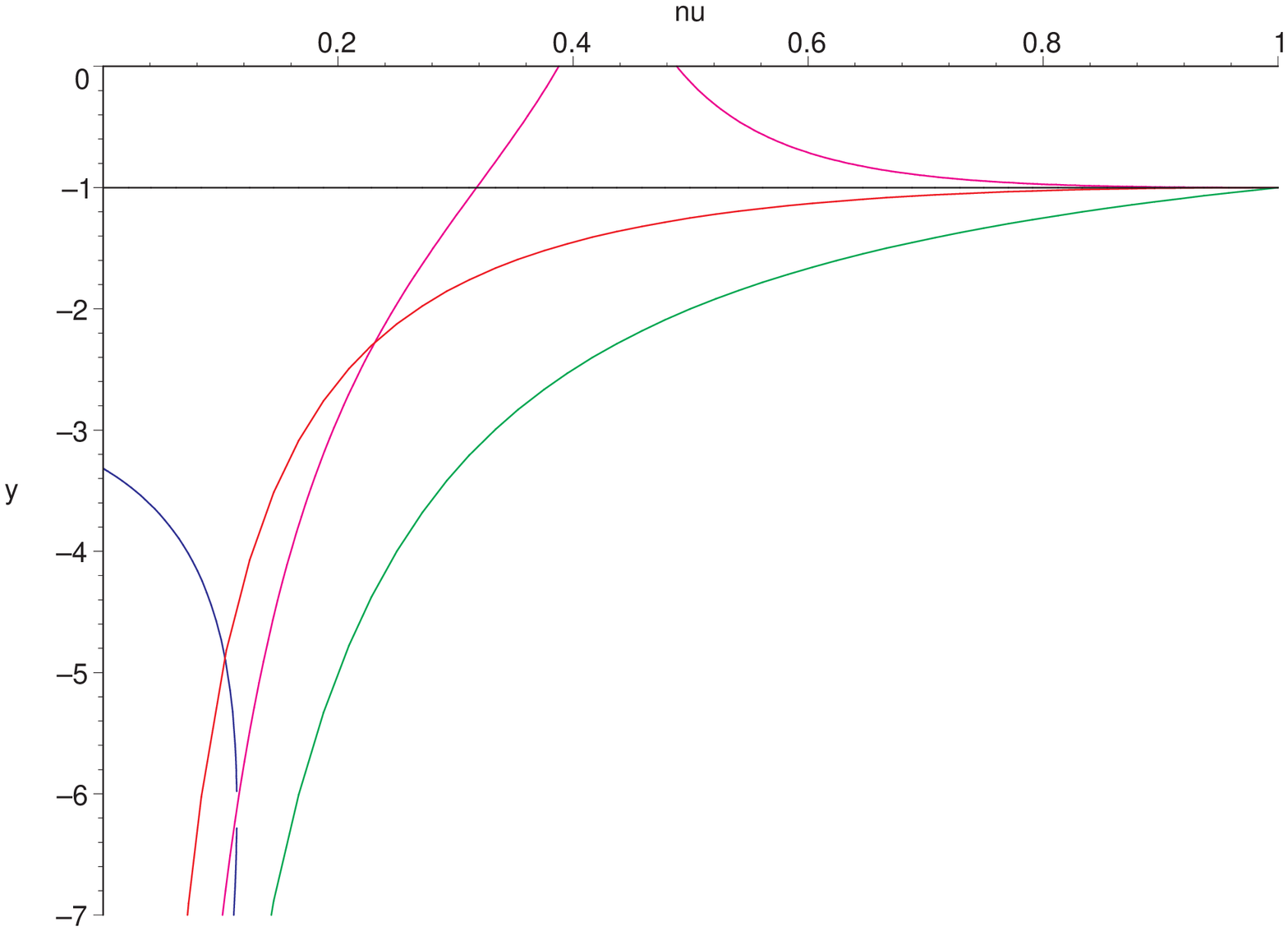}
\caption{In both the left and right hand graphs, the red and purple lines represent the solution curves for $\ddot y=0$ and the blue lines represent the solution curves for $\dot y=0$. The points where circular orbits exist are given by the points of intersection of these lines. The green line gives the position of the event horizon and the black line shows asymptotic infinity at $y=-1$. The left hand plot has $\Psi=2.92$ and the right hand plot has $\Psi=-1.64$. The constants in both plots have been set to $R=1$ and $E=1.20$.}
\label{fig:nullnu}
\end{center}
\end{figure}

The analysis of null geodesics is similar to the timelike case. The solutions to (\ref{circcond}) and (\ref{circcond2}) still give four valid solution curves, two from each equation respectively, with one of the non-trivial solutions to (\ref{circcond2}) giving unphysical positive $y$ solutions. Figure \ref{fig:nullnu} shows the equivalent plots to figure \ref{fig:timeynu} for the null geodesics.

In this figure it is immediately obvious where the curves intersect and thus the points where $\ddot y=\dot y=0$ for the null geodesics. The values of $E$ and $\Psi$ for these plots have been chosen so that the circular orbits exist for the black rings with $\nu=0.11385$, as in figure \ref{fig:timeynu}. This makes it easy to compare the position of the circular orbits for the null geodesics with the timelike ones. The main difference between the two figures is encapsulated by the blue curves in the various plots. The red and purple curves are identical in general for null and timelike geodesics because they represent the second order geodesic equations. The reason that the red and purple lines are slightly different between figures \ref{fig:timeynu} and \ref{fig:nullnu} is because the two figures use different values for $\Psi$, which does affect the second order equations and thus their solution curves.

Figure \ref{fig:nullnu} shows that the null circular orbits, that are off the ergosurface, are closer to the curvature singularity (at $y=-\infty$) than the respective timelike orbits, as one might expect. The difference is particularly pronounced for the negative angular momentum plots, shown on the right of figures \ref{fig:timeynu} and \ref{fig:nullnu}, where the circular orbit is outside the ergosurface for the timelike geodesics but inside for the null case. For any particular shape of black ring, given by fixing the value of $\nu$, the null circular orbits will always be closer to the curvature singularity of the ring than the respective timelike orbits.

The derivation for the relationship between the various conserved quantities is similar to that given for the timelike geodesics. For the null circular orbits, the critical value of $\Psi$ that allows the geodesics to form a circular orbit, is found by substituting $\epsilon=0$ into (\ref{circgeocond}) and solving for $\Psi$. To calculate the position of the circular orbit it is then necessary to substitute this value of $\Psi$ into either (\ref{circcond}) or (\ref{circcond2}) and then solve for $y$. Solving (\ref{circgeocond}) for null geodesics gives
\begin{equation}
\Psi=\pm\frac{RE}{\sqrt{1-\nu}}\left(2\sqrt{\nu}+\sqrt{\left(3\nu+1\pm 2\sqrt{2}\sqrt{\nu(1+\nu)}\right)\left(1+\nu\right)}\right)
\label{nullcircorb}
\end{equation}
As in the analysis of the timelike case, the intermediate roots have been discarded as they are non-physical. The two remaining solutions represent the circular orbits when the null geodesic is orbiting with and against the direction of rotation of the ring respectively. The positive $\Psi$ solution is always outside the ergosurface because of the frame dragging effect caused by the rotation of the ring. The negative solution can be either inside or outside the ergosurface, depending on the magnitude of $\Psi$.

The other null circular orbit is found where the blue curve in figure \ref{fig:nullnu} intersects the red $y=-\frac{1}{\lambda}$ curve. This is easily calculated by substituting $y=-\frac{1}{\nu}$ into (\ref{circcond}) and then solving for $\Psi$ in terms of $E$ and $\nu$. Doing this gives
\begin{equation}
\Psi=-\frac{ER(\nu^2+8\nu+3)}{2\sqrt{2(1-\nu^2)}}
\label{nullergocirc}
\end{equation}
thus allowing the position of the second circular orbit to be calculated as above.

As for the timelike geodesics, there will always be a null circular orbit on the ergosurface but unlike the timelike circular orbits, there will always be a second solution with angular momentum given by (\ref{nullcircorb}). This means that for null geodesics on circular orbits there will always be two possible circular orbits for particular values of $E$ and $\nu$. For the timelike case there will sometimes only be one solution for particular values of $E$ and $\nu$, specifically when $E<1$.

There is one specific instance where the null geodesics can only form one circular orbit for given values of $E$ and $\nu$. This is when the values of (\ref{nullergocirc}) and (\ref{nullcircorb}) are the same. For the null geodesics $E$ and $R$ are only scaling constants, so the angular momenta will only be degenerate for a particular value of $\nu$. Equating (\ref{nullergocirc}) and (\ref{nullcircorb}) shows that a thin ring with $\nu=0.04042$ will have both of the null circular orbits on the ergosurface.

\newsection{Geodesics Orbiting through the Ring}
\label{ch:ringgeos}

\begin{figure}[htbp]
\begin{center}
\includegraphics[viewport=0 5 420 540,width=6.8cm,angle=270,clip]{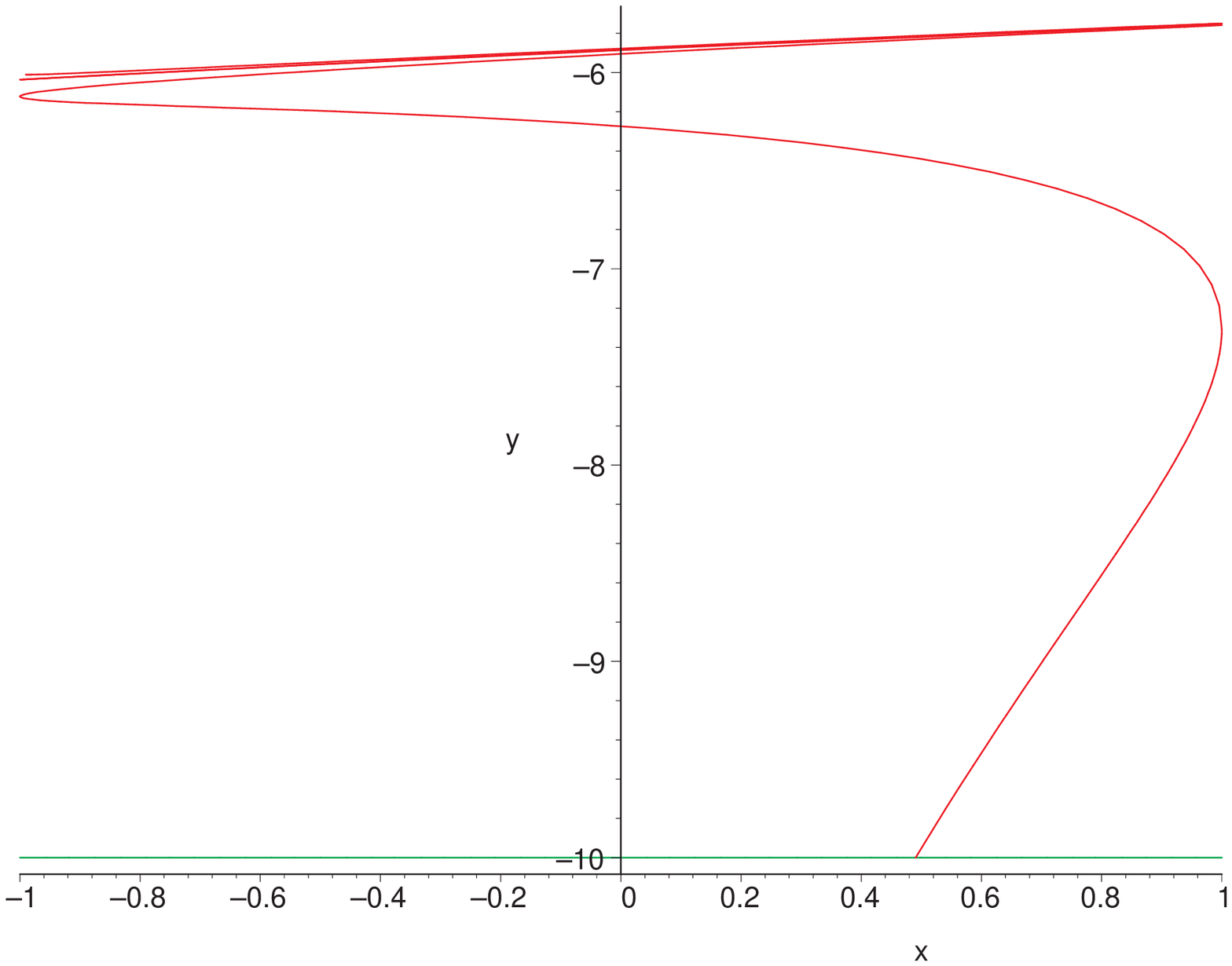}
\includegraphics[viewport=0 0 420 563,width=6.8cm,angle=270,clip]{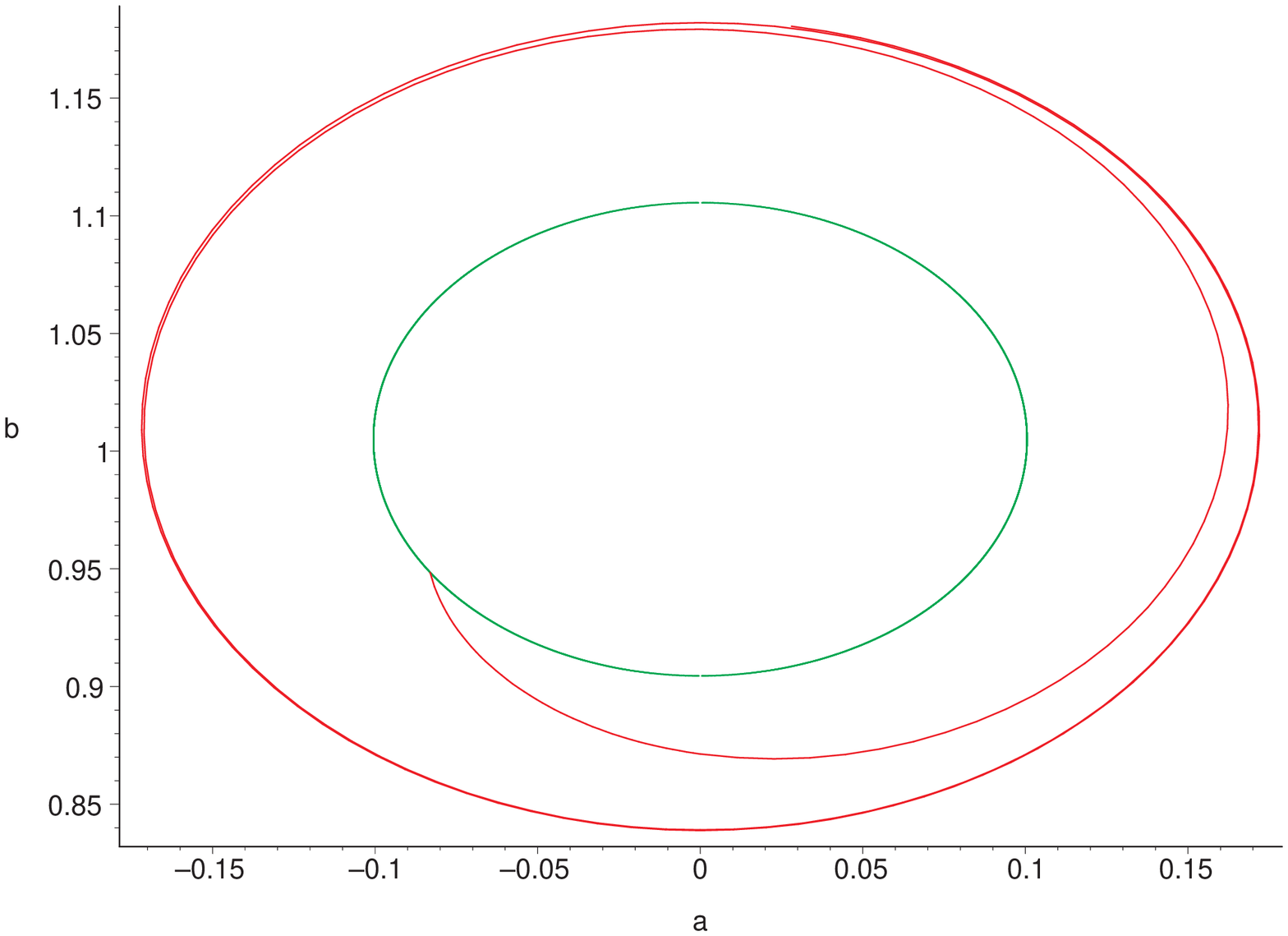}
\caption{These plots show the numerical integration of the equations of motion for a timelike geodesic started at $(x,\dot x,y,\dot y)=(-0.99000,3.03072,-6.01041,0)$, with $E=2$ and $\ell=\Psi=0$. The Black Ring has radius $R=1$ and $\nu=0.1$. The left hand plot shows the orbit of the particle in the toroidal coordinates, with the right hand plot showing the orbit in polar coordinates. The green lines indicate the position of the event horizon.}
\label{fig:BRorbit}
\end{center}
\end{figure}

For geodesics that orbit through the ring at constant $y$, equation (\ref{yELeqn}) becomes
\begin{eqnarray}
\frac{R^2F(y)^2F(x)\ddot y}{G(y)(x-y)^2}-\frac{(x-y)^2\left[ECR(1+y)+\Psi F(y)\right]^2}{2R^2F(x)G(y)}\left[\frac{F(y)G'(y)}{G(y)}+\lambda\right]+\frac{\epsilon F(y)^2}{x-y} \nonumber \\
-\frac{E^2F(x)\left[\lambda(x-y)-2F(y)\right]}{2(x-y)} +\frac{E(x-y)^2C(1-\lambda)\left[RCE(1+y)+\Psi F(y)\right]}{RF(x)G(y)} =0
\label{constyorbs}
\end{eqnarray}
where $\dot y\rightarrow 0$ and $\dot x$ has been eliminated using (\ref{firstint}). To find possible solutions to this equation it is necessary to look for specific values of the constants $\nu,\Psi,\epsilon,y,E$ that cause all of the terms not involving $\ddot y$ to go to zero. In practice this means expanding all of the terms to give a polynomial in $x$, since $x$ is free to vary while $y$ is constrained to be a constant throughout all the motion.

In order to get a feel for the equations without having to look for general solutions it is helpful to look at the special case where $[ECR(1+y)+\Psi F(y)]=0$. This is possible in this case because $y$ is being treated as a constant and all of the other terms are constants. This means that $\Psi$ can be chosen so that
\begin{equation}
\Psi=-\frac{ECR(1+y)}{F(y)}
\end{equation}
Applying this to (\ref{constyorbs}) reduces the equation to
\begin{equation}
\frac{R^2F(y)^2F(x)\ddot y}{G(y)(x-y)^2}=-\frac{\epsilon F(y)^2}{x-y} +\frac{E^2F(x)\left[\lambda(x-y)-2F(y)\right]}{2(x-y)}
\end{equation}
It can be seen straight away that the only way that the terms on the right hand side can be set to zero is by choosing $E=\epsilon=0$, which is the same constraint as was imposed in order to separate the equations of motion in section \ref{ch:geoequs}.

If the constants aren't constrained in any way (other than the physical constraints) then (\ref{constyorbs}) becomes an eighth order polynomial in $x$. This unfortunately doesn't have any solutions for physically applicable values for $\nu,\Psi,\epsilon,y,E$.

Figure \ref{fig:BRorbit} gives an example of a timelike geodesic in the exterior of the black ring, in a reference frame which is rotating in the $\psi$ direction with the particle. The particle's initial angular momentum is carefully chosen so that it doesn't fall straight into the black hole, but it does eventually spiral into the ring when the integration is continued. It is possible to keep fine tuning the initial velocity, so that the particle stays out of the black ring longer but in the end, the particle will either spiral into the event horizon, or escape to asymptotic infinity at $y=-1$.

The right hand plot in figure \ref{fig:BRorbit} converts the orbit into polar coordinates, given in (\ref{rtrans}) and (\ref{thetatrans}), and then plots it using
\begin{eqnarray}
a & = &r\cos\theta \\
b & = &r\sin\theta
\end{eqnarray}
This plot gives a more intuitive picture of what is happening to the particle. As one might expect, the particle initially appears to be in a stable orbit but, after approximately two revolutions, the orbit starts to decay and then rapidly falls through the event horizon. If the initial angular momentum is fine-tuned further, then it is possible to have the particle orbit the ring for a significantly longer period with the radius varying as it orbits. Unfortunately, the orbit always seems to decay eventually.

The orbit shown in the right hand plot of figure \ref{fig:BRorbit} appears to be circular but closer inspection shows that it is slightly elliptical. The eccentricity of the orbit increases as the energy is reduced until $E\sim 0.8$ where it is no longer possible to find a bound orbit. It would appear from the numerical simulations that bound orbits can be found for all values of $E$ greater than $0.8$ though.

\begin{figure}[htbp]
\begin{center}
\includegraphics[viewport=0 0 420 520,width=6.8cm,angle=270,clip]{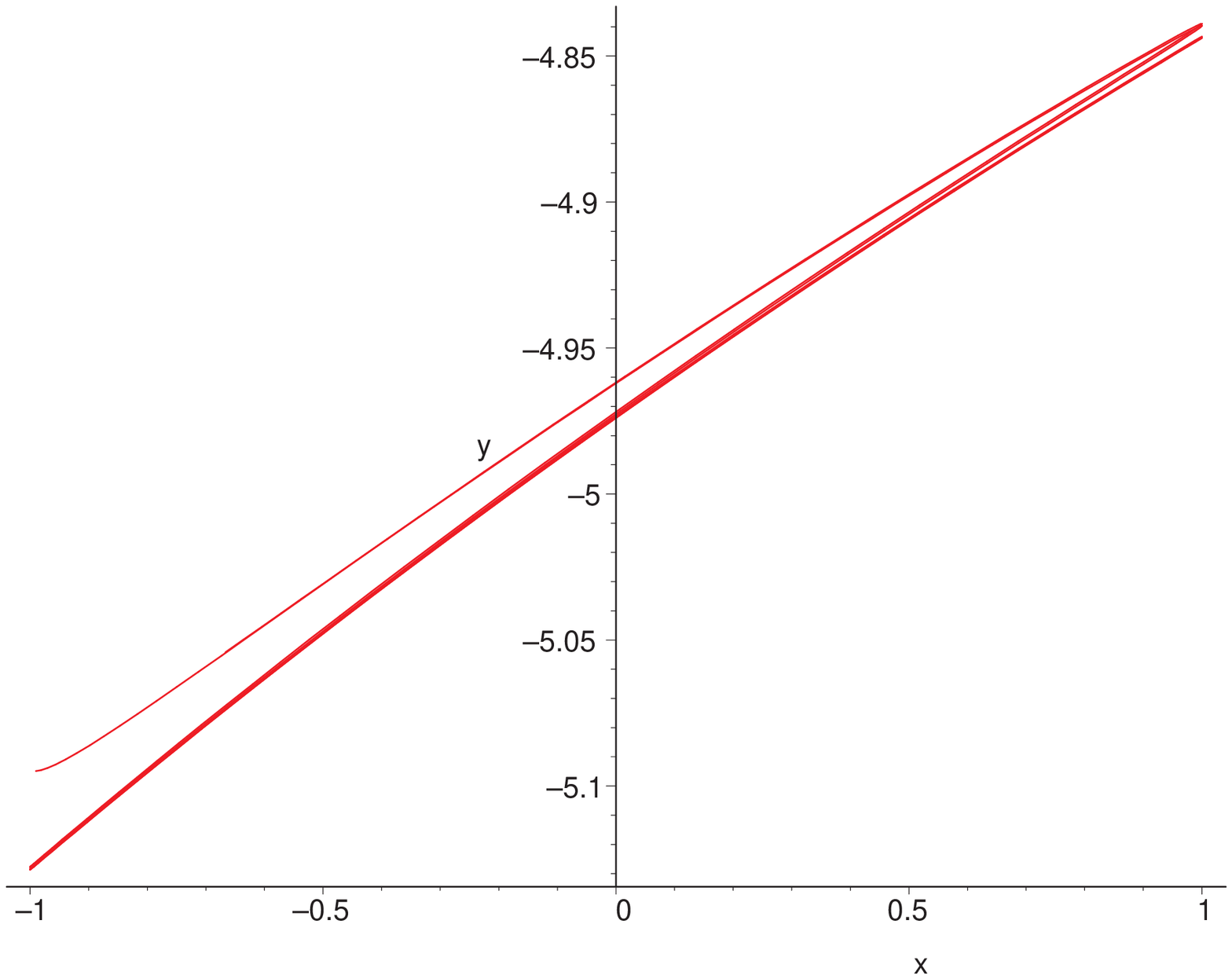}
\includegraphics[viewport=0 0 420 520,width=6.8cm,angle=270,clip]{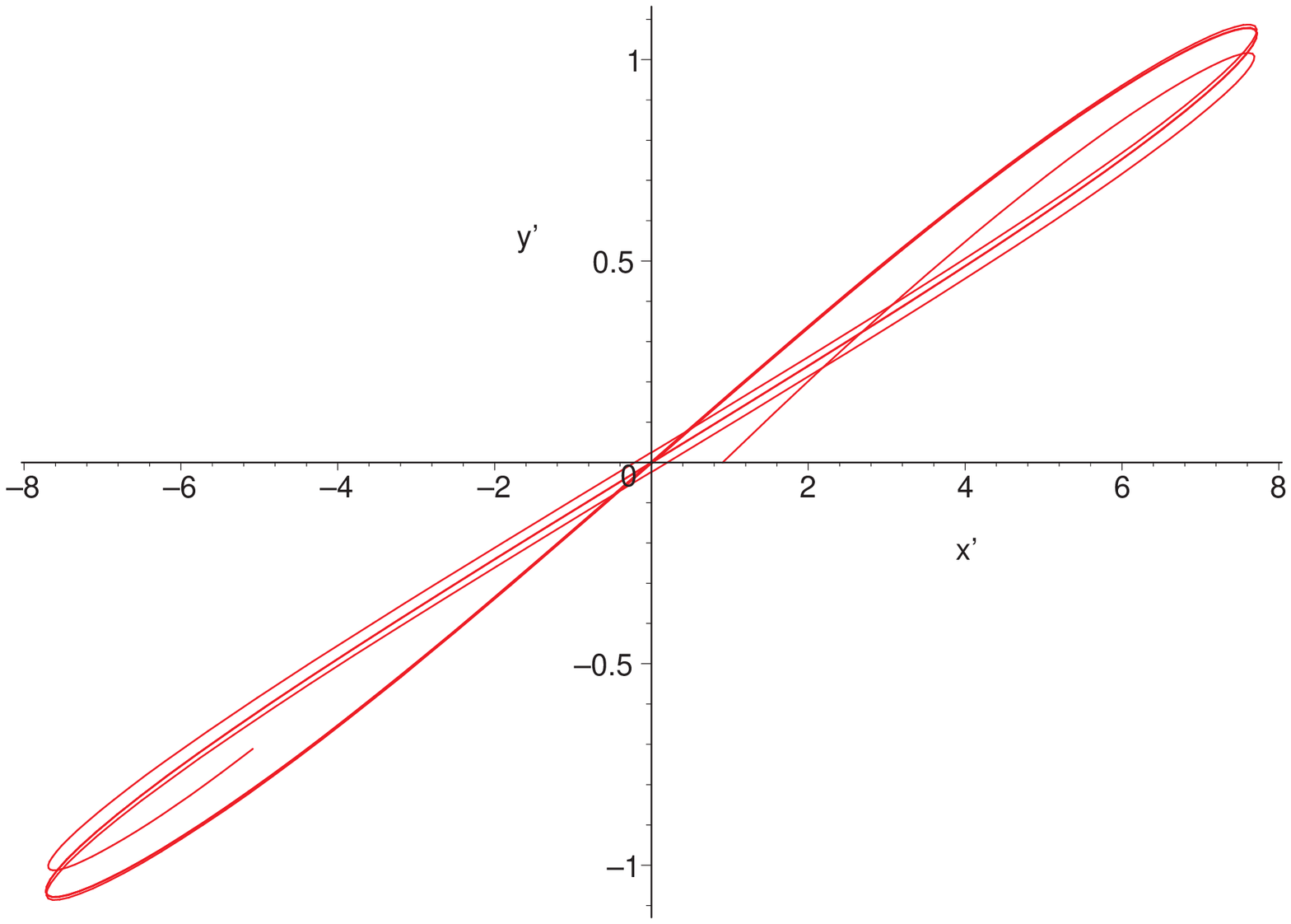}
\caption{These plots show the numerical integration of the equations of motion for a timelike geodesic in the toroidal coordinates. The initial conditions are set to $(x,y,\dot y)=(-0.9900000,-5.0948494,0)$, with $\ell=\Psi=0$, for $R=1$ and $\nu=0.1$. The other initial conditions are $(E,\dot x)=(1,0.9085145)$.}
\label{fig:stabBRorbit}
\end{center}
\end{figure}

If the starting value of $y$ is fine tuned further, then the orbit of the particle looks something like that given in figure \ref{fig:stabBRorbit}. The two plots in this figure show how $y$ varies with $x$ and how $\dot y$ varies with $\dot x$. The left hand plot gives a more detailed view of the periodic motion as the particle orbits in a relatively stable ellipse. Apart from the first revolution, before the particle falls into the stable orbit, it would appear that the particle is moving back and forth along a line with $y$ depending linearly on $x$, although this isn't quite true, since the trace is slightly curved.

The right hand plot gives the phase curve for the motion of the particle. The particle slows down at $x = \pm 1$, with both $\dot y$ and $\dot x$ going to zero at these two points. The plot also shows how the motion is fairly consistent with the curves being well grouped, apart from at the beginning and near the end of the plotted motion, where the curve starts to diverge from the bow-tie shape. It is tempting to conclude from this shape that the motion in the $x$ and $y$ directions is given by some trigonometric function but this unfortunately doesn't appear to be so. The only solvable case (given by substituting $E=0$ and $\epsilon = 0$ into equation (\ref{firstint})) has solutions in terms of elliptic functions, so it seems reasonable to assume that the solutions for the more complicated motion would also be in terms of these elliptic functions.

\newsection{Pseudo Radial Geodesics}
\label{ch:radgeos}

It was mentioned in the introduction that the Kerr metric has some ``pseudo'' radial geodesics, where the azimuthal angle remains constant throughout the motion, so that the geodesic only moves in the $(r,\phi)$ plane. The analogue of this for the Black Ring metric would be to find geodesics that move along lines of constant $x$, as shown in figure \ref{fig:xyXsec}.

As in the Kerr metric, it is impossible to have purely radial geodesics for the Black Ring because there is an analogous frame dragging effect in the Black Ring metric. Combining (\ref{ppsiequ}) and (\ref{ptequ}) gives an expression for $\dot\psi$ in terms of the constants of motion
\begin{equation}
\dot \psi=\frac{(x-y)^2[ECR(1+y)-\Psi F(y)]}{R^2F(x)G(y)}
\end{equation}
In order for $\dot \psi$ to be zero for all values of $y$
\begin{equation}
ECR(1+y)-\Psi F(y)=0
\end{equation}
Expanding this, and collecting in terms of $y$, gives
\begin{equation}
ECR-\Psi+(ECR-\Psi\lambda) y=0
\end{equation}
In order for this to hold for all values of $y$, it would require
\begin{eqnarray}
ECR & = &\Psi \\
ECR & = &\Psi\lambda
\end{eqnarray}
These two equations can only simultaneously be true if $\lambda=1$, in which case the event horizon reduces to a three-sphere, which is a rather trivial solution. This indicates that there are no radial geodesics for the Black Ring metric, where $\dot\psi=0$.

If $\psi$ is allowed to vary throughout the motion, then the situation becomes the opposite of that investigated in the ``Geodesics Orbiting through the Ring'' section. In this case only $x$ is held constant, which means that the equation of motion (\ref{xELeqn}) reduces to
\begin{equation}
\frac{\ell^2(x-y)^3}{2R^2}\left[\frac{G'(x)}{G(x)}-\frac{\lambda}{F(x)}\right]-E^2G(x)-\frac{\epsilon G(x)\left[F(x)+F(y)\right]}{2F(x)G(y)}=0
\label{constxeqn}
\end{equation}
where (\ref{firstint}) has been used to eliminate the $\dot y$ dependence.

Examining the solutions to equation (\ref{constxeqn}) indicates that there are only physically consistent solutions for $x=\pm 1$. This class of solutions has already been examined in some detail in the Planar Circular Geodesics section.

\newsection{Conclusions}

In this paper the equations of motion for the geodesics of the neutral rotating Black Ring metric were set up and numerically integrated for some special classes of solutions. The solutions can be broadly separated into those that are confined to the axis of rotation and those that gave circular orbits in the plane of the ring. It was also shown that there are no circular geodesics that orbit through the ring or any ``pseudo-radial'' geodesics. Although it was shown that there aren't any circular orbits through the ring, some numerical evidence was presented that bound orbits of this form may exist.

The effective potential for the on-axis solutions is very similar to that for a static black hole, with the potential being attractive for the geodesics with zero angular momentum. In this case both the null and timelike particles can pass through the origin of the ring and out to infinity, or in the case of the timelike geodesics, oscillate back and forth. This agrees with the analogous Newtonian case of a massive ring when a small test particle is placed on the axis of symmetry and then displaced slightly.

Increasing the angular momentum of the geodesics causes a centrifugal barrier to appear which stops the geodesics from approaching the origin of the ring, as in the case of the Schwarzschild black hole. Even though timelike and null particles can't reach the origin it is still possible for them to pass through the centre of the ring. This is because the Black Ring is five dimensional, so the axis of rotation is actually a plane, which means the particles can go from one side of the ring to the other without passing through $y=-1$ and $x=+1$. The particle motion in the $x$-$\phi$ plane is similar to a small asteroid moving in the Sun's gravitational field. The particle can either be captured and orbit indefinitely, or it can escape to infinity.

The timelike potential has a local minimum, which allows for a rich array of geodesic motion because it is possible to have a geodesic that is in a stable orbit in the $x$-$\phi$ plane near to the centre of the ring. The shape of the potential well is unsymmetric so the orbit is always elliptical with the period of the orbit depending on the initial radius: the larger the radius, the longer the period of oscillation.

The effective potential for the null geodesics is very similar to that for the timelike ones when the angular momentum is zero, but once the angular momentum is increased the potential becomes totally repulsive for small $\nu$. If $\nu$ is large enough then it is possible to create a small local minimum for values of $\nu>0.653$. This potential is interesting, since it means that it is possible for the Black Ring to have light rays in stable orbits circling through it. If $\nu$ is decreased, then the null geodesic will always go off to infinity, no matter what the angular momenta of the geodesic is.

In the case of the planar circular orbits, the angular momentum in the $\phi$ direction has to be zero in order for them to remain on the plane through the centre of the ring. This means that the geodesics are confined to move in only one spatial dimension. For timelike geodesics it is only possible to have a constant circular orbit on the ergosurface at $y=-\frac{1}{\lambda}$. This requires the energy and angular momentum in the $\psi$ direction to be carefully chosen though. Also, this orbit only exists in the outer equatorial plane. It is impossible to have any circular orbits in the interior of the ring.

For null geodesics there is always at least one solution for $\psi$ that will give a circular orbit for all values of $\nu$. If the angular momentum in the $\psi$ direction is chosen to go against the rotation of the ring, it is also possible to have two static orbits for the same value of $\nu$. These circular orbits do require a certain amount of tuning because for small values of $\Psi$ it is impossible to have any circular orbits, no matter what the shape and size of the ring.

The possibility of the Black Ring metric having geodesics that orbit through the ring at constant $y$ and radial geodesics of constant $x$ was also examined but it was shown that these cannot occur, at least not for these particular toroidal coordinates, where orbits of constant $y$ describe circles. The numerical evidence suggests that there may be elliptical orbits through the ring for at least one value of $y$, but the lack of separability of the equations of motion means that it is impossible to interpret these orbits quantitatively.

It would be interesting to investigate these orbits more thoroughly, to see if the motion of the geodesics reveals any underlying properties of the Black Ring metric that have been thus far overlooked. A more systematic way of doing this might be to look for regions of the space where the geodesics are bounded, by numerically integrating the fully specified equations of motion for varying initial positions. The regions of space close to the $x=\pm 1$ planes and the $y=-1$ axis have properties similar to the results presented here, so this may provide a way of estimating values for the conserved momenta that could give bounded geodesics for some points.

Another related avenue of investigation has also recently opened up due to the discovery of a more general solution for the Black Ring \cite{cite:PomeranskySenkov}, where the ring rotates in both the $\psi$ and $\phi$ directions. It would be interesting to examine the geodesics in this metric to see if the extremal situation where both of the ring's horizons are degenerate will allow the equations of motion to be separated. This may then give some insight into the properties of the singularly rotating Black Ring.

\newsection{Acknowledgements}

I would like to thank Mukund Rangamani for some valuable insights and advice during the preparation of this work. I would also like to thank Edward Teo and Yen Kheng for pointing out an error in an earlier draft of this paper. This research was supported by PPARC grant no. PPA/S/S/2004/03814 and by the University of Durham.

\newsection{Appendix A: Analysis of the Singular Terms in the Equations of Motion}

In certain situations the equations of motion given by (\ref{tequ}) - (\ref{psiequ}) and (\ref{xELeqn}) - (\ref{yELeqn}) break down, such as when $y=-1$ is substituted into (\ref{yELeqn}). In cases such as this, certain terms become indeterminate in the ${x,y}$ coordinates. Once these terms are isolated, they can be analysed by transforming to spherical polar coordinates, as given in (\ref{rtrans}) and (\ref{thetatrans}). In the case of (\ref{yELeqn}) the singularities occur in
\begin{equation}
\frac{\Psi^2}{G(y)} \hspace{2cm} \mbox{and} \hspace{2cm} \frac{\dot y^2}{G(y)}
\end{equation}
In section \ref{ch:axisgeos} it was stated that so long as $\Psi=0$ the $G(y)$ term would not blow up. This is more evident if $G(y)$ is converted into spherical polar coordinates. Doing this gives
\begin{equation}
G(y)=-\frac{4R^2r^2\sin^2{\theta}[P-\nu(R^2+r^2)]}{P^3}
\label{Gytrans}
\end{equation}
where $P=\sqrt{r^4+2R^2r^2\cos{2\theta}+R^4}$.

The rotational axis, given by $y=-1$ is equivalent to $\theta=0$, so it is obvious that $G(y)^{-1}\rightarrow \infty$ as $\theta\rightarrow 0$ because of the $\sin^2 \theta$ term. Fortunately, the $\Psi^2$ term will cancel out the $\sin^2\theta$ term if it is initially chosen to be zero.

Transforming the other problematic term into spherical polar coordinates gives
\begin{equation}
\frac{\dot y^2}{G(y)}=-\frac{4R^2[(r^2-R^2)\dot r\sin\theta - (r^2+R^2)r\dot\theta\cos{\theta}]^2}{P^3[P-\nu(R^2+r^2)]}
\end{equation}
Taking the limit as $\theta\rightarrow 0$ gives
\begin{equation}
\lim_{\theta\rightarrow 0}\frac{\dot y^2}{G(y)}=-\frac{4R^2r^2\dot\theta^2}{(R^2+r^2)^2(1-\nu)}
\label{ydsing}
\end{equation}
In this form it is obvious that this term is not singular, as $R^2$ is always positive and $\nu<1$.

There are similar problems with equation (\ref{xELeqn}) when the geodesics on the equatorial plane are to be considered. The terms in question are
\begin{equation}
\frac{\dot x^2}{G(x)} \hspace{2cm} \mbox{and} \hspace{2cm} \frac{\ell^2}{G(x)}
\label{xprobs}
\end{equation}
Using the same process as above the terms can be transformed as follows
\begin{equation}
\frac{\dot x^2}{G(x)}=\frac{4R^2[r\dot\theta\sin\theta(R^2-r^2)-\dot r\cos\theta(R^2+r^2)]^2}{P^3[P+\nu(R^2-r^2)]}
\label{xdotsin}
\end{equation}
The equatorial plane corresponds to $\theta=\frac{\pi}{2}$, so taking the limit gives
\begin{equation}
\lim_{\theta\rightarrow\frac{\pi}{2}}\frac{\dot x^2}{G(x)}=\frac{4R^2r^2\dot\theta^2}{(R^2-r^2)^2(1+\nu)}
\end{equation}
This term is very similar to (\ref{ydsing}). In this case the denominator is always positive because the $(R^2-r^2)^2$ term is always positive.

Transforming the other term in (\ref{xprobs}) gives
\begin{equation}
\frac{\ell^2}{G(x)}=\frac{S^3\ell^2}{4R^2r^2\cos^2{\theta}[P+\nu(R^2-r^2)]}
\end{equation}
This will obviously blow up for $\theta=\frac{\pi}{2}$ unless $\ell=0$. In a similar manner to the geodesics on the rotational axis, $\ell$ has to be zero for geodesics on the equatorial plane, since the $\phi$ coordinate is measured with respect to the $x=\pm 1$ axis.

Equation (\ref{xdotsin}) also causes a problem when calculating geodesics that pass through the origin. In the toroidal coordinates, the origin corresponds to $x=1$ and $y=-1$, unfortunately the transformation given in (\ref{thetatrans}) becomes undefined. This means that a different coordinate system will have to be used to remove the singularity at this point. A good candidate is Cartesian coordinates.

The transformations between the Cartesian coordinates and the toroidal coordinates are given by
\begin{eqnarray}
z_0 &= & \pm \frac{R\sqrt{1-x^2}}{y-x} \\
z_1 &= & \pm \frac{R\sqrt{y^2-1}}{y-x}
\end{eqnarray}
where $z_0$ and $z_1$ are the Cartesian coordinates on the $(x,y)$ plane. In these coordinates
\begin{eqnarray}
\dot x & = & -\frac{4R^2z_0[R^2{\dot z}_0+2z_0z_1{\dot z}_1+{z_0}^2{\dot z}_0 - {z_1}^2{\dot z}_0]}{Q^3} \\
\dot y & = & -\frac{4R^2z_1[R^2{\dot z}_1-2z_0z_1{\dot z}_0-{z_1}^2{\dot z}_1 + {z_0}^2{\dot z}_1]}{Q^3} \\
G(x) & =& \hspace{0.33cm} \frac{4R^2{z_0}^2\left(\nu R^2-\nu{z_0}^2-\nu{z_1}^2+Q\right)}{Q^3} \\
G(y) & =& \hspace{0.33cm} \frac{4R^2{z_1}^2\left(\nu R^2+\nu{z_0}^2+\nu{z_1}^2-Q\right)}{Q^3}
\end{eqnarray}
where $Q=\sqrt{[(z_1-R)^2+z_0^2][(z_1+R)^2+z_0^2]}$. Expressing the terms that become singular at $x=1$ and $y=-1$ in Cartesian coordinates gives
\begin{eqnarray}
\frac{\dot x^2}{G(x)} &= & \frac{4R^2[R^2{\dot z}_0+2z_0z_1{\dot z}_1+{z_0}^2{\dot z}_0 - {z_1}^2{\dot z}_0]^2}{\left(\nu R^2-\nu{z_0}^2-\nu{z_1}^2+Q\right)Q^3} \\
\frac{\dot y^2}{G(y)} &= & \frac{4R^2[R^2{\dot z}_1-2z_0z_1{\dot z}_0-{z_1}^2{\dot z}_1 + {z_0}^2{\dot z}_1]^2}{\left(\nu R^2+\nu{z_0}^2+\nu{z_1}^2-Q\right)Q^3}
\end{eqnarray}
In Cartesian coordinates the origin is at $z_0=z_1=0$, so substituting these values into the above equations gives
\begin{eqnarray}
\lim_{z_0\rightarrow 0}{\left[\lim_{z_1\rightarrow 0}{\frac{\dot x^2}{G(x)}}\right]}=\frac{4{{\dot z}_0}^2}{R^2(\nu +1)} \\
\lim_{z_0\rightarrow 0}{\left[\lim_{z_1\rightarrow 0}{\frac{\dot y^2}{G(y)}}\right]}=\frac{4{{\dot z}_1}^2}{R^2(\nu -1)}
\end{eqnarray}
It is now manifest that these terms are non-singular at the origin and are dependent on ${\dot z}_0$ and ${\dot z}_1$ respectively.

\newsection{Appendix B: Effective Potential on the Equatorial Plane}

For geodesics confined to the equatorial planes, given by $x=\pm 1$, $\dot y^2$ can be calculated from the first integral equation. Substituting $\dot x=0$ and $\ell=0$ in (\ref{firstint}) gives
\begin{equation}
-\frac{R^2F(x)\dot y^2}{G(y)(x-y)^2} - \frac{E^2F(x)}{F(y)}-\frac{(x-y)^2\left[RE(1+y)C+\Psi F(y)\right]^2}{F(x)F(y)R^2G(y)}=\epsilon
\label{xeffpot}
\end{equation}
In principle, the effective potential can now be calculated but there is a problem caused by the $F(y)$ terms in the denominator. These terms become singular when $y=-\frac{1}{\lambda}$ so a coordinate transformation is required to make sure that the effective potential is continuous across the ergosurface. The transformation
\begin{equation}
z=-\tanh^{-1}\left(\frac{1+\lambda y}{y+\lambda}\right)
\end{equation}
is continuous when $y\rightarrow -\frac{1}{\lambda}$ and approaches infinity as $y\rightarrow -1$.

Expressing (\ref{xeffpot}) in terms of $z$ gives
\begin{equation}
\dot z^2=KE^2+LE+M
\label{E-Vequ}
\end{equation}
where
\begin{eqnarray}
K & = & -\frac{(-\lambda-\tanh z +\nu+\nu\lambda\tanh z)(x\lambda+x\tanh z +1+\lambda\tanh z )^2}{R^2\tanh z(1-\lambda^2)^2(1-\tanh^2 z)} \nonumber \\
&& -\frac{(x\lambda+x\tanh z +1+\lambda\tanh z )^4C^2(\lambda-1)^2}{R^2\tanh z F(x)^2(1-\lambda^2)^3(1+\tanh z)^2(\lambda+\tanh z )} \\ \nonumber \\
L & = & -\frac{2(x\lambda+x\tanh z+1+\lambda\tanh z)^4\Psi C(\lambda-1)(1-\tanh(z))}{R^3F(x)^2(1-\lambda^2)^2(1-\tanh^2 z)^2(\lambda+\tanh z)} \\ \nonumber \\
M & = & -\frac{(-\lambda-\tanh z+\nu+\nu\lambda\tanh z)(x\lambda+x\tanh z+1+\lambda\tanh z)^2\epsilon}{(1-\lambda^2)(1-\tanh^2 z)(\lambda+\tanh z)R^2F(x)} \nonumber \\
&& - \frac{(x\lambda+x\tanh z+1+\lambda\tanh z)^4\tanh z\Psi^2}{R^4F(x)^2(1-\lambda^2)(1-\tanh^2 z)^2(\lambda+\tanh z)}
\end{eqnarray}
The effective potentials for these planar geodesics can be calculated in a similar way to those at the beginning of section \ref{ch:axisgeos}, so solving (\ref{E-Vequ}) for $E$ when $\dot z=0$ gives
\begin{equation}
V_\pm = \frac{-L\pm\sqrt{L^2-4KM}}{2K}
\end{equation}
Technically both effective potentials need to be considered, since the $L$ term is not equal to zero, as was the case for the on axis geodesics considered in section 4, but in practice it is usually possible to consider only $V_+$, since $V_-$ is usually negative for all values of $z$. However, if $\Psi<0$, $V_-$ is positive for some values of $z$, in which case $V_+$ and the portion of $V_-$ that is positive will be considered as the effective potential function.

\providecommand{\href}[2]{#2}\begingroup\raggedright\endgroup

\end{document}